\documentclass[12pt]{article}
\usepackage[margin=1in]{geometry}
\usepackage{siunitx}
\usepackage{amssymb}
\usepackage{amsmath}
\usepackage{amsthm}
\usepackage{latexsym}
\usepackage{amsfonts}
\usepackage{graphicx,lscape,fancybox,shapepar,titletoc}
\usepackage{makecell}  
\usepackage{xstring}   
\usepackage{setspace}
\usepackage{fvextra}
\usepackage{booktabs}

\usepackage{newtxtt}
\usepackage{url}
\usepackage{fullpage}
\usepackage{epstopdf}
\RequirePackage{xcolor}
\definecolor{CardinalRed}{cmyk}{0,1,0.65,0.34}
\usepackage[colorlinks=true, anchorcolor=CardinalRed, citecolor=CardinalRed,linkcolor=CardinalRed, breaklinks=true, urlcolor = CardinalRed]{hyperref}
\usepackage{tikz}
\usepackage{natbib}
\usepackage{array}
\usepackage{todonotes}
\usepackage{xr}
\usepackage{caption}
\usepackage{subcaption}
\usepackage{lscape}
\usepackage[inline, shortlabels]{enumitem}
\usepackage{rotating}
\usepackage{longtable}
\usepackage{placeins}

\usepackage{titling}
\usepackage[flushmargin]{footmisc}
\RequirePackage[online]{threeparttable}

\newcolumntype{H}{>{\setbox0=\hbox\bgroup}c<{\egroup}@{}}
\newcolumntype{L}[1]{>{\raggedright\arraybackslash}p{#1}}

\usepackage{listings}

\bibpunct{(}{)}{;}{a}{}{,}
\usetikzlibrary{arrows,positioning}
\tikzset{
    >=stealth',
    rec_a/.style={
           text width=11.5em,
           minimum height=3em,
           text centered},
    rec_b/.style={
           text width=9.5em,
           minimum height=2em,
           text centered},
    rec_c/.style={
           text width=5.5em,
           minimum height=2em,
           text centered},
    filled/.style={
            fill=circle area,
            draw=circle edge,
            thick},
    filled2/.style={
            fill=circle area2,
            draw=circle edge2,
            thick},
     outline/.style={
            draw=circle edge,
            thick},
     outline2/.style={
            draw=circle edge2}
}

\colorlet{circle edge}{black!70}
\colorlet{circle area}{black!10}
\colorlet{circle area2}{white}
\colorlet{circle edge2}{white}

\def\Put(#1,#2)#3{\leavevmode\makebox(0,0){\put(#1,#2){#3}}}

\newcommand{\tabnote}[1]{\\[3pt]\begin{minipage}{\linewidth}\originalfootnotesize\emph{Note:} #1\end{minipage}}
\newcommand{\fignote}[1]{\par\smallskip\begin{minipage}{0.92\linewidth}\originalfootnotesize\emph{Note:} #1\end{minipage}}

\let\originalfootnotesize\footnotesize
\renewcommand{\footnotesize}{\normalsize}
\title{\bf The Credibility Revolution in Political Science
\thanks{\small Carolina Torreblanca, PDRI-DevLab Postdoctoral Fellow, Department of Political Science, University of Pennsylvania. Email: \href{mailto:catba@sas.upenn.edu}{catba@sas.upenn.edu}.  William Dinneen, Data Scientist and JD at Stanford Law School. Email: \href{mailto:willdinneen@gmail.com}{willdinneen@gmail.com}. Guy Grossman, David M. Knott Professor of Global Politics, Department of Political Science and PDRI-DevLab, University of Pennsylvania, \emph{corresponding author}. Email: \href{mailto:ggros@upenn.edu}{ggros@upenn.edu}. Yiqing Xu, Associate Professor, Department of Political Science, Stanford University. Email: \href{mailto:yiqingxu@stanford.edu}{yiqingxu@stanford.edu}. We thank Yunfei Chen, Ziyi Chen, Honaminto E. Djohi, Chunxuan Ren, Biman Yang, Jiayi Zou, Boran Zhang, and He Zhao for excellent research assistance. We acknowledge the use of AI assistance (Anthropic’s Claude Opus 4.8 and Claude Fable) for copy editing, proofreading, developing and testing prompts, and code review. The online appendix documents extensively the use of AI in the article classification process.}}

\author{Carolina Torreblanca\\(UPenn) \and William Dinneen\\(Stanford) \and Guy Grossman\\(UPenn) \and  Yiqing Xu\\(Stanford)}

\date{\bigskip\today}

\begin{document}
  \maketitle
\thispagestyle{empty}
\setstretch{1}
\vspace{-2em}\begin{abstract}
\noindent How far has the credibility revolution spread through political science? We address this question by classifying 91,632 articles published between 2003 and 2023 across 156 political science journals using large language models, focusing on research design, credibility-enhancing practices, and citation patterns. We find that design-based studies, those leveraging plausibly exogenous variation to justify causal claims, have become increasingly common. In contrast, model-based approaches that rely more heavily on modeling assumptions have declined. Yet the rise of design-based work is uneven: it is concentrated in top journals and among authors at highly ranked institutions, and it is driven primarily by the growth of survey experiments. Other credibility-enhancing practices that help reduce false positives and false negatives, such as placebo tests and power calculations, remain rare. Taken together, our findings point to substantial but selective change, more consistent with a partial reform than a revolution.
\end{abstract}

\newpage
\setstretch{1.75}
\setcounter{page}{1}

\section*{Introduction}\label{sec:intro}

Over the last two decades, political science has been influenced by what many describe as a credibility revolution: an intellectual movement advocating a change in how researchers conceptualize and conduct causal empirical scholarship \citep{angrist2010credibility, dunning2012natural, gerber2012field, morgan2014counterfactuals, samii2016, aronow2019foundations, blair2023research, SloughTyson2024, imbens2025comparing}. In this paper, we examine whether and to what extent political science has moved in the direction the credibility revolution charted.

As an epistemic movement that seeks to change what researchers view as credible evidence when assessing causal claims, the credibility revolution carries transformative potential. Its intellectual core is a reconceptualization of causality. This reconceptualization encourages a shift away from model-based identification strategies, which often rely on strong functional-form assumptions and implicit views of treatment assignment, toward design-based approaches that exploit features of treatment assignment and rest on explicit identification assumptions. This movement views credibility not only as a consequence of adopting particular study designs, but also as requiring clearer statements of identification assumptions and supporting evidence through practices such as pre-analysis plans, power analyses, and placebo tests \citep{sekhon2009opiate}.

Whether, on balance, this movement has produced an epistemic change in political science depends on how widely and deeply these ideas have permeated research practice. Quantifying the depth and breadth of the credibility revolution in political science presents two measurement challenges. The first is scope. What constitutes a sufficiently representative sample of political science to allow statements about the discipline as a whole? Previous research examining methodological changes in other social science disciplines has focused on elite journals or publishing outlets \citep[e.g., ][]{currie2020technology, goldsmith2026tracking}. However, the discipline extends far beyond these outlets, which may publish systematically different research than the rest of the field. Consequently, a systematic assessment of potential changes requires analyzing papers across multiple journal tiers over a substantial period. We address this by analyzing 91,632 articles from 156 political science journals spanning 2003--2023 using large language models (LLMs). This represents, to our knowledge, the most comprehensive analysis of political science research assembled to date.

The second challenge is operationalization. The credibility revolution is an intellectual movement that seeks to transform empirical research. Translating such ideas into concrete research decisions is not straightforward. What changes should we expect to observe if its ideas have taken hold? We focus on three dimensions where change should be evident: (1) the methodological approaches employed in quantitative political science papers, (2) the adoption of practices that enhance credibility beyond research design, and (3) the professional rewards these approaches receive relative to the approaches they aim to substitute.

We leverage advances in LLMs to classify our corpus along these dimensions. We begin by mapping papers by methodology and research goals, then narrow our focus to explanatory empirical quantitative papers—the subset most likely to be shaped by the credibility revolution. Within this subset, we distinguish between papers that have design-based approaches as their primary research design, which most straightforwardly embody the principles of the credibility revolution by leveraging features of the data-generating process for causal identification, and model-based approaches, which ground identification primarily in parametric or other modeling assumptions. We also code the use of several ancillary practices, e.g. power analyses, and placebo tests, which have the potential to further buttress the credibility of causal claims made in quantitative papers.

We first assess the extent to which design-based empirical research has substituted model-based approaches in quantitative studies that aim to explain social phenomena. We find substantial, though uneven, substitution. Between 2003 and 2023, the share of explanatory quantitative papers using design-based methods as their primary research design rose from 12.6\% to 36.5\%, while the share using model-based approaches fell from 87.4\% to 63.5\%. Survey experiments alone formed part of the primary research design in 49.3\% of all design-based papers published in 2023. The shift was much more pronounced in top-20 journals, where the design-based share rose from 13.8\% in 2003 to 50.5\% in 2023, compared with an increase from 11.8\% to 28.7\% in less prominent journals. Taken together, these patterns suggest that methodological change has been partial and uneven. Traditional model-based approaches remain central to explanatory research, especially in lower-visibility outlets.

Second, we examine the adoption of practices that enhance credibility beyond research design. We find that pre-analysis plans, power analyses, and placebo tests have become more common in design-based quantitative research, although their adoption remains uneven. Among experiments, pre-analysis plans rose from less than 1\% in 2013 to 39\% in 2023, while 17\% of experimental papers explicitly mentioned a power analysis. By contrast, only 6.1\% of observational studies published in 2023 reported a pre-analysis plan, and fewer than 5\% mentioned power calculations. Placebo testing shows the opposite pattern: more than 40\% of observational studies report such tests, compared with fewer than 10\% of experiments. Model-based papers rarely reported any of these three practices.

Third, we assess whether design-based research is associated with different professional rewards than model-based approaches. We compare citation counts for design-based and model-based papers published in the same year, thereby holding constant the time available to accumulate citations. In each periods---2003--2008, 2009--2013, and 2014--2018---the average design-based paper has received more citations than the average model-based paper. For any given number of year since publication, this difference is larger for more recently published papers. 

Overall, we find evidence consistent with a moderate influence of the credibility revolution in political science. Design-based methods have expanded as model-based approaches have declined, indicating growth in quantitative research that aligns with the movement's principles. Yet this shift remains incomplete and narrow. Traditional model-based approaches are still deeply embedded in explanatory empirical quantitative research and continute to underpin most of the discipline's causal work. The credibility revolution's limited reach also reflects the fact that much of political science falls outside the domain to which its lessons most directly apply: quantitative studies aimed at identifying causal relationships. Its partial influence therefore reflects not only uneven methodological change, but also the discipline's enduring methodological diversity.

The movement's reach is also not especially deep. Broader credibility-enhancing practices, such as pre-analysis plans and power calculations, remain uncommon even among papers that use design-based methods. Moreover, a substantial share of the shift from model-based to design-based research is driven by the adoption of survey experiments alone. In this light, the credibility revolution is best understood as a reform: a genuine but limited transformation concentrated in relatively small, though consequential, segments of the discipline.

Still, design-based approaches have become more prevalent, consistent with the credibility revolution's prescriptions. It would be tempting to infer from this trend that political science research has become more credible. Our analysis, however, is descriptive rather than prescriptive. We document changes in how scholars conceptualize causality, justify causal claims, and report supporting evidence, taking what papers report at face value. Assessing whether a paper's identifying assumptions are warranted, whether its analyses are well executed, or whether its findings are correct requires a different empirical design and lies beyond the scope of this study.

Furthermore, credibility depends on the validity of a study's assumptions and the quality of its implementation. Both design-based and model-based approaches can rest on defensible assumptions and be executed well, or they can be misspecified, poorly implemented, and ultimately unconvincing. We therefore do not treat one approach as inherently superior to the other. Instead, we interpret the patterns we document as a shift in the conventions political scientists use to justify identification and present causal claims, not as evidence that published findings are now more likely to be correct.

Our paper builds on \citet{currie2020technology} and \citet{goldsmith2026tracking}, which quantify the credibility revolution movement in economics using papers published in elite journals and National Bureau of Economic Research (NBER) working papers series. A key limitation of these studies is that NBER affiliates do not represent the economics discipline as a whole. By contrast, we analyze articles from all political science journals with an impact factor above one, allowing us to examine methodological change well beyond the outlets most closely associated with scholars at elite institutions. 

We also extend \citet{garg2025causal}, which uses LLMs to identify causal claims in economics, by systematically classifying design choices and transparency practices across an entire discipline. Our analysis also complement recent large-scale diagnostics of quantitative political science that assess the inferential properties of published research \citep{arel2026quantitative}. Finally, we contribute to a broader literature on how the credibility revolution has shaped political science, including patterns of knowledge production, graduate training and hiring, and the evolution of methodological norms~\citep{grossman2025political}.

\section*{The Credibility Revolution}\label{sec:crsec}

We use the term ``credibility revolution'' to refer to a broad movement in the social sciences that sought to change how empirical causal research is conducted and evaluated. As an epistemic project, it emphasized explicit and defensible identification assumptions in quantitative causal research and placed particular weight on reasoning about the data-generating process when justifying causal comparisons \citep{morgan2014counterfactuals, imbens2015causal}. It also stressed the importance of making identification strategies transparent and open to scrutiny, while reducing reliance on modeling assumptions that are difficult to defend as the basis for causal claims \citep{aronow2019foundations, samii2016}.

Like many epistemic movements, the credibility revolution is broad, heterogeneous, and internally contested. Our aim is not to adjudicate its exact meaning, but to identify a core distinction that allows us to assess whether and how empirical causal practice has changed in ways consistent with the movement's goals. To that end, we adopt a taxonomy that distinguishes between two broad approaches to causal identification: whether identification rests primarily on features of treatment assignment or on assumptions embedded in statistical models.

\begin{itemize}\itemsep0em
\item \emph{Design-based approaches} ground causal identification in features of the data-generating process that generate plausibly exogenous variation in treatment assignment, such as randomization, institutional rules, or policy thresholds. While estimation may rely on statistical models, the validity of causal claims is understood to rest primarily on assumptions about treatment assignment and counterfactual comparisons rather than on functional-form or distributional assumptions. Natural experiments, randomized controlled trials, instrumental variables, regression discontinuity designs, and difference-in-differences designs are canonical examples.

\item \emph{Model-based approaches}, by contrast, ground causal identification primarily in parametric or statistical modeling assumptions, such as linearity, additivity, conditional independence, and distributional assumptions. Causal interpretation therefore depends on the plausibility of modeling assumptions, which may be supported by theory or features of the data and can be more or less defensible depending on the context. Unlike design-based approaches, these assumptions are typically not anchored in an explicit claim about the treatment assignment process.
\end{itemize}

The taxonomy is imperfect. Methods associated with design-based approaches can also incorproate elements of model-based reasoning, e.g., difference-in-differences.\footnote{In essence, the parallel trends assumption underlying difference-in-differences designs is a mean-independence assumption between the treatment and demeaned or first-differenced non-treated potential outcomes.} Nonetheless, this taxonomy captures a central distinction in where causal identification is anchored.

\section*{The Credibility Revolution and Political Science}\label{sec:ourapproach}

The credibility revolution began in applied microeconomics in the early 1990s, driven by studies that exploited institutional rules, spatial policy changes, and quasi-random variation to identify causal effects \citep[e.g.,][]{angrist1991does,card1994minimumwage}.

Early contributions challenged prevailing practices that relied heavily on parametric regression models for causal interpretation, with causal claims resting largely on modeling choices. Researchers often modeled the conditional expectation of an outcome as a function of treatment variables and covariates---often without clearly distinguishing between the two---and then interpreted regression coefficients as causal effects. Proponents of the credibility revolution argued that this practice placed a heavy burden on modeling assumptions, often without direct scrutiny of the underlying identification strategy.\footnote{Already in the early 1980s, researchers had noted that regression-based causal claims were often fragile to specification choices and sensitive to researcher degrees of freedom~\citep[e.g.,][]{leamer1983}.} Early critiques also showed that conventional model-based adjustments in observational studies could diverge sharply from experimental benchmarks and lead to erroneous conclusions about the effects of policies and programs \citep{lalonde1986}.

These empirical developments also drew on formal frameworks for causal inference that emphasized explicit identification assumptions. The Neyman--Rubin potential outcomes framework provided a common language for defining causal effects \citep{rubin1974, holland1986}, while structural causal models and directed acyclic graphs offered a graphical representation of causal assumptions \citep{pearl2009Causality}. Related work also formalized the assumptions underlying instrumental-variable designs and local average treatment effects \citep{imbens1994identification}.

Several research designs became especially prominent within the movement. Difference-in-differences provided a flexible framework for exploiting policy changes and temporal variation \citep{card1994minimumwage}. Natural experiments leveraged institutional rules, naturally occurring shocks, or geographic boundaries to isolate plausibly exogenous variation in treatment assignment \citep{dell2010persistent}, while instrumental-variable approaches addressed unobserved confounding under explicit exclusion restrictions \citep{acemoglu2001colonial}. Randomized controlled trials gained prominence because average treatment effects can be identified under a minimal set of transparent assumptions, and regression discontinuity designs offered credible identification around policy or electoral thresholds \citep{lee2008rd}. Despite differences in scope and assumptions, these research designs share an emphasis on tying causal identification explicitly to features of the empirical setting.

Political science absorbed these ideas, with some lag, through several channels.\footnote{Within political science, concern with endogeneity and research design predated the movement's consolidation. The treatment in \citet{achen1986statistical} of quasi-experiments and the framework of \citet{king2021designing} for inference and research design both pushed the discipline toward more explicit reasoning about how evidence supports causal claims.} Early field experiments demonstrated the feasibility of randomized interventions in political settings \citep{wantchekon2003clientelism} and showed how randomization could replace statistical adjustment as the basis for causal claims about questions the discipline had long studied observationally \citep{gerber2000field}.\footnote{\citet{rosenstone1993mobilization} estimated the relationship between campaign contact and turnout using observational survey data, adjusting for characteristics associated with both. \citet{gerber2000field} revisited the question by randomly assigning face-to-face canvassing and found that contact substantially increased turnout. The two studies ask the same substantive question but differ in what makes the comparison credible. In the model-based approach, credibility rests on the assumption that adjustment for observed characteristics removes the relevant differences between contacted and non-contacted voters. That assumption is difficult to defend here because campaigns deliberately target people already likely to vote, while the characteristics that make someone a target---such as prior participation, civic engagement, or a stable address---are imperfectly measured at best. Random assignment sidesteps this problem by creating groups that are comparable in expectation on both observed and unobserved characteristics, so the design itself carries the identification.} \citet{dunning2012natural}'s synthesis of natural experiments further helped codify design-based reasoning within the discipline. Regression discontinuity designs became influential in electoral studies \citep{eggers2015validity}, while instrumental-variable approaches grounded in local average treatment effects gained traction in comparative and international research \citep{sovey2011instrumental}. Difference-in-differences designs also became increasingly common, particularly in studies of policy change \citep{bechtel2011lasting}.

The diffusion of credibility-revolution ideas also coincided with greater attention to research practices intended to make empirical claims more transparent and, thereby, more credible. Robustness checks, placebo tests, and sensitivity analyses became more common, alongside the spread of pre-analysis plans and data and code sharing. Although transparency and reproducibility practices have intellectual origins distinct from design-based identification, they share a common emphasis on making assumptions and analytic choices visible and open to scrutiny.

These developments have not gone uncontested. A growing body of scholarship emphasizes the limitations and trade-offs of design-based approaches. One concern is the locality of estimands: credible causal estimates may apply only to narrow subpopulations or institutional settings, limiting their theoretical or policy relevance \citep[e.g.,][]{heckman1995randomization}. A related concern is that emphasizing quasi-random variation can narrow the scope of inquiry. By privileging questions that permit credible identification, researchers may give less attention to substantively important but historically contingent processes---such as democratization or state formation---that resist design-based analysis \citep{mahoney2006tale}.

Other critiques highlight the tension between credible identification and substantive interpretation. Even when causal effects are well identified, their meaning often rests on untested assumptions about mechanisms, scope conditions, and counterfactual relevance. These concerns become especially salient when researchers generalize from localized interventions to broader theories of political behavior or institutional change \citep{findley2021external}. Taken together, these critiques underscore that the credibility revolution reflects a specific epistemological orientation---one that prioritizes internal validity and research design rigor---but does not resolve longstanding tensions among identification, explanation, and generalization in empirical political science.

\subsection*{On Survey Experiments}\label{sec:surveys}

Survey experiments occupy an ambiguous position in the genealogy of the credibility revolution. They emerge from a distinct epistemic lineage---rooted in psychology and behavioral science---that predates the debates over causal identification that shaped applied microeconomics in the 1990s and 2000s. As \citet{samii2025survey} argues, their growth need not therefore represent the direct diffusion of credibility-revolution ideas. Early survey experiments were developed primarily as tools of measurement and elicitation, designed to probe latent attitudes, beliefs, and cognitive processes \citep{mutz2011population}. Their purpose was not necessarily to estimate the causal effects of real-world interventions, but to improve understanding of the psychological mechanisms underlying political behavior. In this sense, survey experiments are not direct descendants of the credibility revolution that coalesced around the identification of causal effects in field settings.

At the same time, survey experiments are experiments and therefore share the credibility revolution's emphasis on exogenous treatment assignment. Their interventions, however, are typically narrow: researchers manipulate information through vignettes or prompts rather than political conditions, policies, or institutional arrangements themselves. This allows them to identify the causal effects of informational treatments in controlled settings, but the resulting estimands often concern short-term attitudinal responses to textual stimuli rather than behavioral or institutional change. These effects may therefore be well identified while remaining substantively constrained,, and their implications for real-world political behavior can be difficult to trace~\citep{barabas2010survey}.

For these reasons, we classify survey experiments as design-based in our analyses, consistent with their widespread treatment in the discipline as causally identified research designs. At the same time, their distinct intellectual origins and the narrower scope of many survey treatments raise questions about how fully they reflect the goals of the credibility revolution. We therefore report results both including survey experiments and separating them from other design-based approaches, allowing us to assess how much they drive the broader pattern of methodological change in political science.

\section*{Quantifying the Credibility Revolution} 

We assess changes in empirical practice in political science along three dimensions: methodological choices, transparency practices, and professional recognition. For each dimension, we ask what patterns we should observe in published research if the credibility revolution has meaningfully taken hold in empirical political science. To be clear, our analysis is not designed to identify the ``causal effect'' of the movement, which would require specifying and estimating an appropriate counterfactual. Rather, we document the breadth and depth of practices associated with the movement’s standards and goals.

First, we examine shifts in research design. If the credibility revolution changed empirical practice, we should expect a move away from approaches that rely primarily on modeling assumptions, such as functional forms, and toward those that ground causal claims in identification strategies that leverage explicit features of the data generating process. Empirically, this should appear as growth in design-based methods relative to model-based approaches.

Second, we examine changes in transparency practices, such as pre-analysis plans, and reporting practices, such as power calculations and placebo tests. Although these practices have distinct intellectual origins, they are now frequently bundled with credibility-revolution standards in contemporary empirical work \citep{blair2023research}, and their proponents emphasize their role in strengthening the credibility of causal claims \citep{sekhon2009opiate}. If these practices have diffused alongside changes in research design, we should observe increasing adoption of practices that make identification assumptions and analytic decisions more explicit.

Third, we assess whether these methodological shifts are reflected in professional rewards. If standards of credible causal inference have changed within the discipline, research using design-based methods or associated transparency practices may receive greater recognition. We examine this descriptively through publication patterns and citation outcomes for design-based and model-based papers, holding constant publication timing and venue.

\section*{Data and Measurements}\label{sec:data}

We evaluate these possible trends using a corpus of political science research from the past two decades. This section describes the data, key measures, and methods used to construct them.

We began with 188 journals classified by \href{https://clarivate.com}{Clarivate} as political science and with a Scientific Journal Rankings (SJR) score of at least one. This threshold roughly corresponds to the upper tier of the field in citation influence and focuses attention on outlets that sit near the center of political science's citation network. From this list, we removed journals for three distinct reasons: indexing (three journals not indexed in Scopus), quality control (three journals that lacked peer review), and language or format (seven journals that were not published in English and one listed as a book series). The resulting sample included 174 English-language, peer-reviewed, Scopus-indexed political science journals. Using Scopus, we extracted journal-level metadata, including citation statistics, and article-level information for all publications in these selected journals. This process produced a dataset of 129,751 articles published between 2003 and 2023. 

\begin{figure}[!th]
    \centering
    \caption{Number of Papers per Year, Overall and by Subfield}
    \label{fig:trends}
    \includegraphics[width=0.7\textwidth, height=0.8\textheight, keepaspectratio, trim={0cm 0cm 0cm 1cm}, clip]{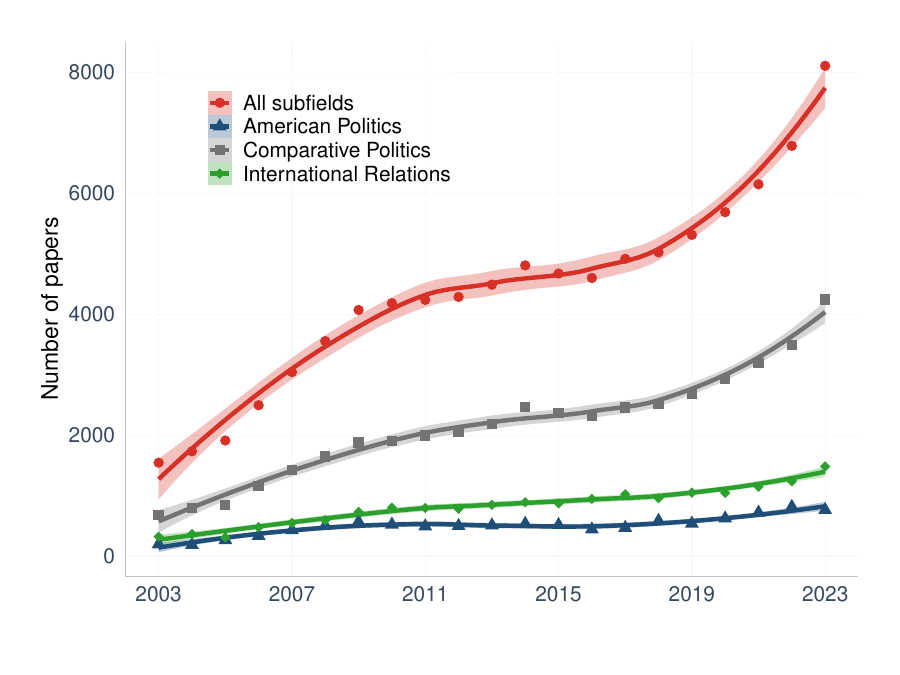}
    \fignote{The full sample contains 91,632 papers published from 2003 to 2023. Lines are loess fits. The red line shows total papers; blue, American Politics; gray, Comparative Politics; and green, International Relations. The subfield labels are generated by GPT-4o.}
\end{figure}

Next, we obtained full-text versions of the articles where possible, successfully retrieving the complete text of 91,632 articles from 156 journals. Full-text availability varies across publishers, so the resulting corpus is broad but not exhaustive. We then used a combination of supervised machine learning and two commercial LLMs, GPT-4o and Claude Sonnet 5, to classify the full-text articles. Figure~\ref{fig:trends} shows the annual number of papers in our full-text sample, both overall and by subfield. The volume of published political science research has grown rapidly: from 1,546 papers in 2003 to 8,109 in 2023, an increase of more than fourfold.\footnote{See \citet{grossman2025political} for an analysis of the factors underlying the dramatic increase in political science publication volume.}

\subsection*{Constructing Measures}

We construct a set of variables describing each paper's methodological features, substantive focus, and transparency practices. Section~\ref{sisec:prompt} of the Supplementary Materials describes the full coding procedure.

Variable construction proceeds in two stages. In the first stage, we use GPT-4o to label every paper in the corpus and retain two labels. The first identifies whether a paper is an \emph{empirical quantitative} study or instead empirical qualitative, normative, formal-theoretic, or a review. We classify a paper as empirical quantitative if it analyzes observational or experimental data, including reanalyses of existing datasets. We exclude simulations and purely methodological discussions that do not analyze data. The second label assigns each paper to one of six subfields: American Politics, Comparative Politics, International Relations, Methodology and Formal Theory, Political Theory and Philosophy, or Public Policy/Administration. This classification reflects the paper's primary substantive contribution and is recorded for the full corpus.

\begin{figure}[!ht]
    \centering
    \caption{Sequential Classification of Type of Studies}
    \label{fig:sankey}
    \includegraphics[width=0.9\linewidth, trim={0cm 2cm 0cm 1.5cm}, clip]{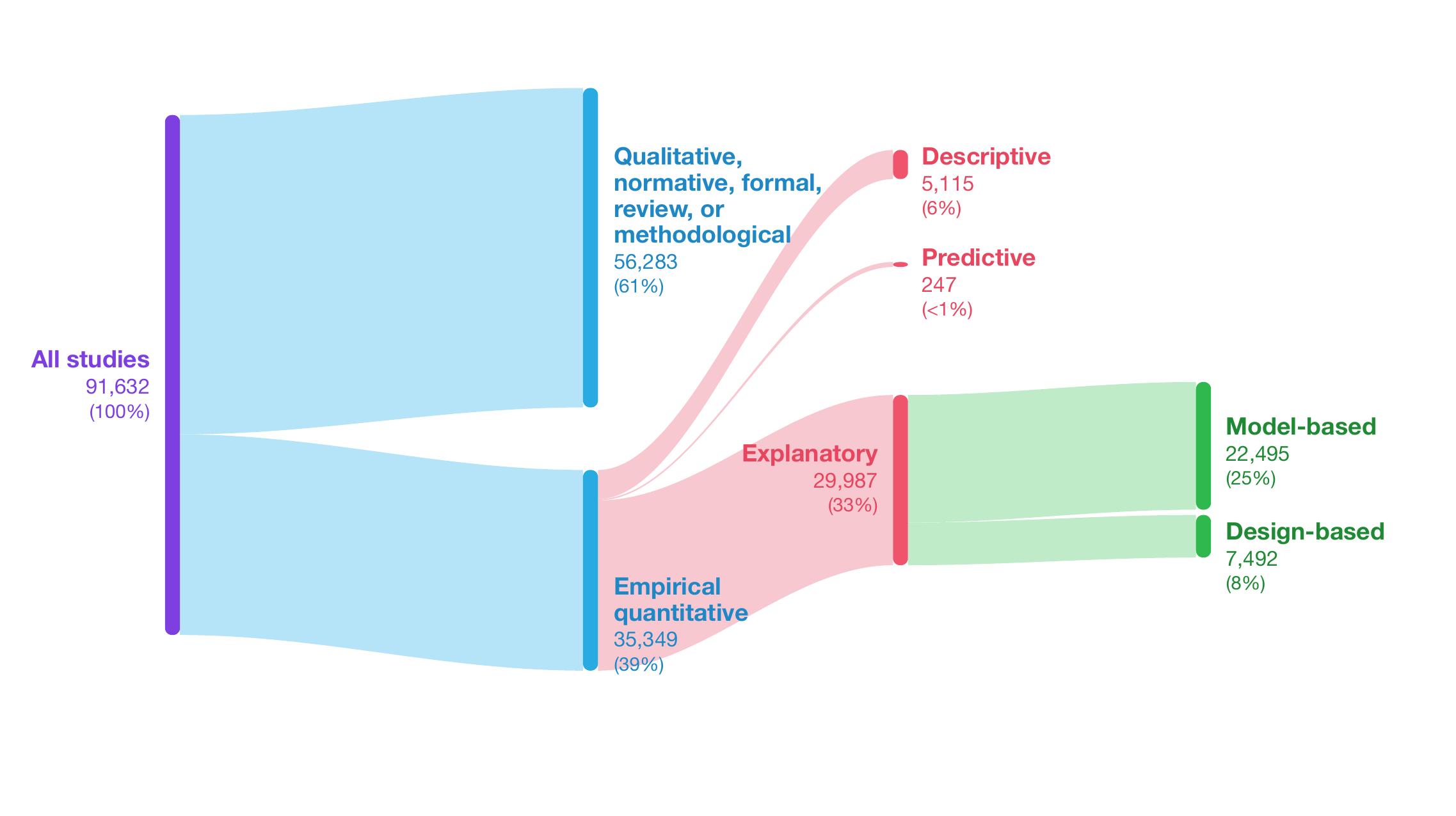}
    \fignote{Flows trace papers from the full sample to quantitative studies, then to descriptive, predictive, and explanatory goals, and finally to design-based versus model-based approaches. The percentage numbers use the full sample of studies, $91,632$, as the denominator.}
\end{figure}

Since the credibility revolution primarily concerns empirical quantitative research, only papers classified as \emph{empirical quantitative} in the first stage proceed to the second. We use Claude Sonnet 5 for this stage because the remaining classifications require finer methodological distinctions. We begin by re-screening the papers that passed Stage~1. We reclassify papers whose main analysis is qualitative or interpretive, papers centered on a formal-theoretic model with only illustrative empirical analysis, methodological work, and reviews that summarize findings from other studies without statistically analyzing them. This re-screening removes 1,814 of the 37,163 papers that passed Stage~1, reclassifying them as qualitative, formal-theoretic, methodological, or review articles. We treat these papers as non-empirical-quantitative in all subsequent analyses.

We then assign each remaining empirical quantitative paper one of three general goals: explanatory, descriptive, or predictive. Explanatory papers investigate the causes or consequences of social phenomena and present evidence consistent with one or more causal relationships, even when they do not use explicit causal language.\footnote{Our ``explanatory'' category includes both causes-of-effects (``explanation'') and effects-of-causes (``causal inference'') in the sense of \citet{lundberg2021your}. Proponents of the credibility revolution call for a shift in the balance between these goals. Our measures do not clearly distinguish between them, and doing so is a promising avenue for future work.} Descriptive papers focus on measurement or characterization. They may introduce new indicators \citep[e.g.,][]{bonica2018inferring} or document differences across groups without developing or testing a causal mechanism \citep[e.g.,][]{blinder2007dissonance}. Predictive papers forecast outcomes or identify the variables that best predict them. They typically evaluate performance out of sample or through cross-validation and do not interpret predictors as causes \citep[e.g.,][]{muchlinski2016comparing}. When a paper serves multiple goals, we classify it according to the goal of its primary analysis.

Figure~\ref{fig:sankey} illustrates this sequential classification. Two patterns stand out. First, empirical quantitative papers are a minority, accounting for 39\% of all papers pooled across years. Second, explanatory papers dominate within empirical quantitative research, comprising 85\% of the total. Descriptive papers account for 14\%, while predictive papers make up less than 1\%.

\begin{figure}[!ht]
    \centering
    \caption{Proportion of Empirical Quantitative Papers over Time}
    \label{fig:quantemptime}
    \includegraphics[width=0.7\textwidth, height=0.8\textheight, keepaspectratio, trim={0cm 0.5cm 0cm 0.5cm}, clip]{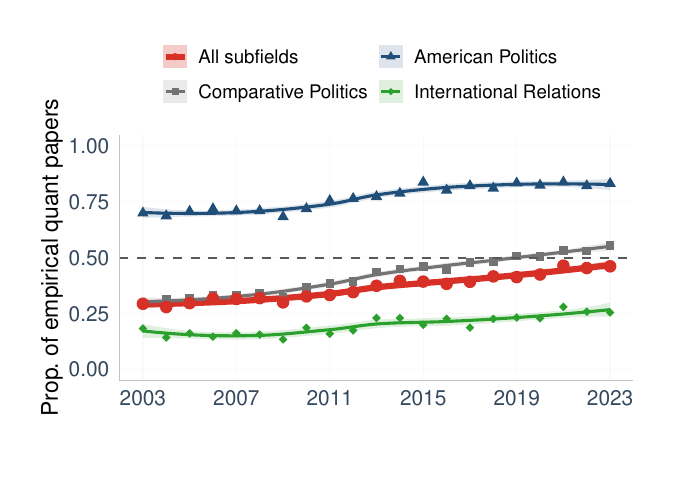}
    \fignote{The red curve shows the overall share, and the blue, gray, and green curves show trends for American Politics, Comparative Politics, and International Relations. Lines are lowess fits with 95\% confidence intervals; points indicate raw annual proportions.}
\end{figure}

Averages over two decades may mask important temporal dynamics, which we examine in Figure~\ref{fig:quantemptime}. The share of empirical quantitative papers rises steadily over time, reaching about 46\% by 2023 (red curve). The remaining curves show the same trend for the three largest substantive subfields: Comparative Politics, which accounts for 49\% of the corpus; International Relations, 19\%; and American Politics, 11\%. All three trend upward, but their levels remain markedly different. In 2023, 26\% of International Relations papers use empirical quantitative methods, compared with 56\% in Comparative Politics and 83\% in American Politics.

For the 29,987 papers classified as explanatory, we identify the primary research design used to support the central causal estimate. We distinguish between two broad families: design-based and model-based approaches. \textit{Design-based approaches} link causal identification to a stated source of exogenous or as-if-exogenous variation in treatment assignment, such as randomized assignment, institutional cutoffs, policy shocks, or plausibly exogenous timing. Their identifying assumptions therefore rest primarily on the assignment process rather than on the model specification alone. Specifically, design-based approaches include field, survey, and lab experiments; selection-on-observable designs; instrumental-variable designs; regression discontinuity designs; difference-in-differences designs;\footnote{This category includes causal panel analyses that rely on some version of the parallel trends assumption \citep{chiu2025causal}: canonical difference-in-differences, two-way fixed effects and event-study designs, and methods for staggered treatment adoption, when they are framed as treated-versus-comparison contrasts over time. Unit fixed effects aimied at adjusting for time-invariant confounders are coded as model-based.} synthetic control methods; and natural and quasi-experiments.\footnote{These studies use treatment variation not controlled by the researcher, such as natural disasters, draft lotteries, or externally imposed policy changes. We classify a paper here when it presents its analysis as a natural experiment or as quasi-experimental, provided it does not fall into one of the aforementioned design categories.}  We record the specific design used by each paper and, when a paper relies on more than one design to support its central empirical claim, record each of them.

\textit{Model-based approaches} identify effects through parametric or other modeling assumptions without a separate design-based source of identifying variation. We classify a paper as design-based if at least one of its recorded strategies belongs to the design-based family. Because linear regression as an algorithm can recover causal effects in both experimental and non-experimental settings \citep{angrist2009mostly, aronow2016does, imbens2025comparing}, we do not treat the use of linear regression itself as diagnostic of model-based research.

For explanatory studies, we also code whether the paper explicitly states the key identification assumption underlying its design, such as unconfoundedness, parallel trends, continuity, or an exclusion restriction. We separately code whether the paper makes an explicit causal claim using terms such as ``causes,'' ``effect,'' or ``impact.'' We initially also sought to assess whether a study clearly articulated an interpretable estimand. Consistent with \citet{lundberg2021your}, however, our pilot study suggested that few papers do so outside experimental, instrumental variables, and regression discontiguity designs, which typically target quantities such as the average treatment effect, average treatment effect on the treated, or local average treatment effect. We therefore dropped this measure.

We additionally code several features of the analysis, including sample size and three practices intended to enhance credibility: (1) whether the paper reports placebo tests, (2) whether it reports conducting a power analysis, and (3) whether its hypotheses or analysis were pre-registered. We focus on these practices because they are commonly used across research designs and address different forms of inferential error. Placebo tests and pre-analysis plans can reduce the risk of false positives, respectively by detecting spurious results \citep{eggers2024placebo} and by limiting undisclosed analytic flexibility, including p-hacking \citep{brodeur2024preregistration}. Power analyses address false negatives by assessing whether a study has sufficient statistical power to detect substantively meaningful effects, a concern that has received increasing attention in social science research \citep{chiu2025causal, arel2026quantitative}. Although sample size is not itself a research practice, we treat it as an indirect indicator of attention to statistical power: greater concern about power may lead researchers to design larger studies even when they do not report formal calculations. Together, these measures allow us to document changes in methodological choices and causal reasoning across quantitative political science.

Two caveats are important. First, we take what papers report at face value. We do not assess whether a research design was implemented well, whether its identifying assumptions are warranted in a particular application, or whether a paper contains errors or misrepresents its methods. Second, our taxonomy is descriptive rather than evaluative or normative. It does not imply that design-based studies are inherently more or less credible than model-based studies. In either family, the assumptions required for identification may hold or fail in practice. Recent replication studies in political science likewise suggest that problems involving assumption validity, statistical power, and execution persist across research designs \citep[e.g.,][]{stommes2023reliability, lal2024much, chiu2025causal}.

\subsection*{LLM Pipeline and Human Validation}

As described above, we label the corpus in two stages. In Stage~1, we pass the full text of every article to GPT-4o, which screens out papers that are not empirical and quantitative and assigns each paper to a subfield. In Stage~2, we pass each of the 37,163 papers classified as empirical quantitative to Claude Sonnet~5 using a fixed prompt. Claude Sonnet 5 re-screens the full text, classifies each paper as explanatory, descriptive, or predictive, and assigns a research design to explanatory papers. Sections~\ref{sisec:prompt} and~\ref{sisec:prompt_v8e} reproduce both prompts in full.

We evaluated each stage against independently established reference labels. For Stage~1, two research assistants coded the sample and a third adjudicated disagreements. For Stage~2, two language models independently coded each paper. An author adjudicated disagreements and audited a sample of agreements without access to the classifier’s labels

We find that the Stage~1 screen for empirical quantitative papers matches the reference label for 97.5\% of a random sample of 200 papers. In a second random sample of 200 papers, Stage~2 correctly identifies both the paper's purpose (explanatory, descriptive, or predictive) and its research design in 90.5\% of cases. Because these random samples contain too few papers with rare designs to assess design-specific accuracy, we also drew a sample of 244 papers balanced across designs and periods. Among the 240 papers the classifier retains as explanatory, it correctly distinguishes design-based from model-based work in 93.3\% of cases and matches the exact design in 87.1\%. Accuracy is at least 71\% for each of the ten single-design categories.

The classifier also performs close to the practical ceiling implied by ambiguity in how some papers describe their research designs. Four human coders independently classified 60 papers using the same codebook and without seeing the classifier's output. Scored against the same reference labels, the human coders and classifier perform equally well, while the coders disagree with one another on 11.7\% of papers. Some remaining error therefore reflects ambiguity in the papers rather than limitations specific to the classifier. The codebook can standardize how such cases are resolved, but it cannot eliminate that ambiguity. Section~\ref{sisec:prompt_validation} reports results by design and the full cross-tabulation of reference and assigned labels.

Three further results suggest that classification error does not drive our findings. First, the classifier measures our main distinction---design-based versus model-based research---more accurately (93.3\%) than exact research design (87.1\%). Second, the accuracy of the former measure is similar across periods: 94.4\% in 2003--2009, 92.6\% in 2010--2016, and 93.1\% in 2017--2023. Third, the errors are asymmetric: fifteen design-based papers are classified as model-based, compared with only one error in the opposite direction, implying that the uncorrected estimates slightly understate the design-based share. As a further check, Section~\ref{sisec:dslcorr} applies the design-based supervised learning correction of \citet{egami2024dsl} to adjust the corpus estimates for measured classification error. The corrected estimates show the same upward trend in design-based research, with a slightly flatter slope.

\section*{Main Findings}

To assess how empirical quantitative research in political science has changed, and whether those changes are consistent with the aspirations of the credibility revolution, we present descriptive evidence on the scope and pace of methodological change across the discipline. We examine shifts in research design, empirical practices, and scholarly rewards associated with design-based versus model-based approaches.

\begin{figure}[!ht]
    \centering
    \caption{Trends in the Use of Design-based Methods}
    \label{fig:basictrends}
    \begin{subfigure}[b]{0.47\textwidth}
        \centering
        \includegraphics[width=\linewidth]{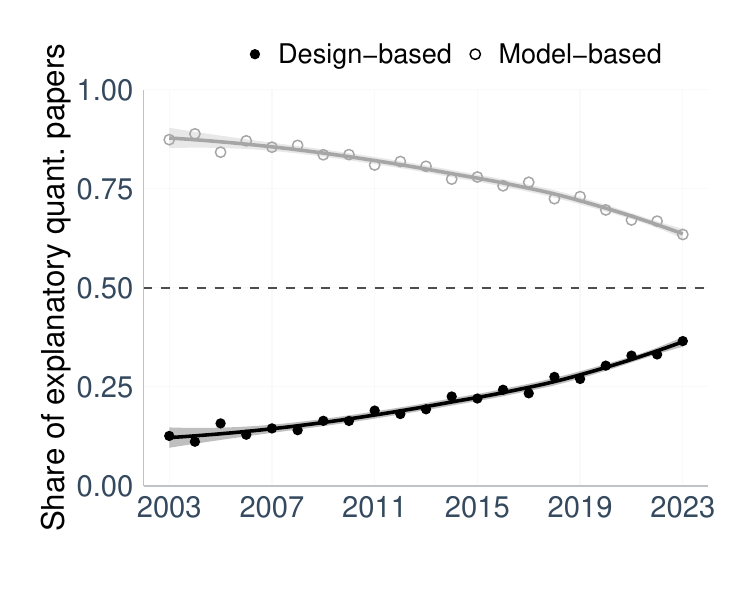}
        \caption{Design-based vs. model-based studies}
        \label{fig:design_overall}
    \end{subfigure}    \hfill
    \begin{subfigure}[b]{0.47\textwidth}
        \centering
        \includegraphics[width=\linewidth]{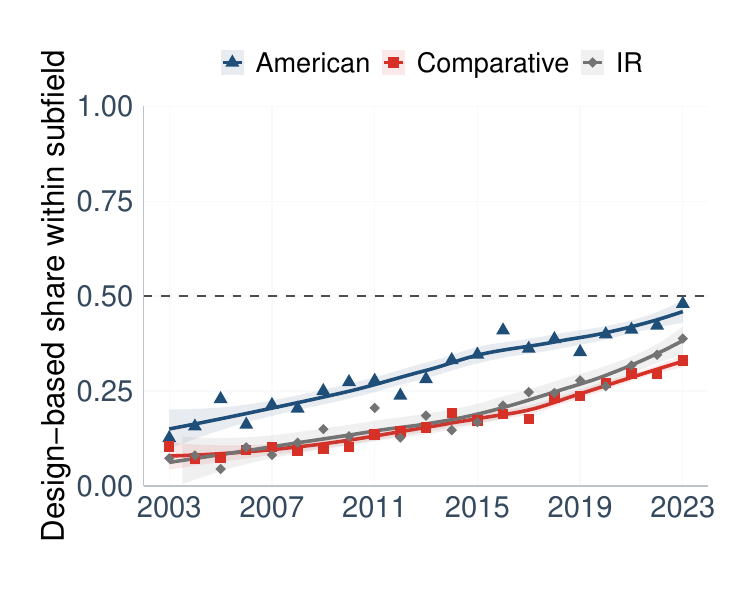}
        \caption{Design-based studies by subfield}
        \label{fig:design_subfields}
    \end{subfigure}
    \fignote{The sample comprises all explanatory empirical quantitative papers.}
\end{figure}

\paragraph*{Design-based methods rose steadily but unevenly.} How have methodological practices changed within explanatory quantitative research? If empirical practice has moved in the direction advocated by the credibility revolution, the share of design-based work should rise relative to model-based work. Figure~\ref{fig:basictrends}(a) traces these trends from 2003 to 2023. Model-based methods (gray) remain dominant throughout the period. Design-based approaches (black), however, nearly tripled their share, rising from 12.6\% in 2003 to 36.5\% in 2023. Even after two decades of growth, model-based methods still accounted for nearly two-thirds of explanatory quantitative research in 2023.

Within this overall shift, differences across subfields are relatively modest. As Figure~\ref{fig:basictrends}(b) shows, American Politics adopted design-based approaches earlier, but by 2023 all three major subfields show substantial uptake: 48\% in American Politics, 39\% in International Relations, and 33\% in Comparative Politics.

Figure~\ref{fig:basictrends_nosurvey}(a) in the Supplementary Materials reports the overall trends excluding survey experiments. Design-based work is considerably less common in this subset, rising from 10\% of explanatory quantitative papers in 2003 to 22\% in 2023, while model-based methods still account for more than three-quarters of papers in 2023. Figure~\ref{fig:basictrends_nosurvey}(b) reports subfield trends under the same exclusion. Removing survey experiments reduces the 2023 design-based share most in American Politics, by 23 percentage points, followed by International Relations at 16 points and Comparative Politics at 11 points.

\begin{figure}[!ht]
    \centering
    \caption{Popularity of Design-based Methods by Year}
    \label{fig:bymethods2}
    \begin{subfigure}[b]{0.9\textwidth}
        \centering
        \includegraphics[width=\linewidth]{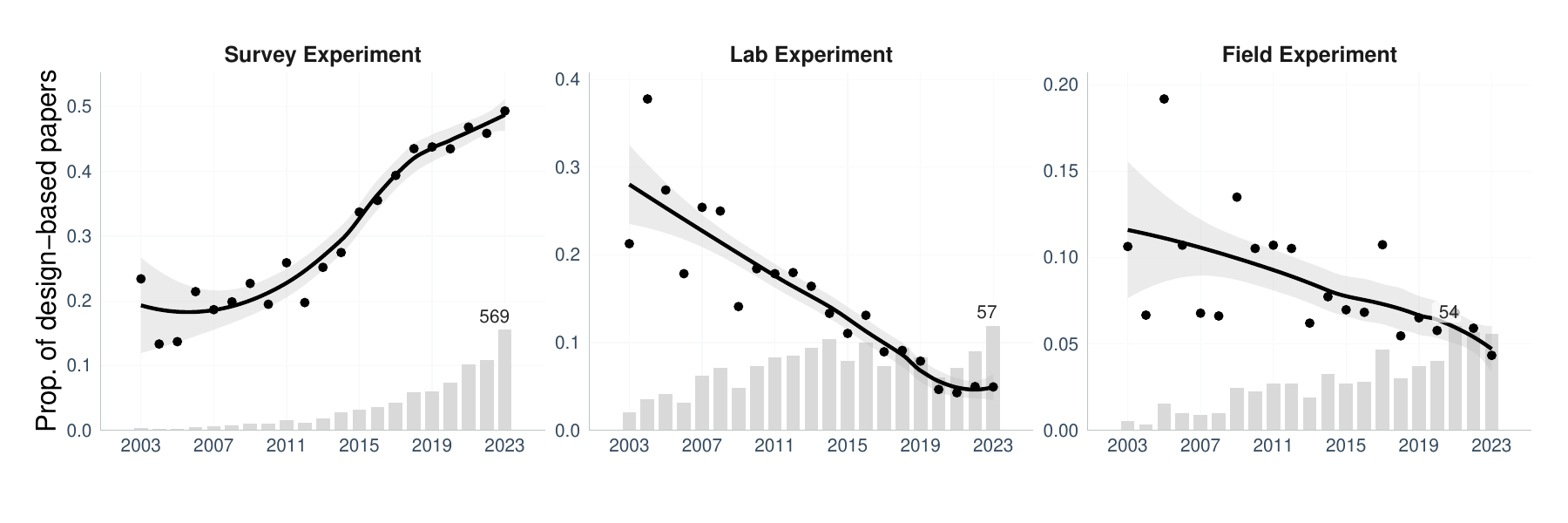}
        \caption{Experimental}
        \label{fig:bymethods_experimental}
    \end{subfigure}
    \hfill
    \begin{subfigure}[b]{0.9\textwidth}
        \centering
        \includegraphics[width=\linewidth]{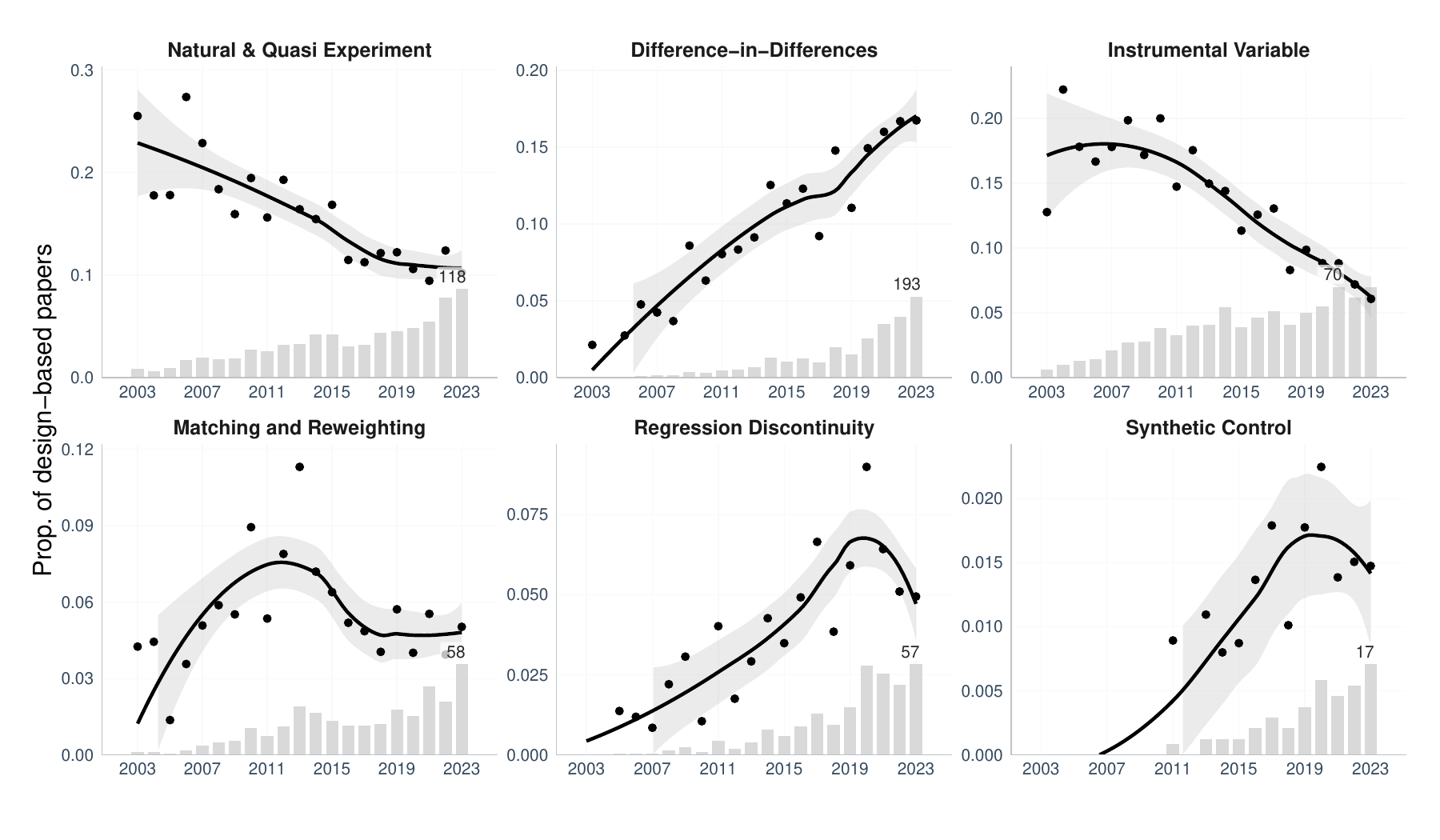}
        \caption{Non-experimental}
        \label{fig:bymethods_nonexperimental}
    \end{subfigure}
    \fignote{Points show the share of design-based papers using the method in a given year. The line is a loess fit with a 95\% confidence band. The gray bars show the numerator of the yearly share.}
\end{figure}

Next, we examine which specific designs drive this growth and how their use has changed over time. Figure~\ref{fig:bymethods2} traces these trajectories for experimental and non-experimental approaches. Each panel plots the number of papers using a method in a given year divided by the total number of design-based papers published that year.\footnote{A small group of papers, 3.4\% of all design-based papers, combines two or more design-based methods. These papers appear in the panel for each method they use, so the yearly shares sum to slightly more than one.}

Figure~\ref{fig:bymethods2}(a) reports trends for survey, field, and lab experiments. In each panel, the line shows the share of design-based papers using the method in a given year,\footnote{By fixing the denominator to the design-based papers published in a given year, we can track the composition of design-based research separately from its growth, which Figure~\ref{fig:basictrends} reports.} while the gray bars show the numerator of that share. Survey experiments rise from about one-quarter of design-based papers in 2003 to nearly one-half in 2023, when 569 papers use the method. Lab and field experiments follow a different pattern. Their absolute numbers remain roughly stable, with 57 lab experiments and 50 field experiments in 2023, but their shares decline as design-based research expands. Lab experiments fall from more than one-fifth of design-based papers in 2003 to 5\% in 2023, while field experiments fall from 11\% to 4\%. Survey experiments therefore account for much of the overall growth in design-based research. The expansion of online survey platforms may have contributed to this shift by lowering the cost of experimentation and making larger and more diverse samples easier to recruit.

Figure~\ref{fig:bymethods2}(b) reports the same quantities for six non-experimental methods: difference-in-differences, natural and quasi-experiments, instrumental variables, matching and reweighting, regression discontinuity, and synthetic control. Difference-in-differences grows almost monotonically, from 2\% of design-based papers in 2003 to 17\% in 2023. By 2023, it is the most common non-experimental design in the discipline, appearing in 193 papers. Natural and quasi-experiments decline from nearly one-quarter of design-based papers early in the period to 10\% in 2023, despite slow growth in absolute numbers. This may partly reflect their role as a residual category, as rising standards increasingly push researchers to specify the exact design they use \citep{sekhon2012natural}. Instrumental-variable designs peak at nearly one-fifth of design-based papers in the late 2000s and decline to 6\% by 2023. Selection-on-observable designs using matching and reweighting methods rise to about 11\% around 2013 before settling near 5\%. Regression discontinuity designs grow gradually and level off near 5\%, consistent with the limited availability of suitable discontinuities. Synthetic control methods appear in the mid-2010s and remain below 2\% of all design-based papers.

Taken together, these patterns show that methodological practice has shifted in the direction charted by the credibility revolution, but that the shift is concentrated in a few research designs. Survey experiments and difference-in-differences are the only methods whose shares of design-based work rise substantially over the period, and together they account for roughly two-thirds of design-based papers by 2023. Designs that require relatively rare empirical circumstances, such as regression discontinuity and synthetic control, remain uncommon. The use of methods that rely on comparatively demanding assumptions, such as instrumental variable designs, has declined over time.

\paragraph*{Design-based studies are concentrated in influential journals.} We find that the adoption of design-based methods has increased slowly but steadily over the past two decades. In 2003, explanatory papers using model-based approaches outnumbered those using design-based approaches by nearly seven to one. By 2020, the ratio had fallen to just over two to one. This aggregate trend, however, may conceal important differences across publishing venues. If higher-impact journals adopt design-based methods earlier or more extensively, discipline-wide averages may understate how far methodological change has progressed in the most visible outlets, which are also more likely to shape research standards. We therefore examine whether adoption differs between top journals and the rest of the field.

\begin{figure}[!ht]
    \centering
    \caption{Design-based and Model-based Studies by Journal Ranking}
    \label{fig:weighted}
    \begin{subfigure}[b]{0.47\textwidth}
        \centering
        \includegraphics[width=\linewidth]{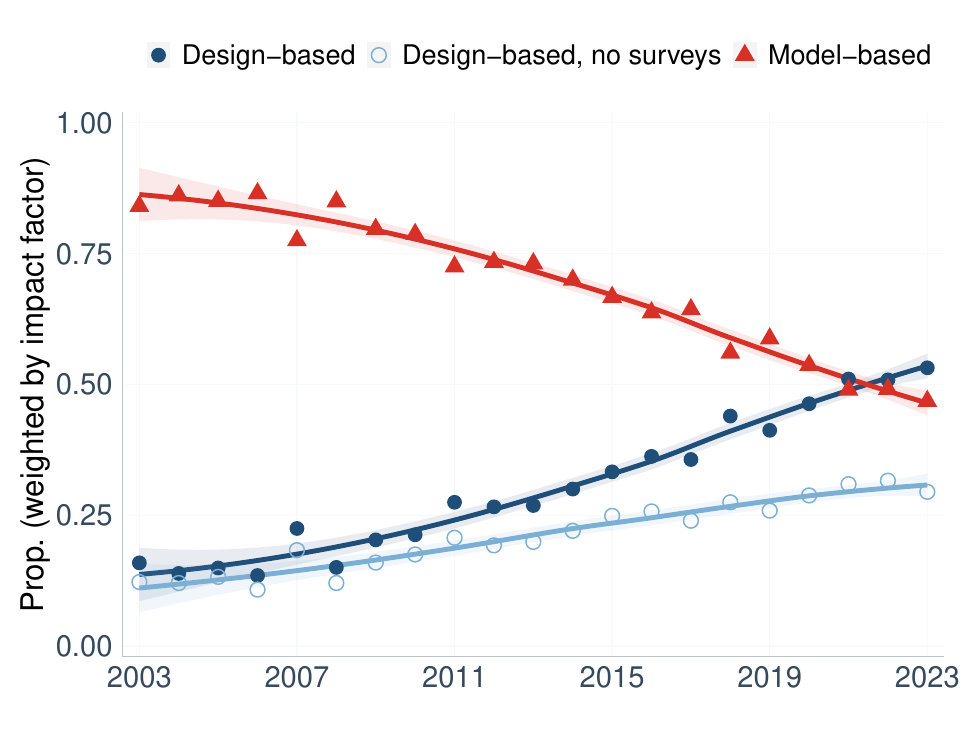}
        \caption{Studies published in top 20 journals}
        \label{fig:top20weighted}
    \end{subfigure}
    \hfill
    \begin{subfigure}[b]{0.47\textwidth}
        \centering
        \includegraphics[width=\linewidth]{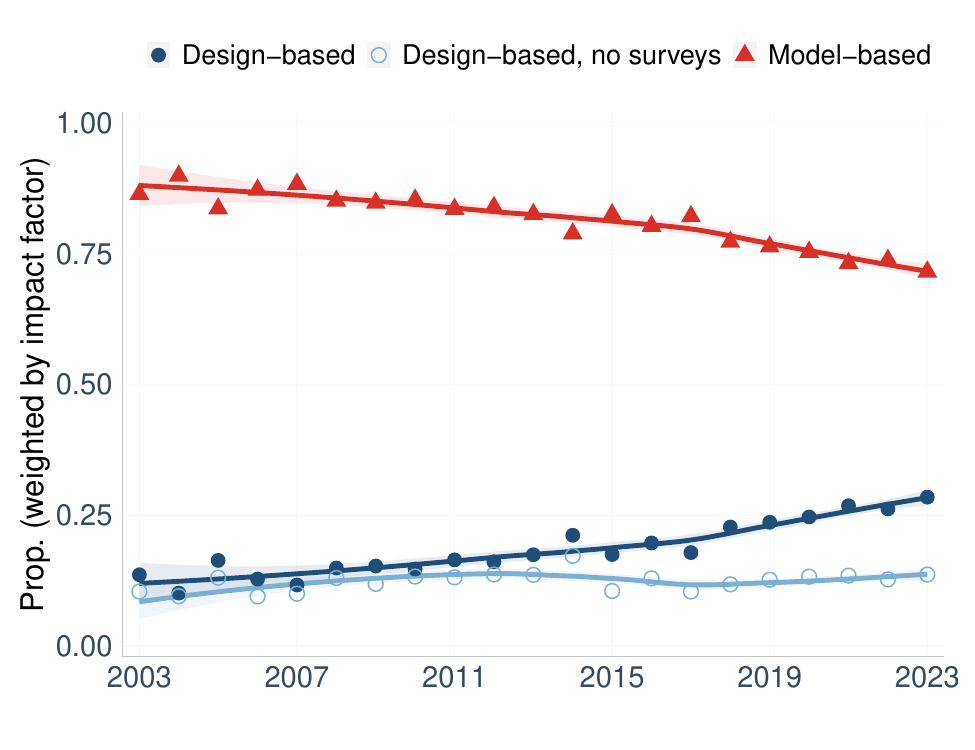}
        \caption{Studies published in other journals}
        \label{fig:bottomrest}
    \end{subfigure}
    \fignote{Shares are shown both including and excluding survey experiments. The denominator is the total number of explanatory empirical quantitative studies published each year, weighted by SJR (a measure of journal impact). The left panel shows papers in the top 20 journals; the right panel shows papers in other journals. Curves with 95\% confidence intervals are lowess fits.}
\end{figure}

Figure~\ref{fig:weighted} plots the share of design- and model-based studies among explanatory quantitative articles, weighting each article by the SJR score of its publishing journal.\footnote{We weight by SJR, a predetermined attribute of the outlet, rather than by each article's own citation count, because article citations are the outcome we analyze in the next section; weighting by them would fold that later comparison into this one. The weighted share therefore describes the methodological composition of the literature as encountered by likely readers, giving greater weight to more influential outlets.} The left panel reports trends for the 20 highest-impact journals, while the right panel reports all remaining outlets. Darker shades include survey experiments, while  lighter shades exclude them. After weighting, design-based approaches rise more rapidly in the highest-impact journals and, when survey experiments are included, surpass model-based approaches in 2021. In the remaining journals, model-based approaches remain predominant throughout the period. Excluding survey experiments, the design-based share in these outlets rises only from 10\% in 2003 to 14\% in 2023. Survey experiments therefore account for much of the growth outside the top journals.\footnote{Figure~\ref{fig:unweightedtiers} in the Supplementary Materials shows the corresponding unweighted trends. The same journal-tier pattern remains: design-based work grows faster in the top 20 journals than elsewhere, although it reaches parity with model-based work only in 2023 rather than 2021.} Overall, methodological change has been concentrated in the most visible segment of the discipline.

\begin{figure}[!ht]
    \centering
    \caption{Design-based Studies among the Top-20 Journals}
    \label{fig:journals}
    \includegraphics[width=1\linewidth, trim={0cm 0.5cm 0cm 0.5cm}, clip]{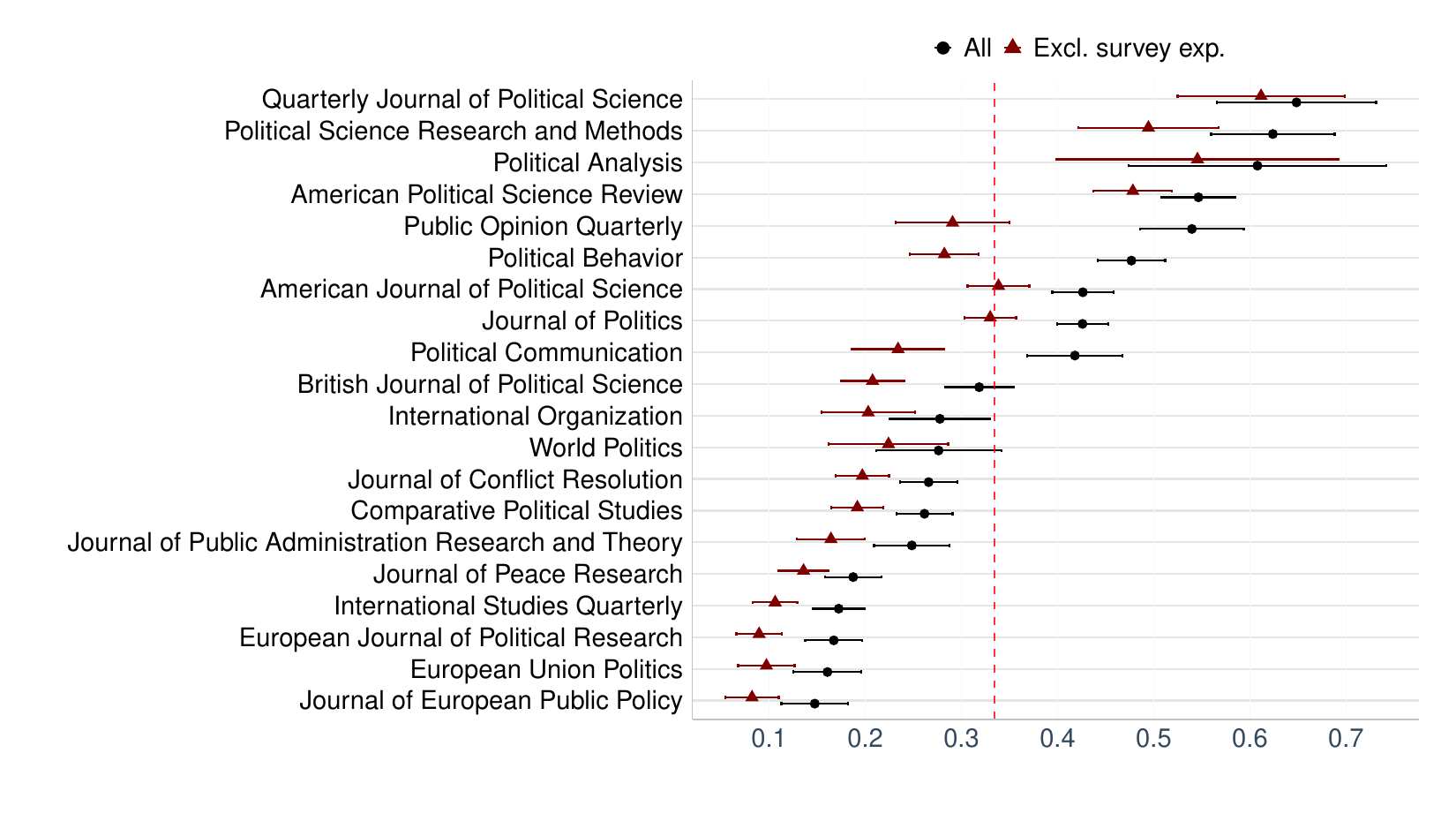}
    \fignote{Journals are the 20 highest-ranked political science outlets by SJR score. Estimates carry 95\% confidence intervals. The numerator is the number of design-based studies published between 2003 and 2023. The denominator is the total number of explanatory empirical quantitative papers published during the same period. Circles show proportions including survey experiments. Triangles show proportions excluding survey-experiments from both the numerator and the denominator.}
\end{figure}

Impact factor weighting captures differences in journal influence but does not show where those differences originate. Figure~\ref{fig:journals} reports the share of design-based studies in the top 20 political science journals between 2003 and 2023, both including survey experiments (black circles) and excluding them (red triangles). The figure shows substantial heterogeneity across journals. Design-based research accounts for more than 60\% of papers in \emph{Quarterly Journal of Political Science} and \emph{Political Science Research and Methods}, but less than 20\% in outlets such as the \emph{Journal of European Public Policy}.

The figure also reveals substantial variation in how much survey experiments contribute to each journal's design-based share. The gap is especially large in public opinion- and behavior-focused journals such as \emph{Public Opinion Quarterly} (54\% versus 29\%) and \emph{Political Behavior} (48\% versus 28\%), consistent with the roots of survey experiments in social psychology and survey research. The gaps in general-interest journals are smaller but still meaningful. Excluding survey experiments reduces the design-based share from 55\% to 48\% in the \emph{American Political Science Review}, from 43\% to 34\% in the \emph{American Journal of Political Science}, and from 43\% to 33\% in the \emph{Journal of Politics}.

\FloatBarrier

\paragraph*{Authors from higher-ranked institutions are more likely to adopt design-based methods.} 

Design-based methods appeared earlier and are more common in the discipline's top journals. Next, we ask whether adoption also varies across institutions. We extract all authors in our corpus, identify their institutional affiliations at the time of publication, and match those institutions to the Academic Ranking of World Universities (ARWU). We then assign each paper the mean rank of its matched authors. This procedure links 111,657 paper--author pairs to universities appearing in ARWU. At the paper level, 77.6\% of explanatory quantitative papers---the subset for which we classify research designs---have at least one author at a university ranked in the year of publication. The remaining papers have authors only at universities outside the ranking, at research institutes or organizations outside academia, or with affiliations that Scopus records incompletely or not at all. Section~\ref{rankingmeth} of the Supplementary Materials details the matching procedure, describes unmatched affiliations, and examines incomplete coverage further.

\begin{figure}[!ht]
    \centering
    \caption{Adoption of Design-based Methods by Institutional Rank}
    \label{fig:institutional}
    \includegraphics[width=.7\linewidth, trim={0cm 0.5cm 0cm 0.5cm}, clip]{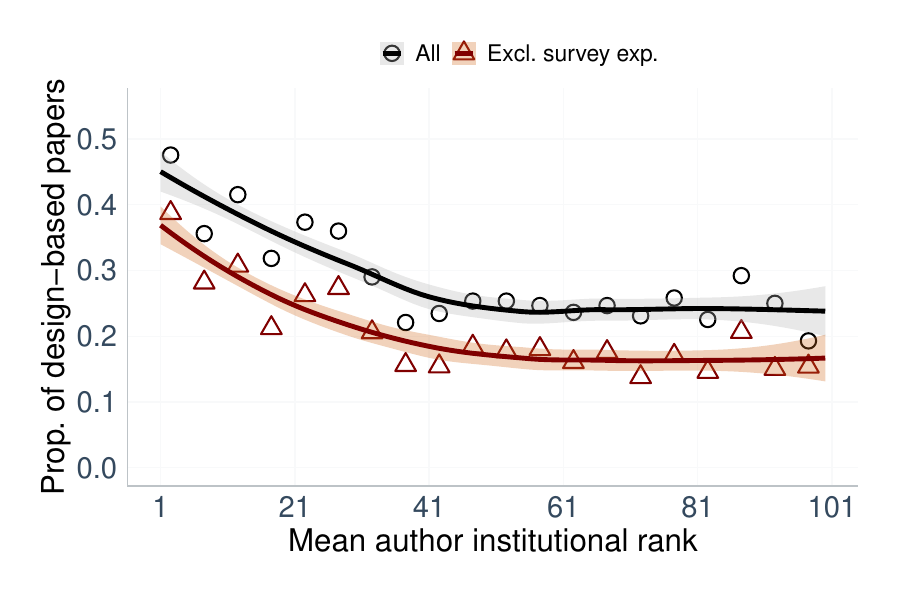}
    \fignote{Each paper enters at the mean ARWU rank of the authors matched to a university ranked in the year of publication. Lower ranks indicate more prestigious institutions. Points show the share of design-based papers within bins of five ranks, with circles for the full sample and triangles for the sample excluding survey experiments. The denominator is the number of explanatory empirical quantitative papers in the bin. Lines show loess fits with 95\% confidence bands.}
\end{figure}

Figure~\ref{fig:institutional} plots the design-based share of explanatory quantitative papers against mean author rank for the 9,457 papers with a mean rank of 100 or better, including survey experiments in black and excluding them in red. Design-based methods are more common at higher-ranked institutions under both definitions. Among papers from the highest-ranked institutions, 47.5\% are design-based, or 38.7\% when survey experiments are excluded. Both shares decline steadily as rank worsens through rank 40. Between ranks 41 and 100, the relationship flattens, with the design-based share fluctuating around 25\%, or 17\% without survey experiments.\footnote{The negative gradient over ranks 1--40 is robust to restricting the sample to papers for which every author's rank is observed, using ranks from the nearest available ARWU year, or adversarially imputing missing ranks.} The institutional pattern therefore mirrors the journal results: design-based methods are disproportionately concentrated among authors at highly ranked institutions.

\paragraph*{Credibility-enhancing practices progressed modestly.}\label{errorates}

Beyond research design, scholars may adopt practices intended to reduce inferential risks and strengthen the credibility of explanatory claims. We group these practices according to whether they primarily address false positives or false negatives. We examine four indicators: placebo tests and pre-analysis plans, which primarily address false positives, and sample size and power analysis, which primarily address false negatives. Because norms differ across research designs, we report trends separately for experiments, design-based observational studies, and model-based studies.

\begin{figure}[!ht]
    \centering
    \caption{Methodological Practices by Research Design Type}
    \label{fig:errorates}
    \begin{subfigure}[b]{0.90\textwidth}
        \centering
        \includegraphics[width=\linewidth]{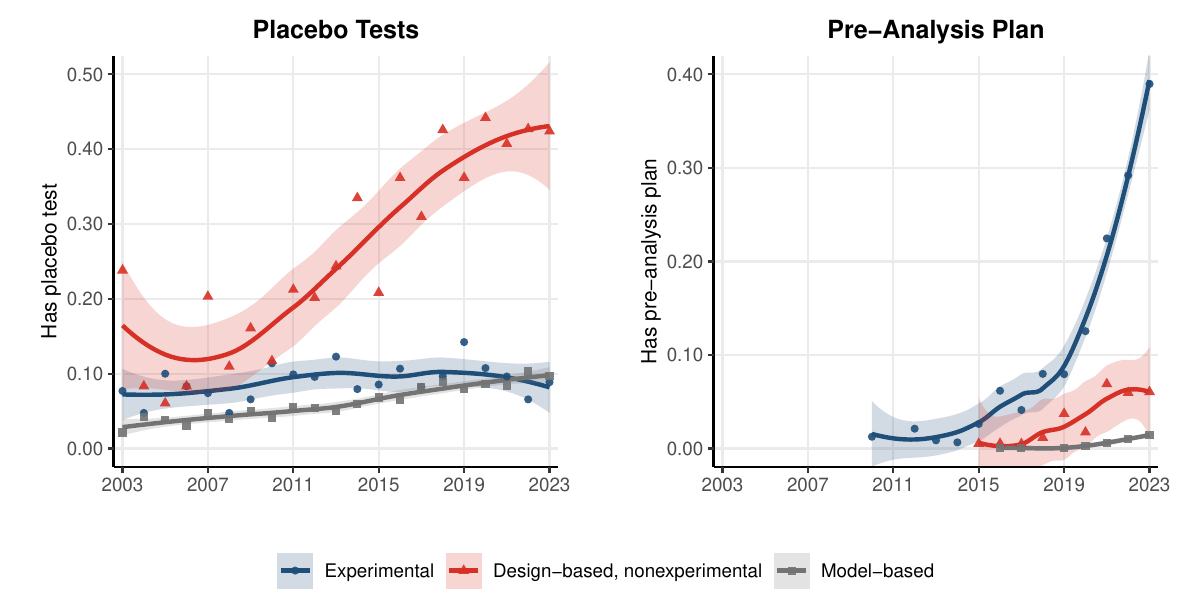}
        \caption{False positive inference}
        \label{fig:type1}
    \end{subfigure}
    \hfill
    \begin{subfigure}[b]{0.90\textwidth}
        \centering
        \includegraphics[width=\linewidth]{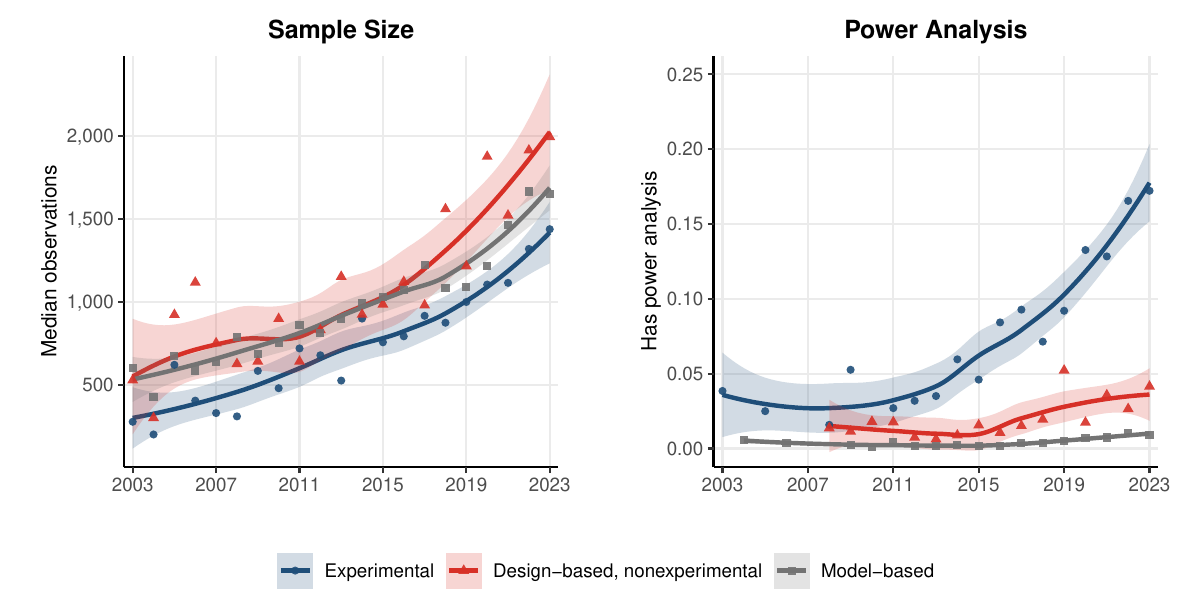}
        \caption{False negative inference}
        \label{fig:top2}
    \end{subfigure}
    \fignote{Panel (a) shows practices that reduce false positives: placebo tests and pre-analysis plans. Panel (b) shows practices that reduce false negatives: sample size and power analysis.}
\end{figure}

Figure~\ref{fig:errorates} presents the results. Practices aimed at limiting false positives have expanded mainly within design-based research, but in different ways across experimental and observational studies. Placebo tests are concentrated in design-based observational work, rising from 12\% of papers in the early 2000s to 42\% in 2023. Pre-analysis plans, by contrast, are concentrated in experiments. None of the papers in our corpus referenced one before 2010, compared with 39\% of experimental papers in 2023. They remain much less common in design-based observational studies (6\%) and nearly absent in model-based work (1.4\%).

Practices aimed at reducing false negatives show broader change. Most notably, median sample sizes increased substantially across all research designs. This trend is consistent with growing attention to statistical power. Sample size, however, is a noisy indicator of such attention. Increases may reflect mechanical features of data accumulation over time, such as longer panel series, as well as shifts in empirical focus from national to sub-national units of analysis that mechanically generate more observations. Power analyses themselves remain largely confined to experimental work, rising from 2\% in the early 2000s to 17\% in 2023. They remain rare in design-based observational studies (4\%) and model-based research (under 1\%), despite their relevance for interpreting null findings across designs \citep{arel2026quantitative}.

Taken together, these patterns point to gradual but uneven diffusion of practices aimed at reducing false-positive and false-negative findings. Efforts to limit false positives have expanded mainly within design-based research, but through different tools: experiments increasingly use pre-analysis plans, while observational studies rely more on placebo tests. Model-based studies show little uptake of either practice. Efforts to limit false negatives, such as power analysis, remain concentrated in experiments, even as sample sizes have grown substantially across all research designs.

\FloatBarrier

\subsection*{Citation Differences}\label{sec:impact} 

If the credibility revolution changed what the discipline values, design-based research may receive greater professional rewards than model-based work. We examine citations as one descriptive measure of those rewards. Despite their limitations, citations provide a useful proxy for scholarly attention and influence: they capture what gets read, built upon, and recognized within the field.

We define the \emph{citation gap} as the difference in average citations between design-based and model-based papers published in the same year. This gap is descriptive rather than causal. Design-based and model-based papers differ systematically in topic, subfield, journal placement, author characteristics, and research quality, any of which may contribute to citation differences. For example, more productive or better-resourced scholars may both adopt design-based methods earlier and receive more citations for reasons unrelated to method. We therefore interpret the citation gap as evidence about how scholarly attention is distributed across the two types of research, not as the causal effect of research design on citations.

\begin{figure}[!ht]
    \centering
    \caption{Citation Gap between Design-based and Model-based Studies}
    \label{fig:citemeansoverall}
    \begin{subfigure}[b]{0.49\textwidth}
        \centering
        \includegraphics[width=\linewidth]{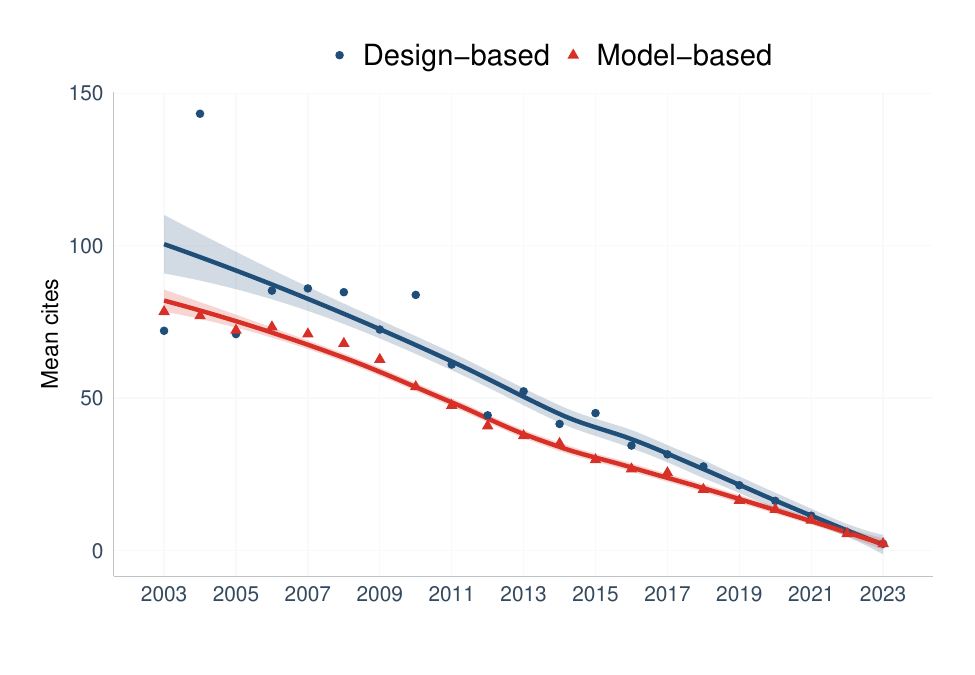}
        \caption{Citation count in 2023 by publication year}
        \label{fig:cite.raw}
    \end{subfigure}
    \hfill
    \begin{subfigure}[b]{0.49\textwidth}
        \centering
        \includegraphics[width=\linewidth]{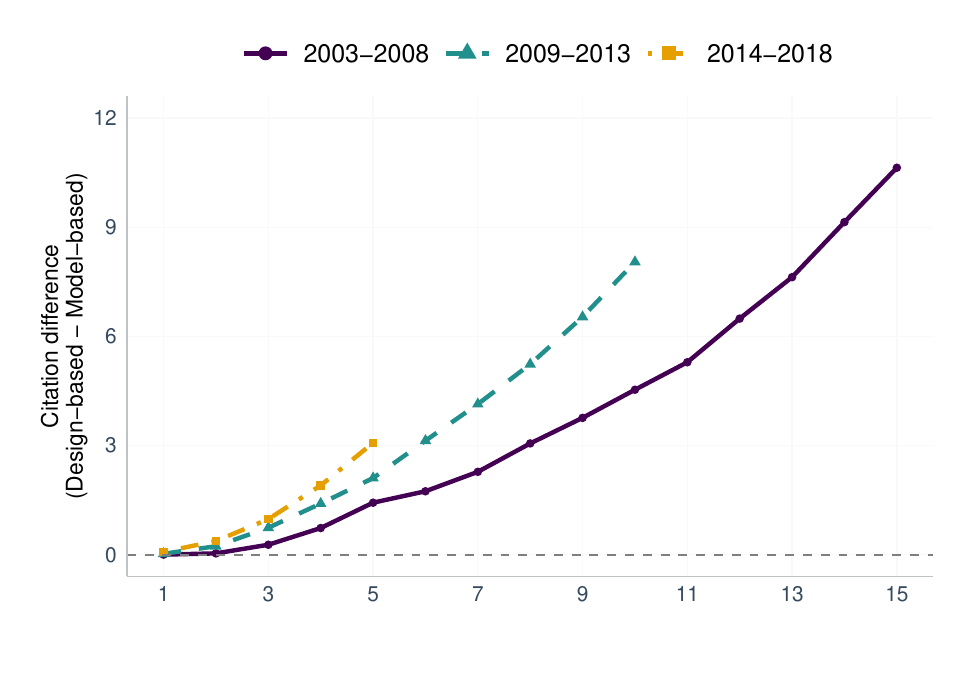}
        \caption{Citation gap by publication year}
        \label{fig:cite.sd}
    \end{subfigure}
    \fignote{The left panel shows the average citation count for design-based (blue) and model-based (red) studies by publication year in 2023. The right panel shows the citation gap (the difference in average citations between design-based and model-based studies) measured relative to the publication year for three cohorts of papers defined by their year of publication.}
\end{figure}

Comparing citation levels across years is complicated because citations accumulate over time: more recent papers have had fewer years to be cited, while design-based methods have become more common in later cohorts. We therefore examine both raw citation counts by publication year and cohort-specific citation trajectories.Figure~\ref{fig:citemeansoverall}(a) reports average citations for design-based and model-based papers by publication year. Between 2006 and 2021, design-based papers receive more citations than model-based papers published in the same year. The gap is largest for papers published around 2010, at roughly thirty citations, and narrows toward zero in the most recent years, when papers of both types have had little time to accumulate citations. Before 2004, the design-based cohorts contain fewer than fifty papers per year, making the yearly means correspondingly noisy.

Figure~\ref{fig:citemeansoverall}(b) plots the citation gap for three publication cohorts---2003-2008, 2009--2013, and 2014--2018---as a function of time since publication. For each cohort and each year since publication, we compute the difference between the mean citations accumulated by design-based and model-based papers. We follow each cohort only for years in which all papers are observable. The earliest cohort therefore extends fifteen years after publication, while the most recent extends five. The citation gap is positive in every cohort and grows with time since publication. It also grows faster in more recent cohorts. Five years after publication, the gap is 1.4 citations for papers published between 2003 and 2008, 2.1 for those published between 2009 and 2013, and 3.1 for those published between 2014 and 2018.

We complement these raw comparisons with regression-based comparisons of design-based and model-based papers published in the same year and subfield, reported in the Supplementary Materials Section~\ref{sisec:citeprem}. We regress the citation count of each paper on an indicator for whether it uses a design-based method, adding publication-year and subfield fixed effects. Publication-year fixed effects account for differences in the time available to accumulate citations, while subfield fixed effects account for broad differences in citation patterns across areas of the discipline. We do not condition on journal placement in the main specification because publication venue may itself be part of the professional rewards associated with methodological choice. The resulting estimates remain descriptive rather than causal, but provide a more constrained comparison, among papers published at the same time and within the same broad area of political science. With this strategy, we again find that design-based papers receive, on average, more citations than model-based papers published in the same year and subfield (Table~\ref{tab:citation_regression}). The adjusted difference is 11.9 citations, on average, larger than the raw gap of 6.2 citations. Further, we find that all period-specific estimates are positive and statistically indistinguishable from each other.

\FloatBarrier

\section*{Discussion}\label{sec:discussion}

Our analysis paints a mixed picture of the credibility revolution in political science. On the one hand, design-based methods have moved from the margins into the mainstream of explanatory empirical quantitative research, underpinning more than one-third of such papers. Design-based papers also tend to accumulate more citations than model-based papers published in the same year. On the other hand, this transformation remains highly stratified. It is concentrated in top journals and among authors at highly ranked institutions. Credibility-enhancing practices beyond research design, such as pre-analysis plans, power analyses, and placebo tests, remain far from universal even among design-based studies. In this sense, the credibility revolution in political science is better described as a substantial but partial reform than as a wholesale reordering of empirical practice.

Several developments are nonetheless incontestable. First, the discipline has become less reliant on parametric modeling as the default route to causal claims. The growth of experiments, regression discontinuity, and other designs has shifted attention from functional-form assumptions toward explicit claims about treatment assignment and identifying variation. Second, researchers are now more explicit about the assumptions required for causal interpretation: more than 77\% of design-based papers state an identification assumption. Third, median sample sizes have increased across methodological families, which, all else equal, should improve statistical power and reduce the risk that findings are driven by small-sample noise. Together, these changes move the discipline closer to standards emphasized by proponents of the credibility revolution.

At the same time, the evidence suggests that the credibility revolution is not especially deep in at least two respects. First, the adoption of design-based methods is highly concentrated in survey experiments, which account for nearly half of all design-based papers by the end of the period (49\%), and, to a lesser extent, difference-in-differences (17\%). Other designs central to the credibility revolution remain rare, such as regression discontinuity and synthetic control, or show non-monotonic patterns of use, such as  instrumental variables. This concentration is not hard to explain. Relative to survey experiments, selection on observables, regression discontinuity, and instrumental variable designs often require more demanding empirical conditions or identifying assumptions. Field experiments face a different constraint: their substantial costs in time and money naturally limit how widely they can be adopted.

Second, beyond identification assumptions, other credibility-enhancing practices remain weakly institutionalized. Placebo tests appear in only about one-third of design-based observational studies and are rare elsewhere. Pre-analysis plans and power analyses remain concentrated in experimental work and are still the exception even there. If the goal is to make published evidence more diagnostic of causal claims, these practices would need to become more routine across research designs, not just within a subset of high-profile experiments. 

The reach of the credibility revolution is also uneven. Design-based methods were adopted earlier and at higher rates in the discipline's most prestigious journals and among authors at top-ranked institutions. Regression discontinuity designs and field experiments, for example, are disproportionately concentrated in top-10 outlets and among authors at highly ranked institutions. These patterns are descriptive and may reflect differences in training, resources, submission strategies, and the extent to which credibility-revolution principles have been internalized. They nonetheless raise the possibility of a two-tiered discipline in which standards for credible causal inference vary by institutional location---for a similar conclusion, see \citet{grossman2025political}.

Importantly, a descriptive account of changing praxis, such as the one presented here, can only document changes in how researchers justify causal claims. It cannot establish whether published causal claims are more likely to be correct. Several existing research programs address this latter question through replication and reproducibility audits that re-execute published analyses \citep{stommes2023reliability, lal2024much, chiu2025causal, xu2026scaling}; benchmarking exercises that compare observational estimates with experimental ground truth \citep{imbens2025comparing}; and meta-analytic diagnostics of selective reporting and low power \citep{brodeur2024preregistration, arel2026quantitative}. The common logic of these efforts is to subject published claims to external tests rather than infer credibility from methodological choice alone. Findings from this research caution against equating a credible design with a credible result: assumption failures, low power, and execution errors occur across methods, including design-based work.

A second, and in our view unresolved, question is whether design-based research provides more credible answers only to tractable questions, thereby narrowing the scope of the discipline. Methods that identify effects cleanly under stringent conditions may systematically avoid substantively important but identification-resistant problems, such as state formation or democratization \citep{mahoney2006tale}. Assessing whether such a trade-off exists, and to what extent, is difficult because it requires a measure of the importance of a research question. Any such metric would embed a contested judgment about what political science should value. Imperfect proxies exist---such as the locality of the estimands a literature identifies \citep{SloughTyson2024}, or the distribution of design-based methods across substantive questions, which our corpus can characterize---but none resolves the underlying normative issue. Evaluating the overall epistemic payoff of the credibility revolution therefore requires both a tractable program for assessing accuracy and sustained argument about relevance and importance that measurement alone cannot settle. Our contribution is more limited: we document how far these methods have spread. Whether that spread has made the discipline more trustworthy or narrowed its agenda remains a separate question that this descriptive baseline may help others pursue.

Our findings warrant an important interpretive caveat. Because our measures rely on how authors describe their designs and aims, we cannot fully separate changes in research practice from changes in how scholars present their work. Evolving reporting norms could lead us to overstate or understate the influence of the credibility revolution. For example, if researchers have become more cautious in making causal claims, model-based studies may now use fewer or more qualified causal statements even without substantive changes in practice, causing us to understate change. Conversely, if scholars increasingly describe their work as ``descriptive'' rather than ``explanatory,'' the pool of explanatory studies would shrink, potentially inflating the apparent rise of design-based methods within that category. We cannot directly test these dynamics, but any intellectual movement that redefines standards of causality is likely to affect both how causal claims are evaluated and how they are articulated. Descriptive evidence in Section~\ref{sisec:altexps} of the Supplementary Materials suggests that these processes are not the main drivers of our results: the share of descriptive quantitative studies remains essentially flat over time, and the rate at which model-based papers discuss identification assumptions remains low and stable throughout the period.

In sum, the credibility revolution is reshaping, but not displacing, other modes of inquiry. Explanatory quantitative research remains a minority, comprising about 33\% of articles in our corpus. Although its share has grown, it has not supplanted the discipline's methodological diversity. Our findings instead point to an emerging equilibrium in which design-based research coexists with qualitative, historical, ethnographic, and interpretive approaches, as well as formal, methodological, and normative theory. As standards for causal inference evolve, these other traditions remain essential for studying meaning, context, complexity, and contingency. Qualitative scholars, for example, have long developed tools for causal inference, including process tracing and within-case analysis, that rely on different logics and forms of evidence. Descriptive work likewise continues to play a vital role in mapping new phenomena, measuring key constructs, and shaping research agendas. From this perspective, the credibility revolution may be most productive when it contributes to a broader division of labor across methods rather than elevating any single methodological paradigm.


\vspace{3em}
\singlespacing
\bibliography{CausalClaimsBib}
\clearpage
\onehalfspacing

\appendix
\onehalfspacing
\setcounter{page}{1}
\setcounter{table}{0}
\setcounter{figure}{0}
\setcounter{equation}{0}
\setcounter{footnote}{0}
\renewcommand\thetable{A\arabic{table}}
\renewcommand\thefigure{A\arabic{figure}}
\renewcommand{\thepage}{A-\arabic{page}}
\renewcommand{\theequation}{A\arabic{equation}}
\renewcommand{\thefootnote}{A\arabic{footnote}}
\let\footnotesize\originalfootnotesize

\vspace{0em}
\section{Online Supplementary Materials}
\bigskip
\noindent\vspace{0em}{\bf \underline{Table of Contents}}

\startcontents[appendix]
\printcontents[appendix]{}{2}{\setcounter{tocdepth}{3}}
\setlength{\parskip}{0.3em}

\clearpage

\subsection{Data and sample}

\subsubsection{Data collection and access}

Our analysis draws on a comprehensive corpus of political science research assembled for a companion study \citep{grossman2025political}. We briefly describe the dataset and its construction here; full details appear in that paper's appendix.

We began with 188 journals classified as political science by Clarivate's Web of Science platform that have a Scientific Journal Rankings (SJR) score of at least one. An SJR below one substantively means that papers in a journal occupy significantly less visibility  and  impact  than  the  average  paper  in  the  discipline. From this list, we removed journals that were not indexed in Scopus, not peer-reviewed, or not published in English, yielding a final sample of 174 journals. Using the Scopus API, we extracted journal-level metadata---including citation statistics---and article-level information (authors, title, abstract, publication date, and DOI) for all 129,751 articles published in these journals between 2003 and 2023.

Wherever possible---and wherever retrieval was consistent with the text and data mining provisions in the terms of service of the publishers (e.g. Taylor \& Francis states: ``If you or your institution subscribes to content from Taylor \& Francis you can carry out text and data mining activities on this content, as well as open access content, without any additional charge, provided this is on a non-commercial basis.'')---we obtained the full text of these articles, accessing them through institutional library subscriptions. We successfully retrieved the complete text for 91,632 articles from 156 journals. Importantly, our use of article text is limited to computational analysis; we do not (and will not) redistribute, republish, or publicly share any of the raw text we retrieved. Table \ref{tab:insample_text} shows the universe of journals included in our full-text sample. 

\setlength\LTleft{0pt}
\setlength\LTright{0pt}


{
\footnotesize
\begin{longtable}{p{0.32\textwidth}p{0.32\textwidth}p{0.32\textwidth}}
\caption{Journal Sample Composition: In Sample (With Text)} 
\label{tab:insample_text} \\
\hline
\textbf{Journal Title} & \textbf{Journal Title} & \textbf{Journal Title} \\
\hline
\endfirsthead
\hline
\textbf{Journal Title} & \textbf{Journal Title} & \textbf{Journal Title} \\
\hline
\endhead
Acta Politica & African Affairs & American Journal of Political Science \\
American Political Science Review & American Politics Research & Annals of the American Academy of Political and Social Science \\
Armed Forces and Society & Australian Journal of Political Science & Australian Journal of Politics and History \\
British Journal of Political Science & British Journal of Politics and International Relations & British Politics \\
Business and Politics & Canadian Journal of Political Science & Citizenship Studies \\
Communist and Post-Communist Studies & Comparative European Politics & Comparative Political Studies \\
Comparative Politics & Contemporary Political Theory & Contemporary Security Policy \\
Contemporary Southeast Asia & Cooperation and Conflict & Critical Review \\
Current History & Democratization & East European Politics \\
East European Politics and Societies & Economics and Politics & Electoral Studies \\
Environmental Politics & Ethics and Global Politics & Ethics and International Affairs \\
Europe-Asia Studies & European History Quarterly & European Journal of Political Economy \\
European Journal of Political Research & European Political Science & European Political Science Review \\
European Security & European Union Politics & Forum (Germany) \\
Geopolitics & German Politics & Global Environmental Politics \\
Global Policy & Governance & Government and Opposition \\
Historical Materialism & Human Rights Quarterly & Intelligence and National Security \\
International Affairs & International Environmental Agreements: Politics, Law and Economics & International Feminist Journal of Politics \\
International Journal of Conflict and Violence & International Journal of Press/Politics & International Journal of Public Opinion Research \\
International Organization & International Political Science Review & International Political Sociology \\
International Politics & International Studies Quarterly & International Studies Review \\
International Theory & Irish Political Studies & Japanese Journal of Political Science \\
Journal of Chinese Political Science & Journal of Common Market Studies & Journal of Conflict Resolution \\
Journal of Contemporary European Studies & Journal of Current Southeast Asian Affairs & Journal of Democracy \\
Journal of Elections, Public Opinion and Parties & Journal of European Integration & Journal of European Public Policy \\
Journal of Human Rights & Journal of Information Technology and Politics & Journal of International Relations and Development \\
Journal of Peace Research & Journal of Policy History & Journal of Political Philosophy \\
Journal of Politics & Journal of Public Administration Research and Theory & Journal of Public Policy \\
Journal of Strategic Studies & Journal of Theoretical Politics & Journal of Women, Politics and Policy \\
Latin American Perspectives & Latin American Politics and Society & Legislative Studies Quarterly \\
Local Government Studies & Mediterranean Politics & Nationalities Papers \\
Nations and Nationalism & New Left Review & New Political Economy \\
PS - Political Science and Politics & Parliamentary Affairs & Party Politics \\
Perspectives on Politics & Philosophy and Public Affairs & Policy Studies Journal \\
Policy and Internet & Political Analysis & Political Behavior \\
Political Communication & Political Geography & Political Psychology \\
Political Quarterly & Political Research Quarterly & Political Science \\
Political Science Quarterly & Political Science Research and Methods & Political Studies \\
Political Studies Review & Political Theory & Politicka Ekonomie \\
Politics & Politics and Gender & Politics and Governance \\
Politics and Religion & Politics and Society & Politics, Philosophy and Economics \\
Polity & Post-Soviet Affairs & Presidential Studies Quarterly \\
Problems of Post-Communism & Public Administration & Public Choice \\
Public Opinion Quarterly & Publius: The Journal of Federalism & Quarterly Journal of Political Science \\
Regulation and Governance & Research and Politics & Review of African Political Economy \\
Review of International Organizations & Review of International Political Economy & Review of Policy Research \\
Revista Brasileira de Politica Internacional & Scandinavian Political Studies & Scottish Journal of Political Economy \\
Social Movement Studies & Social Science Quarterly & Socio-Economic Review \\
South European Society and Politics & Studies in American Political Development & Studies in Comparative International Development \\
Studies in Conflict and Terrorism & Survival & Swiss Political Science Review \\
Telos & Territory, Politics, Governance & Terrorism and Political Violence \\
The International Journal of Transitional Justice & West European Politics & World Politics \\
\hline
\end{longtable}
}

\clearpage

\subsubsection{SCImago Journal Rank}

SJR (SCImago Journal Rank) is a metric used in Scopus to assess the impact and prestige of
academic journals. A key feature of SJR is that it is a Prestige-Weighted Metric: it measures
the scientific influence of journals by considering both the number of citations they receive
and the prestige of the journals where those citations come from. Unlike raw citation counts,
SJR assigns higher value to citations from more influential journals. Specifically, SJR is the
number of citations in a specific year (e.g., 2023) to articles published in the previous three
years (e.g., 2020–2022), considering the prestige of the citing journals.

\subsubsection{Sample construction}

Table \ref{tab:sankey} reports the sample construction, as shown in Figure \ref{fig:sankey} in the main text. Each panel corresponds to a classification stage. Of the 91,632 papers in our corpus, 35,349 (38.6\%) are classified as empirical quantitative. Among these, 29,987 (84.8\%) are coded as having an explanatory goal. Last, Panel C breaks down the explanatory papers by method family.

\begin{table}[htbp]
\centering
\caption{Sample Construction}
\label{tab:sankey}
\small
\setlength{\tabcolsep}{8pt}
\renewcommand{\arraystretch}{1.08}
\begin{tabular}{@{}lrr@{}}
\toprule
\textbf{Sample category} & \textbf{N} & \textbf{\%} \\
\midrule
\multicolumn{3}{@{}l}{\textbf{Panel A: All studies} \hfill \textit{N = 91,632}} \\
\addlinespace[2pt]
\quad Empirical quantitative & 35,349 & 38.6 \\
\quad Qualitative, normative, formal, review, or methodological & 56,283 & 61.4 \\
\addlinespace[6pt]
\multicolumn{3}{@{}l}{\textbf{Panel B: By goal} \hfill \textit{N = 35,349}} \\
\addlinespace[2pt]
\quad Explanatory & 29,987 & 84.8 \\
\quad Descriptive &  5,115 & 14.5 \\
\quad Predictive  &    247 &  0.7 \\
\addlinespace[6pt]
\multicolumn{3}{@{}l}{\textbf{Panel C: By method} \hfill \textit{N = 29,987}} \\
\addlinespace[2pt]
\quad Design-based &  7,492 & 25.0 \\
\quad Model-based  & 22,495 & 75.0 \\
\bottomrule
\end{tabular}
\end{table}

\clearpage

\subsubsection{Descriptive statistics}

Table \ref{tab:summstats} reports summary statistics for the main variables used in the analysis.\\

{
\fontsize{9pt}{9pt}\selectfont

\setlength{\tabcolsep}{4pt}
\renewcommand{\arraystretch}{1.1}
\setlength{\aboverulesep}{1pt}
\setlength{\belowrulesep}{1pt}
\setlength{\LTpre}{0pt}
\setlength{\LTpost}{0pt}

\begin{longtable}[c]{@{}lrcc@{}}
\caption{Summary Statistics}
\label{tab:summstats}\\[-0.5em]

\toprule
 & Obs. & Mean & SD \\
\midrule
\endfirsthead

\toprule
 & Obs. & Mean & SD \\
\midrule
\endhead

\midrule
\multicolumn{4}{r@{}}{\textit{Continued on next page}}\\
\endfoot

\bottomrule
\endlastfoot

\multicolumn{4}{@{}l}{\textbf{Panel A: Full corpus} $(N=91{,}632)$}\\[-1pt]
\multicolumn{4}{@{}l}{\textit{Subfield (first-stage classification)}}\\[-1pt]
\quad Comparative Politics            & 45,221 & 0.49 & 0.50 \\
\quad International Relations         & 17,240 & 0.19 & 0.39 \\
\quad American Politics               & 10,529 & 0.11 & 0.32 \\
\quad Public Policy/Administration    &  8,162 & 0.09 & 0.28 \\
\quad Political Theory and Philosophy &  7,424 & 0.08 & 0.27 \\
\quad Methodology and Formal Theory   &  1,830 & 0.02 & 0.14 \\
\quad Other                           &  1,226 & 0.01 & 0.11 \\[1pt]

\multicolumn{4}{@{}l}{\textit{Paper classification}}\\[-1pt]
\quad Empirical quantitative          & 91,632 & 0.39 & 0.49 \\[1pt]

\multicolumn{4}{@{}l}{\textit{Goal of analysis (emp.\ quant., $N=35{,}349$)}}\\[-1pt]
\quad Explain                         & 29,987 & 0.85 & 0.36 \\
\quad Describe                        &  5,115 & 0.14 & 0.35 \\
\quad Predict                         &    247 & 0.01 & 0.08 \\[1pt]

\multicolumn{4}{@{}l}{\textit{Citations}}\\[-1pt]
\quad Citation count                  & 91,623 & 25.76 & 64.32 \\

\midrule
\multicolumn{4}{@{}l}{\textbf{Panel B: Credibility practices} $(N=29{,}987)$}\\[-1pt]
\quad Sample size (thousands)         & 26,986 & 254.4 & 27,400.8 \\
\quad Placebo test                    & 29,987 & 0.11 & 0.31 \\
\quad ID assumptions stated           & 29,987 & 0.26 & 0.44 \\
\quad Power analysis                  & 29,987 & 0.02 & 0.13 \\
\quad Pre-analysis plan               & 29,987 & 0.02 & 0.15 \\
\quad Claims significant results      & 29,987 & 0.95 & 0.22 \\[1pt]

\multicolumn{4}{@{}l}{\textit{Data source}}\\[-1pt]
\quad Public or existing data         & 19,593 & 0.65 & 0.48 \\
\quad Original survey                 &  4,657 & 0.16 & 0.36 \\
\quad Experimental                    &  3,882 & 0.13 & 0.34 \\
\quad Restricted or proprietary       &  1,595 & 0.05 & 0.22 \\
\quad Other                           &    260 & 0.01 & 0.09 \\

\midrule
\multicolumn{4}{@{}l}{\textbf{Panel C: Impact and stratification} $(N=29{,}987)$}\\[-1pt]
\quad Citation count                  & 29,987 & 30.81 & 60.60 \\
\quad Journal impact factor           & 29,617 & 2.07 & 1.60 \\
\quad Published in top-20 journal     & 29,987 & 0.36 & 0.48 \\
\quad Single-country study            & 29,828 & 0.60 & 0.49 \\

\midrule
\multicolumn{4}{@{}l}{\textbf{Panel D: Causal claims} $(N=29{,}987)$}\\[-1pt]
\quad Explicit causal claim           & 29,987 & 0.73 & 0.45 \\

\bottomrule
\end{longtable}
}
\clearpage

\subsection{Measures and the LLM pipeline}

We label the corpus in two stages, using a different model at each stage. In the first, we send the full text of every article to GPT-4o, together with the prompt and JSON schema reproduced in Section~\ref{sisec:prompt}. The prompt asks the model to code nineteen dimensions, all of which were used in the original submission. The current manuscript retains two: whether the paper is an empirical quantitative study and the subfield to which it belongs.

The second stage replaces all subsequent classification steps in the original submission. Each of the 37,163 papers that passes the first screen is sent to Claude Sonnet 5 with the instructions reproduced in Section~\ref{sisec:prompt_v8e}. The model first reapplies the screen to the full text, excluding 1,814 papers as qualitative, methodological, formal-theoretic, or review articles. It then classifies the purpose of each remaining paper as explanatory, descriptive, or predictive and, for explanatory papers, identifies the research design from a fixed menu. The model also records ten additional attributes, which Section~\ref{sisec:prompt_validation} describes and validates. 

In the original submission, the model returned a residual \texttt{Other} category for the research design, which we subsequently remapped using \texttt{gpt-3.5-turbo}. With the improved prompt used in this revision, we have retired that step: there is no residual category left for a second pass to resolve.

\subsubsection{First-stage classification prompt and schema}\label{sisec:prompt}

The full classification prompt sent to GPT-4o is provided below. All logic, definitions, variable names, conditional requirements, and enumeration options are preserved exactly as used during classification. Instructions typeset in gray correspond to variables whose outputs are no longer used in the current version of the manuscript.

{\scriptsize
\begin{Verbatim}[breaklines=true]
TASK OVERVIEW

You are an expert-level political science paper analysis engine, a domain‑expert engine that reads political‑science manuscripts and classifies them into a series of variables. You must: interpret complex academic content, focusing on research methods, causal claims, data usage, measurements, identification strategies, use domain knowledge to classify subfields, and more.

Your tasks:

-   Read and understand the entire provided paper -- title, abstract, main text, tables/figures, footnotes, etc.

-   Identify the strongest evidence and framing -- focus on the headline findings that the authors highlight.

-   Extract and return the metadata described below, obeying the interpretation rules for each, and responding according to the JSON schema provided.

Note: If a property's type includes "null" (e.g., ["boolean","null"] or ["array","null"]), use null if and only if information in the paper is missing or it the specific variable is not applicable to the study design (e.g. parallel trends assumption isn't relevant to a survey experiment)

Specific instructions and definitions for each property are outlined below:

0. error_in_raw_text: first, before anything else, confirm that the Paper Full TEXT makes sense in regards to the Paper Title. Specifically, pay attention to the fact that the raw text might be specified or corrupt. Select from the following options.

    - No Error — this is if the raw text seems okay.

    - Missing/Corrupt — this is if the raw text is totally missing or unreadable. Note, it is okay if the text is just messy; this variable is reserved for totally broken texts and NaNs.

    - Title/Text Mismatch — this is if it is *obvious* that the raw text does not match the title. For example, if the paper title is talking about Elections in Africa but the paper is about Gender in US Congress, obviously something has gone wrong. It should be very clear if this type of mismatch has occurred.

1.  subfield: Select of the enumerated canonical Political Science subfields that the paper best fits into. Most papers should fall into one of these, but choose 'Other' if truly uncertain. Canonical definitions:

    - American Politics: Empirical or historical study of political behavior, institutions, or processes inside the United States (e.g. Congress, presidency, courts, elections, parties, public opinion, state policy, race/ethnicity and politics, federalism, etc.).

    - Comparative Politics: Comparative Politics: Systematic comparison across sovereign states, sub-national units (states, provinces, districts, etc.), non-state actors that play an important political role (churches, NGOs, traditional authorities like chiefs, etc), or general inference drawn from non-U.S. politics (e.g. regime type, democratization, comparative institutions, political economy, civil conflict within states, elections outside the U.S., etc.).

    - International Relations: Analysis of cross-border political interactions such as interstate conflict and cooperation, trade, international organizations (such as WHO, WTO, IMF), global governance, global public goods, diplomacy, international political economy, transnational actors, security studies (including, but not limited to nuclear proliferation, military alliances, and technology and weaponry). Also include public opinion surveys and survey experiments that seek to capture opinions of the public or elites toward foreign policy and how they influence foreign policy making.

    - Methodology and Formal Theory – Research that develops or critiques analytical methods, measurement, statistics, or formal (game-theoretic) models for broad disciplinary use.

    - Political Theory and Philosophy – Normative reasoning, conceptual analysis, or interpretation of classic and contemporary texts.

    - Public Policy/Administration – Study of policy design, implementation, and evaluation, and bureaucratic behavior within governmental systems. Public administration generally focuses on practical aspects of managing and implementing public policies, programs, and services. For example, it explores the pros and cons of different ways to organize the bureaucracy and its implication for effective delivery of public services, and ensuring that laws and regulations are effectively executed. 

    - Other – Assigned when paper doesn't fit into any subfield or the work lies entirely outside political science.

2.  is_empirical_quant_paper: true if the paper contains any original data analysis using quantitative methods, descriptive, predictive, causal, or other. False if the paper uses empirical qualitative methods (e.g. case studies, process tracing, etc), normative, theory, etc. If this is false, most of the subsequent variables are no longer applicable and should be null.

3.  general_goal_of_analysis: is_empirical_quant_paper == FALSE, this is "null". Otherwise, read the paper and interpret its high level approach. Specifically, code what the 'goal' of the analysis in the paper is in terms of the three distinct categories below. Note, this is not about the design of the paper per say, but more about the question it is trying to answer / its purported motivation.

    - Describe: A paper which seeks to describe a phenomenon that occurs in the world, understand its baseline levels, differences across groups etc. It doesn't seek to explain it, but merely describes something, its nature, its degree, etc. A descriptive analysis.

    - Predict: A paper which seeks to predict an outcome based on a set of variables or determine which variables are most predictive of a perceived relationship, albeit not assuming those variables are the cause/consequence of the relationship. A predictive analysis.

    - Explain: A paper which seeks to actually explain a phenomenon/relationship and identify the causes or consequences of the said phenomenon/relationship. The paper is motivated by a desire to identify a causal effect in the universe (regardless of whether it actually succeeds in this). Its goal is to make an inference. A causal analysis.
\end{Verbatim}
\begin{Verbatim}[breaklines=true, formatcom=\color{gray}]
4a. single_country_study: is the main analysis of this paper focused on a single country?  Select one of: single_country, multiple_countries, or NA if not relevant (e.g. a paper talking only about methods).

4b. single_region: if single_country_study is “multiple_countries”, then select one of: single_region, multiple_region, otherwise leave as null if not “multiple_countries”. single_region should be selected if all countries that are the focus of the main analysis of the paper are from the same global region. Regions are defined as follows: East Asia & Pacific, Europe & Central Asia, Latin America & Cuba, Middle East & North Africa, North America, South Asia, Sub-Saharan Africa.

5. countries_of_focus: enumerate the countries examined in the primary analysis of the paper, their names separated by semicolons. If there is a single country, just return the name of that country. If there are greater than 20 distinct countries, then return “20+ Countries” instead of a complete list.

6. paper_uses_survey_data: does the paper use survey data for its primary analysis? Return one of: no_survey_data, runs_original_survey, uses_public_available_survey (e.g. Afrobarometer, ANES, data from a previously published paper, etc.)

7. uses_original_dataset: if the authors use an original dataset for their paper’s primary analysis, write the type used here. Importantly, the focus here is only the primary independent or dependent variables in the main analysis. Null if the paper doesn't use any data at all. Options follow:

    - original_survey: if the researchers design and field an original survey

    - field_experiment: if the researchers run a field experiment with a treatment produced by the researchers

    - field_study: non-experimental, original, primary data gathering in the field, thus creating a new dataset that didn't exist before

    - structure_systematize: if the researchers take unstructured information and structure/systematize it into usable data (e.g. text-as-data). Note this does not include the manipulation of variables in existing datasets, but rather it must begin with unstructured, non-data information that the authors themselves parse/structure

    - procure_original_data: if the researchers procure, for example, via a FOIA request, data/information heretofore private or unused by other researchers

    - other_original_data: if the researchers use truly original data that doesn't fall into the above categories (e.g. biometric data collected via wearable devices, sensor data from custom-built hardware,  behavioral lab setups not qualifying as a field experiment, etc)

    - not_original: if the researchers use existing datasets, publicly available surveys, combining datasets to create new measures, etc. it does not count as original

8. seeks_determinants: true if a) the paper is an "Explain" paper (null if "Predict" or "Describe") and b) the purported goal of the paper is to investigate which observable factors explain variation in a specific outcome variable observed in the world — in other words, the determinants of a given variable. In other words, if the paper’s primary goal is to quantify how factors jointly or individually drive an outcome, distinct from estimating the effect of a researcher‐defined treatment. Note: do not include papers that specifically explore determinants to treatment assignment in an effort to make a more credible causal argument (often combined with a selection-on-observables assumption); these are ancillary balance or selection checks rather than a substantive determinants inquiry.

9.  main_causal_research_design: select the research design the papers uses to generate the primary causal claim in the paper. Treat "Other" as catch‑all when none of these specific designs apply. If there are multiple categories from the given list, select "Multiple Designs" and list them in "other_research_design". If general_goal_of_analysis is "Describe" or "Predict", then leave this field "null". Use the following canonical research design definitions

    - Field Experiment: Researcher‐implemented random assignment of treatment in participants’ real-world environment (schools, firms, markets, governments, etc.). Setting is natural. Exclude survey-based or lab settings.

    - Survey Experiment: Random assignment is embedded inside a survey instrument (question wording, vignettes, information treatments, conjoint tasks). Sampling frame is a survey panel or similar; the experiment exists only within the survey session. Exclude field experiments that merely measure outcomes with a survey.

    - Lab Experiment: Random assignment conducted in an artificial, researcher-controlled environment (physical lab or online platform designed to mimic a lab). Tight control over information, incentives, timing; participants act under awareness of being studied.

    - Difference-in-Differences (Diff-in-Diff): Quasi-experimental panel or repeated-cross-section design comparing changes over time between treated and untreated groups. Identification comes from the interaction of post period and treated groups, under the parallel-trends assumption. Implementations may include two-way fixed effects, event studies, staggered adoption corrections, or synthetic DID weights, but a common-trend contrast is the core.

    - Instrumental Variable (IV): Uses an exogenous instrument ZZZ that shifts treatment DDD while affecting the outcome YYY only through DDD (relevance + exclusion). Primary estimator is two-stage least squares, control-function, or equivalent; the estimand is a Local Average Treatment Effect for compliers. Exclude designs where the “instrument” is actually a cutoff (RDD) or time variation (Diff-in-Diff).

    - Regression Discontinuity Design (RDD): Exploits a discrete jump in treatment assignment/assignment probability at a known cutoff of a running variable. Identification comes from continuity of potential outcomes at the threshold; effect is estimated via local linear/polynomial or bias-corrected methods. Exclude kink designs (slope change) and IV interpretations of cutoffs.

    - Regression Kink Design (RKD): Exploits a deterministic change in the slope (kink) of the treatment rule at a threshold. Identification uses the change in the derivative of the outcome with respect to the running variable. Requires smoothness of potential outcomes’ first derivative. Distinct from RDD because treatment probability is continuous at the threshold.

    - Synthetic Control: Constructs a weighted combination of control units to replicate the treated unit’s pre-intervention path; the post-intervention gap yields the causal effect. Key features: donor-pool weighting constrained to pre-period fit, usually a single treated unit or small set. Exclude simple matching or DiD unless weights are chosen to replicate pre-trends without parametric regression.

    - Matching/Balancing/Weighting: explicitly invoke the unconfoundedness assumption (also known as ignorability or selection-on-observables) with explicit construction of comparable treatment and control groups through matching, weighting, or balancing techniques. Such methods include: covariate matching, exact matching, generic matching, coarsened exact match, nearest-neighbor matching, entropy balancing, kernel balancing, propensity score matching, propensity score balancing, propensity score matching, inverse propensity score weighting (IPW), augmented inverse propensity score weighting (IPW), and covariate-balancing propensity score. Identification hinges on both unconfoundedness and the overlap assumption. Overlap means the similarity of the treatment and control groups in terms of their covariates. Assessing overlap is often based on propensity scores or a balance table. Exclude designs that also exploit an IV, cutoff, staggered timing, or panel differential trends—those fall under IV, RDD, DiD, etc. This is distinct from the Kitchen Sink Linear Model given the critical thought required to implement the selection-on-observables w/ matching/weighting/balancing techniques & assessing overlap vs a “throw-everything-in” approach.

    - Kitchen Sink Linear Model: OLS/Logit/etc. with a treatment indicator and an extensive vector of covariates chosen ad-hoc or “throw-everything-in”, but without an explicit, theory-based identification design (no instrument, cutoff, policy shock, or parallel-trend justification). Causal claim relies solely on conditional-independence/selection-on-observables.

    - Multiple Designs: Paper deploys two or more distinct causal designs (from this list) as complementary, co-equal strategies for the same causal question (e.g., RDD + IV robustness, or RDD plus Diff-in-Diff).

10.  other_research_design: If general_goal_of_analysis == "Describe" or "Predict" or main_causal_research_design = "Other" / "Multiple Designs", enter a concise label for the design used here; else null.

11. instrumental_variable_instrument: If main_causal_research_design is "Instrumental Variable", give a concise name for the variable used as an instrument in the analysis. If not an "Instrumental Variable" analysis, then null.

12. placebo_test: TRUE if the article engages in a formal, quantitative placebo test using data. A placebo test is a statistical test in which researchers either (a) use an outcome known not to be affected by the treatment and test the treatment’s effect on this placebo outcome, or (b) use a treatment known to have zero effect on the outcome and test its effect on the outcome. Ideally, the placebo test should return a null effect. The goal of placebo tests is to assess whether the key assumption is plausible. FALSE otherwise. 

13.  independent_variables and dependent_variables: List the key variables used in main research design. Each must have a *variable_name* and a *variable_description*. For variable names, be as concise and general. For example, if the paper is using incumbent vote-share from municipal elections in Brazil in 2016, just call that 'Vote Share' and leave the details for the variable_description. If this IV/DV framework is not relevant to the paper in question, leave as "null" instead of as an array.

14.  main_variable_relationship: Create one element per primary IV→DV pair discussed in the headline findings. Use names consistent with the variables above. Identify the direction of the relationship (relationship_type: "Positive", "Negative", "Non-Monotonic", "null", or "Unknown") and whether it was statistically_significant / substantively_significant (to determine significance significance, rely on authors' language rather than on the numbers, since the numbers/tables in the paper's raw text is unlikely to be parsed correctly). If this framework is not relevant to the paper in question, leave as "null" instead of as an array.

15.  makes_explicit_causal_claim: a boolean value coding whether the paper makes an *explicit* causal claim regarding the primary independent variable. Note, this is specific to the primary analysis/main result of the paper; do not consider ancillary analyses or statements that reference known causal relationships outside the scope of the paper (e.g. citing a difference paper or theorizing in the introduction). "Explicit" means that the paper states the causal relationship explicitly (e.g. "A causes B", "the [key independent variable] caused an x% increase to the [DV]", "the impact of the [key independent variable] on the [DV] is x", "the effect of the [key independent variable] on the [DV] is x", “This article shows the effects of...",  “Table X shows the effect of...”, “___ seems to have been a stronger determinant of ___”, etc.) Other expressions that indicate an explicit causal claim when used to describe the relationship between the focal variables include: "increase", "decrease", "reduce", "diminishes", "lead to", "result in", "impact", "influence", "affect", "elevate", and similar terms.

16.  makes_implicit_causal_claim: a boolean value coding whether the paper makes an *implicit* causal claim regarding the primary independent variable. This is often the case in papers where the authors try to tone down causal language by using non-causal language for the key explanatory variable despite the fact that the paper is organized/oriented as if it were exploring a causal relationship between two variables. Even though these papers avoid causal language, the paper relies on implicit causality to support the theory or mechanism. FALSE if makes_explicit_causal_claim is TRUE.

17. strong_non_causal_causal_qualification: a boolean value coding whether the paper explicitly states that the relationship between the primary independent variables and the outcomes of interest are not causal. Note: strong_non_causal_causal_qualification and makes_implicit_causal_claim causal claim can both be TRUE if the paper implicitly still frames itself causally (i.e. the qualification is not totally sincere). For example:
    - “The analysis is correlational and should not be interpreted as causal”
    
    - “These associations do not imply causation.”
    
    - “Findings are not evidence of causal impact.”
    
    - “Results cannot establish causal direction.”
    
    - “Observed linkages are descriptive rather than causal.”
    
    - “The estimates indicate association; claims of causality would be premature.”
    
    - “Data limitations preclude definitive causal conclusions.”
    
    - “This research identifies patterns, not causal effects.”
    
    - etc.

18. sample_size: if relevant to the paper, try to identify the sample size used in the main analysis. If there are multiple, do not try to combine/aggregate the sample sizes, but instead, use your judgement to identify the primary sample N. Often this is the sample size reported in the table or figure showing the paper's main result. Note: sample size is defined at the unit of observation (e.g. a study of 50 countries across 10 years would probably be a 500 country-year sample size, not 50). Do NOT make up any numbers; it is better to return 'null' than a false value.

19. sample_size_quote: to prove that you are not hallucinating the sample_size value above, please include the exact one-sentence quote/line from the paper which you used to retrieve the sample size. If you are unable to identify this line, then reconsider whether your sample_size coding above is legitimate.

20. claims_any_statistically_significant_results: true if the main result is claimed by the author to be statistically significant. false if not.

Design‑clarity and rigor flags:

21. references_power_analysis: true if the authors reference/report a power analysis. Note: this is true even if the power calculations are only reported in the appendix/pre-analysis plan, as long as the author indicates that they considered statistical power. Else false.

22. clearly_defined_explanatory_variable: when the paper is seeking to Explain, this is true if the paper clearly states which variable(s) it is focusing on to explain the focal relationship and if it discusses the theoretical underpinning behind using said variable(s).

23. clear_causal_quantity_of_interest: write out the causal quantity of interest if the paper states it directly. If the paper does not clearly state a causal quantity of interest listed below, return "FALSE". If the paper is not causal and thus a causal quantity of interest is irrelevant, return "null". You can ONLY select from the list of causal quantities below. Do not make up/write any other causal quantities.
    
    - ATE – Average Treatment Effect
    - ATT – Average Treatment Effect on the Treated
    - ATC – Average Treatment Effect on the Controls
    - SATE – Sample Average Treatment Effect
    - SATT – Sample Average Treatment Effect on the Treated
    - SATC – Sample Average Treatment Effect on the Controls
    - PATE – Population Average Treatment Effect
    - CATE – Conditional Average Treatment Effect
    - ITE – Individual Treatment Effect
    - LATE – Local Average Treatment Effect
    - CACE – Compiler Average Causal Effect
    - LATT – Local Average Treatment Effect on the Treated
    - LATC – Local Average Treatment Effect on the Controls
    - LIV Effect – Local Instrumental-Variable Treatment Effect
    - PRTE – Policy-Relevant Treatment Effect
    - MTE – Marginal Treatment Effect
    - QTE – Quantile Treatment Effect
    - CQTE – Conditional Quantile Treatment Effect
    - LQTE – Local Quantile Treatment Effect
    - TOT – Treatment-on-the-Treated Effect (IV context)
    - ITT – Intention-to-Treat Effect
    - ITTD – Intention-to-Treat Effect on the Difference in outcomes
    - Per-Protocol Effect
    - As-Treated Effect
    - NDE – Natural Direct Effect
    - NIE – Natural Indirect Effect
    - ACME – Average Causal Mediation Effect
    - CDE – Controlled Direct Effect
    - PSE – Path-Specific Effect
    - SACE – Survivor Average Causal Effect
    - PSACE – Principal-Stratum Average Causal Effect
    - APE – Average Partial Effect
    - ASF – Average Structural Function
    - DRF – Dose-Response Function (marginal causal curve)
    - MSM Effect – Marginal Structural Mean Effect
    - RD – Risk Difference (causal)
    - RR – Risk Ratio (causal)
    - OR – Odds Ratio (causal)
    - HR – Hazard Ratio (causal)
    - DiD-ATE – Difference-in-Differences Average Treatment Effect on the Treated
    - Dynamic Treatment Effect (DTE) – Time-varying causal effect path
    - Event-Study ATT – Dynamic DiD causal path
    - SCATE – Synthetic-Control Average Treatment Effect
    - RDD Cut-off Effect – Regression Discontinuity Treatment Effect at threshold
    - RKD Slope Effect – Regression Kink Treatment Effect
    - Spillover Effects – Direct Spillover Effect (DSE), Indirect Spillover Effect (ISE), Total Spillover Effect (TSE)
    - NATE – Network Average Treatment Effect
    - CLATE – Cluster Average Treatment Effect

24. specifies_estimate_equations: true if the paper states/references the actual equation that they are using to run their analysis, making it clear to the reader exactly what they are estimating. False otherwise.

22. discusses_threats_to_causality: true if clear and substantive discussion of possible threats to identifying a causal relationship between the primary IV and DV. Specifically in terms of why it might be, in general, difficult to identify a causal relationship between the primary variables of interest. For example, discussions of exogeneity/endogeneity, self-selection, selection effects, omitted variable bias, reverse causality, survivorship, spillover effects, etc.

23a. statement_of_identification_assumptions_quote: extract the exact quote/sentence-snippets from the text where the identification assumptions defined in 23b below are mentioned. Write 2-3 sentences/sentence snippets maximum. Focus on just the snippet(s) where the specific assumptions are mentioned.

23b. statement_of_identification_assumptions: code true or false whether the paper clearly stated and explicitly discussed the appropriate identification assumptions for the paper’s design. The specific assumptions for each relevant design are enumerated below.

- Field Experiment: Random assignment, Unconfoundedness, SUTVA, Low attrition, No/limited spillover

- Survey Experiment: Random assignment of survey stimulus, No systematic non-response, SUTVA

- Lab Experiment: Random assignment; SUTVA, No uncontrolled environmental or demand effects influencing assignment or outcomes

- Difference-in-Differences (Diff-in-Diff or DID): Parallel trends, Treated and control groups would have identical time-path changes in outcomes, No anticipation effect

- Instrumental Variable (IV): Relevance (or strong instrument), Exclusion Restriction, Independence/unconfoundedness, Monotonicity for LATE

- Regression Discontinuity Design (RDD): Continuity of potential outcomes at cutoff, No precise manipulation of running variable around cutoff, Balanced on covariates around the cutoff

- Regression Kink Design (RKD): Continuity of the slope of potential outcomes in running variable, Kink in treatment rule occurs only at known threshold, No manipulation near kink

- Synthetic Control: Weighted combination of controls replicates treated unit’s pre-treatment outcomes, low-rank structure, No uncontrolled time-varying confounders, No post-treatment shocks unique to controls, No anticipation effect, No major structural changes 

- Matching/Balancing/Weighting: Unconfoundedness given observed covariates, Common support of covariate distributions (also known as overlap or positivity)

- Kitchen Sink Linear Model: Exogeneity of the error term, No perfect multicollinearity, Correct functional form for identification

24. effort_to_explore_mechanisms: Check if the paper devotes a discrete section—or equivalent substance—to exploring the channel/mechanism through which the main IV affects the DV. Note, this does NOT include theoretical discussions at the beginning of the paper, as it must be a discussion which occurs after the main analysis has already yielded results. Options for coding follow:
 
   - "No Mention of Mechanisms/Channels": No mention of mechanisms/channels.

    - "Mechanisms/Channels Mentioned But Not Explored": Mention of mechanisms/channels as passing remarks, conjecture, or future-work notes without data or structured argument.

    - "Mechanisms/Channels Mentioned With Substantial Exploration": (a) a fully articulated discussion of mechanisms/channels to explain the result (more substantial than a passing mention) or (b) a full empirical or formal analysis, such as mediation analysis or exploration of subgroup heterogeneity designed to discriminate mechanisms. 

25. mentions_pre_registered_design_and_analysis_plan: true if the paper mentions that it has been pre-registered or that a pre-analysis plan exists.
\end{Verbatim}
\begin{Verbatim}[breaklines=true]
<!--begin excerpt-->
Paper Title: 
{TITLE}
==============================
Paper Full TEXT: 
{FULL_TEXT}
<!--end excerpt-->

You are given the full text of a political science paper above. Read it carefully. Then extract the requisite fields following the system guidelines, returning them in valid JSON that satisfies every element of the provided schema. 

\end{Verbatim}
}
\vspace{1em}

{\scriptsize
\begin{Verbatim}[breaklines=true]
json_schema = {
    "json_schema": {
        "type": "json_schema",
        "json_schema": {
            "name": "paper_data_extraction",
            "strict": True,
            "schema": {
                "type": "object",            
                "properties": {
                    "error_in_raw_text": {
                        "type": "string",
                        "enum": [
                            "No Error",
                            "Missing/Corrupt",
                            "Title/Text Mismatch",
                        ]
                    },
                    "subfield": { 
                        "type": "string",
                        "enum": [
                            "American Politics",
                            "Comparative Politics",
                            "International Relations",
                            "Methodology and Formal Theory",
                            "Political Theory and Philosophy",
                            "Public Policy/Administration",
                            "Other"
                        ]
                    },
                    "is_empirical_quant_paper": {"type": "boolean"},
                    "general_goal_of_analysis": {
                        "type": "string", 
                        "enum": [
                            "Describe",
                            "Predict",
                            "Explain",
                            "null"
                        ]
                    },
                    "single_country_study": {
                        "type": "string", 
                        "enum": [
                            "single_country",
                            "multiple_countries",
                            "null"
                        ]
                    },
                    "single_region": {
                        "type": "string", 
                        "enum": [
                            "single_region",
                            "multiple_region",
                            "null"
                        ]
                    },
                    "countries_of_focus": {"type": "string"},
                    "paper_uses_survey_data": {
                        "type": "string", 
                        "enum": [
                            "no_survey_data",
                            "runs_original_survey",
                            "uses_public_available_survey"
                        ]
                    },
                    "uses_original_dataset": {
                        "type": "string", 
                        "enum": [
                            "original_survey",
                            "field_experiment",
                            "field_study",
                            "structure_systematize",
                            "procure_original_data",
                            "other_original_data",
                            "not_original",
                            "null"
                        ]
                    },
                    "seeks_determinants": {"type": ["boolean", "null"]},
                    "main_causal_research_design": {
                        "type": "string", 
                        "enum": [
                            "Field Experiment",
                            "Survey Experiment", 
                            "Lab Experiment",
                            "Diff-in-Diff", 
                            "Instrumental Variable", 
                            "Regression Discontinuity Design",
                            "Regression Kink Design",
                            "Synthetic Control",
                            "Matching/Weighting/Balancing",
                            "Kitchen Sink Linear Model",
                            "Multiple Designs",
                            "Other",
                            "null"
                        ]
                    },
                    "other_research_design": {"type": ["string", "null"]},
                    "instrumental_variable_instrument": {"type": ["string", "null"]},
                    "placebo_test": {"type": ["boolean", "null"]},
                    "independent_variables": {
                        "type": ["array", "null"],
                        "items": {
                            "type": "object",
                            "properties": {
                                "variable_name": {"type": "string"},
                                "variable_description": {"type": "string"}
                            },
                            "required": ["variable_name", "variable_description"],
                            "additionalProperties": False
                        }
                    },
                    "dependent_variables": {
                        "type": ["array", "null"],
                        "items": {
                            "type": "object",
                            "properties": {
                                "variable_name": {"type": "string"},
                                "variable_description": {"type": "string"}
                            },
                            "required": ["variable_name", "variable_description"],
                            "additionalProperties": False
                        }
                    },
                    "main_variable_relationship": {
                        "type": ["array", "null"],
                        "items": {
                            "type": "object",
                            "properties": {
                                "iv_var_name": {"type": "string"},
                                "dv_var_name": {"type": "string"},
                                "relationship_type": {"type": "string", "enum": ["Positive", "Negative", "Non-Monotonic", "Null", "Unknown"]},
                                "statistically_significant": {"type": "boolean"},
                                "substantively_significant": {"type": "boolean"},
                            },
                            "required": ["iv_var_name", "dv_var_name", "relationship_type", "statistically_significant", "substantively_significant"],
                            "additionalProperties": False
                        },
                    },
                    "makes_explicit_causal_claim": {"type": ["boolean", "null"]},
                    "makes_implicit_causal_claim": {"type": ["boolean", "null"]},
                    "strong_non_causal_causal_qualification": {"type": ["boolean", "null"]},
                    "sample_size": {"type": ["integer", "null"]},
                    "sample_size_quote": {"type": ["string", "null"]},
                    "claims_any_statistically_significant_results": {"type": ["boolean", "null"]},
                    "references_power_analysis": {"type": ["boolean", "null"]},
                    "clearly_defined_explanatory_variable": {"type": ["boolean", "null"]},
                    "clear_causal_quantity_of_interest":  {"type": ["string", "null"]},
                    "specifies_estimate_equations": {"type": ["boolean", "null"]},
                    "discusses_threats_to_causality": {"type": ["boolean", "null"]},
                    "statement_of_identification_assumptions_quote": {"type": ["string", "null"]},
                    "statement_of_identification_assumptions":{"type": ["boolean", "null"]},
                    "effort_to_explore_mechanisms": {
                        "type": "string", 
                        "enum": [
                            "No Mention of Mechanisms/Channels",
                            "Mechanisms/Channels Mentioned But Not Explored", 
                            "Mechanisms/Channels Mentioned With Substantial Exploration",
                            "null"
                        ]
                        },
                    "mentions_pre_registered_design_and_analysis_plan": {"type": ["boolean", "null"]},
                },
                "required": [
                    "error_in_raw_text",
                    "subfield",
                    "is_empirical_quant_paper",
                    "general_goal_of_analysis",
                    "single_country_study",
                    "single_region",
                    "countries_of_focus",
                    "paper_uses_survey_data",
                    "uses_original_dataset",
                    "seeks_determinants",
                    "main_causal_research_design",
                    "other_research_design",
                    "instrumental_variable_instrument",
                    "placebo_test",
                    "independent_variables",
                    "dependent_variables",
                    "main_variable_relationship",
                    "makes_explicit_causal_claim",
                    "makes_implicit_causal_claim",
                    "strong_non_causal_causal_qualification",
                    "sample_size",
                    "sample_size_quote",
                    "claims_any_statistically_significant_results",
                    "references_power_analysis",
                    "clearly_defined_explanatory_variable",
                    "clear_causal_quantity_of_interest",
                    "specifies_estimate_equations",
                    "discusses_threats_to_causality",
                    "statement_of_identification_assumptions_quote",
                    "statement_of_identification_assumptions",
                    "effort_to_explore_mechanisms",
                    "mentions_pre_registered_design_and_analysis_plan"
                ],
                "additionalProperties": False
            }
        },
    }
}

\end{Verbatim}
}



\subsubsection{Second-stage classification prompt}\label{sisec:prompt_v8e}

The complete text sent with each paper to Claude Sonnet 5 in the second stage is reproduced below, taken directly from the classification code and preserved exactly as used. This prompt generates the corpus labels and underlies every accuracy estimate reported in Section~\ref{sisec:prompt_validation}.

\begin{Verbatim}
You are an expert political-methodology coder. Classify ONE paper. Work through the steps in order and stop as soon as a step resolves -- EXCEPT that STEP 6, the subfield, is always completed, for every paper, including papers that stopped at STEP 1A or STEP 1B. Whatever else you do, subfield is never null.

STEP 1A - POPULATION. Decide whether this paper is an original quantitative empirical study. Take the FIRST that applies.
GOVERNING RULE for this step: a paper that conducts its OWN statistical analysis of data in support of a substantive claim about politics is IN the population, whatever else it also does -- a regression alongside case studies, an empirical test of a formal model, an application that carries the paper's substantive argument. Apply the exclusions below only when the paper's contribution does not rest on such an analysis.
- Its main contribution is a statistical or econometric METHOD -- a new estimator, test, procedure, measurement technique, or data-construction method -- rather than a substantive empirical finding. An empirical application included to demonstrate the method does not change this -> scope="methods". Stop.
- Its main contribution is a formal or game-theoretic model: equilibrium analysis, comparative statics, or theorem-proving about strategic interaction, with any empirical content illustrative at most (motivating anecdotes, descriptive plots, calibration to stylized numbers). A formal model whose parameters are statistically estimated (structural estimation, EITM) is NOT this and stays in the population -> scope="formal". Stop.
- Its main contribution is a review or synthesis of existing literature: a review essay, a research agenda or state-of-the-field piece, a stocktaking or lessons-learned article, with no original data analysis of its own. Reporting numbers, findings, or effect sizes taken from other studies in the text or a summary table is still a review. A QUANTITATIVE synthesis that statistically analyzes results collected from other studies (a meta-analysis, a pooled re-estimation across published studies) is NOT this and stays in the population -> scope="review". Stop.
- Its main analysis is not statistical: qualitative or interpretive work (case studies, process tracing, ethnography, interviews, historical narrative), set-theoretic or configurational analysis (QCA, fsQCA, necessary-and-sufficient-condition analysis), or normative or conceptual argument -- whether or not it argues causally -> scope="qualitative_or_non_statistical". Stop.
- Otherwise the paper is in the population. Continue to STEP 1B.

STEP 1B - GOAL. Every paper reaching this step receives exactly one of three goals. There is no residual and no default. Choose the one the paper's MAIN ANALYSIS is built to deliver, judged by what the paper sets out to establish and by what its central tables and figures report. Motivation, theory sections, and closing discussion do not settle this; the main analysis does.
- explanatory: the main analysis is organized to establish how one thing affects another. The paper puts a specific explanatory variable or treatment and a specific outcome at the center, and its central results report the relationship between them. This holds whether the paper estimates the effect of one defined treatment or asks which factors account for variation in an outcome, and it holds when the paper's language is hedged or associational rather than explicitly causal.
- descriptive: the main analysis is organized to measure or characterize something -- introducing an indicator or dataset, establishing baseline levels, mapping a distribution, documenting differences between groups or change over time. A paper stays descriptive when it mentions or speculates about causes in its motivation or discussion while its central results remain measurements rather than a relationship it set out to establish.
- predictive: the main analysis is organized to anticipate an outcome, or to rank which variables best anticipate it, without interpreting them as causes. Out-of-sample or cross-validated performance is a common cue but is NOT required: a forecasting paper reporting only in-sample fit is still predictive. What decides it is that the paper's stated aim is accurate anticipation rather than an account of what produces the outcome.
BOUNDARIES. Explanatory vs descriptive: ask whether a relationship between a named explanatory variable and a named outcome is what the main analysis exists to establish. Documenting that two groups differ is descriptive; arguing that membership in one of them accounts for the difference is explanatory. Explanatory vs predictive: a paper asking which variables best anticipate an outcome, declining to interpret them as causes, is predictive even when it runs the same regressions an explanatory paper would run. A new measure, indicator, or dataset is descriptive when the measure IS the contribution, and explanatory when the measure was built in order to estimate a relationship the paper then reports as its main result. With elements of more than one goal, choose by the paper's primary contribution. If the choice was close, lower "confidence".
Copy into scope_quote the ONE sentence that best shows what the paper set out to do, usually from the abstract or the end of the introduction. If there is no such sentence, put "".
Only explanatory papers continue to STEP 2. For descriptive and predictive papers, stop here.

STEP 2 - Find the primary estimate. Identify the headline causal relationship the paper is built around (main independent variable -> outcome). Nearly always there is one. A paper has two only when it advances two claims that each stand on their own, not when it advances one claim plus an account of why that claim holds; in that case identify both. Ignore robustness checks, appendix analyses, secondary sub-questions, mechanism and moderator analyses, and methods named only in the literature review.

STEP 3 - Quote the identification. Copy the ONE sentence that states how the primary estimate is identified into evidence_quote. If there is no such sentence, the design is probably Model-based.

KEY PRINCIPLE. If the paper's PRIMARY estimate is identified by a specific device -- an excluded instrument, a cutoff on a running variable, random assignment, or a donor-pool synthetic counterfactual -- name that device, even when the paper ALSO reports control variables, fixed effects, or extra robustness regressions. Control variables and fixed effects are not identification devices; their presence does not make a paper Model-based. A randomized experiment (field, survey, or lab) whose effect is estimated with a covariate-adjusted regression remains that experiment -- never Model-based. Choose Model-based ONLY when the primary estimate rests on a covariate-adjusted regression with NO such device. Classify by the identification strategy, not by the estimator or a method word that merely appears in the text.

DESIGN MENU. There is NO residual option here: papers that do not belong were already removed at STEPS 1A and 1B. Every explanatory paper receives exactly one label.
- Field Experiment: the researcher randomly assigns treatment in a real-world setting.
- Survey Experiment: randomization embedded in a survey instrument (vignette, conjoint, information treatment, question wording). An online questionnaire or vignette study is a Survey Experiment.
- Lab Experiment: randomization in a controlled, researcher-built environment. An online platform built to imitate a lab (interactive economic games, timed tasks) is a Lab Experiment. Lab-in-the-field is also a Lab Experiment: structured games or tasks (public-goods, ultimatum) run under researcher-controlled protocols in a naturalistic site (e.g. a village) -- the controlled session, not the physical location, decides. A questionnaire-delivered stimulus (vignette, cue) administered within a lab session is still a Lab Experiment, not a Survey Experiment. [Decide Field vs Lab vs Survey by WHERE and HOW subjects are treated. Cues: "online", "MTurk/Prolific", "questionnaire", "vignette", "laboratory", "students", "in the field".]
- Difference-in-Differences: the authors CONSCIOUSLY adopt a difference-in-differences design -- a treated-vs-comparison contrast tracked over time and identified by parallel trends, typically two-way (unit and time) fixed effects or an event study around a treatment date. The paper must actually set this design up; do NOT infer DiD from panel data alone, from one-way fixed effects, from a two-group comparison, or from a bare before/after change. One-way fixed effects, or fixed effects used only to adjust for confounders rather than as a treated/control-over-time design, are Model-based. When the paper explicitly frames its OWN identification as difference-in-differences, classify it here even if the treatment is a one-off policy change -- do not demote it to Natural Experiment/Quasi Experiment.
- Instrumental Variables: use when an excluded instrument identifies the PRIMARY treatment for the headline estimate. Do not judge the instrument's strength -- a debatable instrument the paper relies on for its main result is still IV. Demote to Model-based ONLY if the paper itself presents the instrument as a secondary or robustness analysis, or if the headline estimate is a plain regression and the instrument is never used for it. Not IV if the instrument is for a control rather than the treatment, if the "instrument" is a cutoff (that is RDD) or pure time variation (that is DiD), or if instruments appear only in the literature review. A FUZZY regression discontinuity (threshold used as the instrument) is Regression Discontinuity, not IV.
- Regression Discontinuity Design: a genuine running (forcing) variable with a known cutoff and local continuity. Geographic RD qualifies when distance to a boundary is the running variable and the estimate compares units close to the border on both sides. Do NOT use RDD for an across-the-border comparison with no distance-based running variable, a loose "regression-discontinuity" framing, or a quasi-experiment with no running variable -> those are Natural Experiment/Quasi Experiment.
- Synthetic Control: the primary estimate compares a treated unit to a donor-pool-weighted synthetic counterfactual. This holds even when the treatment is a natural policy shock -- the synthetic counterfactual is the identification device. Choose the surrounding design only if the synthetic control is explicitly a supplementary check rather than the headline method.
- Matching/Balancing/Weighting: use ONLY when the paper's PRIMARY identification is selection-on-observables, argued explicitly, with overlap or balance assessed (propensity-score matching, IPW, entropy balancing, coarsened exact matching, as the MAIN strategy). Matching or weighting used as preprocessing or a robustness check on top of a regression -> Model-based. A survey weight, a "matched sample", or the word "match"/"propensity" alone does NOT qualify.
- Natural Experiment/Quasi Experiment: exogenous variation the researcher did not control (lotteries, weather, disasters, redistricting, externally imposed policy changes, arbitrary administrative thresholds) used for identification, when it does not slot cleanly into DiD, RDD, IV, Synthetic Control, or Matching/Balancing/Weighting. FACE VALUE governs this category: when the alternative label would be Model-based (or matching/weighting machinery not argued as the identification), take the researchers at their word about their own study. Classify here when the paper anywhere -- title, abstract, or body -- describes its OWN study, data, analysis, or research design as a natural experiment or as quasi-experimental; hedged or approximate forms count ("mimics/approximates a natural experiment", "quasi-experimental study/data", "creates a natural experiment"), and the framing may attach to only one part of the analysis. Do NOT demote such a paper to Model-based because the variation seems endogenous, a comparison group is missing, or the estimation is an ordinary covariate-adjusted regression. Three exceptions, in order: treatment randomly assigned by the researchers is a Field/Survey/Lab Experiment; identification that is more specifically DiD, RDD, or IV per those bullets (a consciously set-up treated-vs-comparison-over-time design, a running-variable cutoff, an excluded instrument) takes that label; and identification that rests on selection-on-observables -- matching, balancing, or weighting argued as the identification with overlap or balance assessed -- is Matching/Balancing/Weighting, exactly as DiD, RDD, and IV take precedence, even when the paper calls its own study a natural experiment or quasi-experimental. ABSENT that self-description, this is a residual with a HIGH BAR: only treatment variation that is arguably exogenous -- as-good-as-random for reasons outside the actors' control -- belongs here; ordinary observational variation adjusted with controls is Model-based, and "natural experiment(s)" used merely to characterize variation that a fixed-effects or covariate-adjusted regression exploits -- without the study itself being presented as one -- stays Model-based.
- Model-based: the primary estimate rests on a covariate-adjusted regression or structural model with NO explicit identification device (no instrument, cutoff, randomization, or synthetic counterfactual). Includes kitchen-sink OLS/logit with controls, fixed-effects regressions used for adjustment rather than as a design, structural estimation (Cox/survival, Heckman selection, Tobit, GMM, SEM, multilevel/hierarchical models), and time-series models (VAR, ARIMA, error-correction, Granger causality).
- Multiple Designs: the label for a paper with two or more designs that count, decided in STEP 4. It is not a design in itself and never a default.

STEP 4 - NAME THE DESIGN. Nearly every paper has one design. Name it under "one design" below and stop, UNLESS the paper plainly builds on a second analysis using a DIFFERENT design from the menu above -- several studies sharing one design are still one design. Only then work through the rest of this step.
A second design counts only if its claim would still stand as a contribution had the first analysis never been run. It does NOT count if it CHECKS the first result: robustness checks, endogeneity checks, alternative specifications, placebo or falsification tests, appendix analyses, first-stage or descriptive groundwork, or anything the paper itself calls a check, a supplementary analysis, or an additional test. It does NOT count if its claim PRESUPPOSES the first result -- an analysis of why, when, where, or for whom that result holds. Mechanism, channel, moderator, and heterogeneity analyses are of this second kind however prominent they are, and even when the abstract announces them.
For each design that counts, you must be able to copy ONE sentence in which the paper presents it as main evidence. If you cannot, it does not count. When in doubt, it does not count.
Model-based is judged the same way, in BOTH directions. A covariate-adjusted regression counts when the paper presents it as its own route to a central claim; it does not when it extends a design-based result to a wider sample, generalizes it, or describes the setting. Conversely, when the paper's central claim rests on an observational regression and an experiment is added to probe the mechanism, the regression carries and the experiment does not: that paper is Model-based, not Multiple.
- One design: put it in "design" and leave "components" empty.
- Two or more designs that count: put "Multiple Designs" in "design" and list them in "components", in the order the paper presents them, leading with the one the abstract leads with. This holds whether they bear on the SAME causal relationship (two routes to one claim) or on DIFFERENT but related ones (a multi-study paper whose argument rests on more than one study). Copy into multiple_quote the sentence showing they stand together.
- If the designs that count reduce to ONE label, that label is the design, not Multiple.

COMMIT. Do not lead with a broader or safer label: Model-based and Multiple Designs are not defaults, and choosing either when the paper names a specific identification device is an error. Components are drawn from this same menu. A Multiple Designs paper may or may not contain a design-based component -- do not assume it does; two model-based strategies combined are still Multiple Designs with no design-based component.

AMBIGUITY. Always commit to a primary design; never leave "design" empty for an explanatory paper. If the choice was close, lower "confidence". Report confidence honestly: "low" when two labels were nearly tied, "medium" when the evidence is thin, "high" when the paper states its identification plainly.

STEP 5 - AUXILIARY CODING. Only when scope="explanatory": also code the auxiliary fields below. They DESCRIBE the paper; they must not change your STEP 1A-4 answers. Treat the design you chose in STEP 4 as given -- do not revisit or revise it while coding these. The primary estimate is the one you identified in STEP 2; when STEP 2 found two, use the one belonging to the first design listed in "components". makes_explicit_causal_claim, claims_significant_results, sample_size, estimand, and data_source anchor to this estimate. id_assumption_stated, placebo_test, references_power_analysis, and has_pre_analysis_plan are paper-level (evidence anywhere in the paper counts). subfield is NOT an auxiliary field -- it is STEP 6, applies to every paper, and is never null. When scope is NOT "explanatory", return null for every auxiliary field below and go straight to STEP 6.

- id_assumption_stated: true if the paper EXPLICITLY states the core identifying assumption behind its primary design -- parallel trends or no anticipation (DiD), exclusion restriction or instrument exogeneity (IV), continuity of potential outcomes at the cutoff (RDD), unconfoundedness / selection on observables / conditional independence (matching or regression adjustment), random assignment (experiments), or an equivalent explicit statement for another design. Merely NAMING the design, running it, or citing its literature is NOT stating the assumption -- the paper must articulate the assumption itself, formally or in words.
- id_assumption_quote: the sentence stating the assumption; "" when id_assumption_stated=false.
- placebo_quote: the sentence reporting the placebo test; "" when placebo_test=false. If you cannot copy a sentence in which THIS paper reports running one, the answer is false.
- placebo_test: true if the paper reports a formal, quantitative placebo test on data: an outcome the treatment should not affect, or a treatment that should have no effect, tested to show a null. The test must be one THIS paper runs on its own data. NOT placebo tests: balance tests and generic robustness regressions; a placebo or control ARM in an experiment (that is how the design is built, not a test of its results); a placebo test described in the literature review or attributed to another study; and a statement of the identifying assumption -- an exclusion restriction, including any claim that an instrument "should have no effect" on the outcome except through the treatment, parallel trends, continuity at the cutoff -- which belongs to id_assumption_stated, and one sentence must never set both fields.
- makes_explicit_causal_claim: true if the paper states the causal relationship of the primary estimate in so many words ("causes", "the effect of", "impact", "leads to", "increases/reduces"). Primary analysis only; hedged or purely associational language -> false. Hedging is about the strength of the CLAIM, not the verb introducing it: "we explore / examine / investigate / study the effect of X on Y" is EXPLICIT -- "the effect of" is the claim and the verb only says what the paper does with it. Hedged means the relationship itself is qualified: "may be associated with", "suggests a possible link", "is correlated with". The paper's explanatory AIM is already established; this field codes only its explicit LANGUAGE, so false is a normal, expected value for a paper with a causal aim stated in hedged terms.
- references_power_analysis: true only when the paper's own text states that a power analysis was conducted or says where to find one ("power calculations are reported in the pre-analysis plan" counts); never infer one from preregistration alone.
- has_pre_analysis_plan: true if the study states that its hypotheses or analysis were preregistered or follow a pre-analysis plan (OSF, EGAP, AsPredicted, AEA registry, "preregistered", "PAP").
- claims_significant_results: true if the authors claim the headline estimate is statistically significant; false if they do not claim significance for it. Rely on the authors' language, not on reading numbers out of tables.
- sample_size: the N of the headline estimate -- the table or figure column carrying the paper's main result -- at that estimate's unit of observation (50 countries over 10 years is 500 country-years). EXCEPTION: for conjoint experiments, code the number of RESPONDENTS, not respondent-task or profile observations. With multiple analyses, the N of the main specification, never a sum. Never invent a number; null when not stated.
- sample_size_quote: the verbatim sentence or table note stating that N; "" if none.
- estimand: the most precise statement the paper makes of the target quantity behind its primary estimate. Grade the paper's BEST statement and return the highest level it reaches (precise > causal_contrast > statistical_only > none). "precise" = names a defined causal quantity -- ATE, ATT, LATE, ITT, CACE, a subgroup or effect-modification estimand (CATE, causal moderation), or a causal mediation quantity (ACME, natural or controlled direct/indirect effects) -- or states in words BOTH the counterfactual contrast AND the population averaged over ("the average effect of X versus no X on Y among compliers"). A named quantity counts only when applied to the paper's OWN primary estimate -- bare "we estimate the ATT of X on Y" is precise (a named estimand defines its own population), and a named estimand counts whether it is written as an acronym or spelled out -- "the average treatment effect of X on Y" is ATE and is precise, exactly as "the ATE of X on Y" would be, but ATE/LATE mentions in methodological discussion or the literature review never count. "causal_contrast" = states the target only as the effect of X on Y (treatment and outcome named, causal direction clear) without the quantity or population. For words-only statements, the test: can you tell from the paper's own words WHICH average of WHOSE outcomes is targeted? If yes -> "precise"; if only X->Y -> "causal_contrast". "statistical_only" = the stated target is a model parameter ("our coefficient of interest is beta") never mapped to a causal quantity or contrast. "none" = no stated target quantity.
- estimand_quote: the sentence stating the target quantity; "" when estimand="none".
- data_source: the primary data source of the headline estimate, ONE label. "experimental" = data generated by the researchers' own experiment (field, survey, or lab). "original_survey" = a non-experimental survey the researchers fielded themselves. "restricted" = administrative records, government microdata, or industry/proprietary data whose access is genuinely limited -- a secure enclave, named-researcher approval, or a commercial licence. Registration, a standard data-use agreement, or "conditional access" does NOT make data restricted: if any researcher can obtain it by signing up, it is "public" (BHPS, UK Data Service, ICPSR-style archives). "public" = publicly available data (ANES/Afrobarometer-style series, cross-national indices, replication archives, published statistics, scraped public text or records). "other" = none of the above. If several sources feed the headline estimate, pick by precedence: experimental > original_survey > restricted > public.
- data_source_quote: the sentence identifying the data; "" if the source is never described.

STEP 6 - SUBFIELD. Code this for EVERY paper, including papers that stopped at STEP 1A or STEP 1B. A methods paper, a formal model, a review essay, and a qualitative study all receive a subfield; it is the one field that is never null. This step DESCRIBES the paper and must not change any answer you have already given -- treat your STEP 1A-4 answers as final. Choose the ONE subfield the paper best fits, using these labels verbatim. "American Politics" = political behavior, institutions, or processes inside the United States. "Comparative Politics" = comparison across countries or sub-national units, non-U.S. domestic politics, regime type, democratization, or civil conflict within states. "International Relations" = cross-border interactions: interstate conflict and cooperation, trade, international organizations, diplomacy, security; includes public or elite opinion toward foreign policy. "Methodology and Formal Theory" = develops or critiques methods, measurement, statistics, or game-theoretic models for disciplinary use. "Political Theory and Philosophy" = normative reasoning, conceptual analysis, or interpretation of texts. "Public Policy/Administration" = policy design, implementation, evaluation, or bureaucratic behavior. "Other" = none of the above. "Unknown" = truly uncertain.

Return ONLY JSON (auxiliary fields null when scope is not "explanatory"; subfield is always filled, whatever the scope):
{"scope": "explanatory"|"descriptive"|"predictive"|"methods"|"formal"|"review"|"qualitative_or_non_statistical",
 "scope_quote": <verbatim sentence or "">,
 "design": <one menu label, or null when scope is not "explanatory">,
 "components": [<menu labels>, ...],
 "multiple_quote": <verbatim sentence, or "" when design is not "Multiple Designs">,
 "confidence": "high"|"medium"|"low",
 "evidence_quote": <verbatim sentence or "">,
 "id_assumption_stated": true|false,
 "id_assumption_quote": <sentence or "">,
 "placebo_test": true|false,
 "placebo_quote": <verbatim sentence or "">,
 "makes_explicit_causal_claim": true|false,
 "references_power_analysis": true|false,
 "has_pre_analysis_plan": true|false,
 "claims_significant_results": true|false,
 "sample_size": <integer, or null>,
 "sample_size_quote": <verbatim sentence or "">,
 "estimand": "precise"|"causal_contrast"|"statistical_only"|"none",
 "estimand_quote": <sentence or "">,
 "data_source": "experimental"|"original_survey"|"restricted"|"public"|"other",
 "data_source_quote": <sentence or "">,
 "subfield": "American Politics"|"Comparative Politics"|"International Relations"|"Methodology and Formal Theory"|"Political Theory and Philosophy"|"Public Policy/Administration"|"Other"|"Unknown"}
\end{Verbatim}

\clearpage


\subsection{Human validation}\label{sisec:prompt_validation}

This section assesses the accuracy of the LLM-generated labels. We begin by describing the samples in which humans, rather than the model, settled the labels. We then use those samples to evaluate each stage of the pipeline: first the two screens, and then research design, the label on which our findings rest. Because a single overall accuracy rate can mask poor performance for particular designs, we also report accuracy separately for each design, the full cross-tabulation of settled and assigned labels, and separate results for each of the three periods. We then compare the classifier with four human coders who read the same papers using the same codebook. The section concludes by explaining what this validation can and cannot establish. Section~\ref{sisec:dslcorr} uses the same samples to correct the corpus estimates for the errors measured here.

\subsubsection{The validation samples}

We classify each paper in two stages and evaluate each stage against papers whose labels we treat as correct. We refer to these labeled samples as \emph{gold standards}. Table~\ref{tab:val_design} summarizes the five samples we use, who assigned their labels, and which stage of the pipeline each sample evaluates.

\begin{table}[!t]\centering
\caption{Validation Samples}
\label{tab:val_design}
\footnotesize
\begin{tabular}{@{}lL{3.4cm}L{4.6cm}lL{3.2cm}@{}}
\toprule
Sample & Papers & Who produced the gold standards & Checks & What it tells us \\
\midrule
G1 & 200, random & two RAs; a third settled disagreements &
Stage 1 & how accurate the quantitative-empirical screen is \\
\addlinespace
G2 & 200, random among quantitative-empirical & two language models read every paper; one author resolved all 15
disagreements and re-checked 20 agreements without seeing the classifier's label & Stage 2 & overall
accuracy among quantitative-empirical papers \\
\addlinespace
G3 & 244; up to eight of each design in each period & two language models read every paper; one author resolved
every disagreement and re-checked 30 agreements without seeing the classifier's label; four human coders
re-read 60 papers, two coders per paper & Stage 2 & accuracy among quantitative-empirical papers by design, and whether
it is stable over time \\
\midrule
\multicolumn{5}{@{}l}{\emph{Development samples.}}\\
\addlinespace
G2$'$ & 199, random & two RAs; we later re-checked the
identification-assumption labels & --- & the other variables the submitted
version recorded \\
\addlinespace
G3$'$ & 280 scoreable; picked so that every label the initial submission
assigned is represented & two RAs; we later re-read the papers
and corrected their design labels using the codebook & --- & which codebook rules
to write, and which prompt version to keep \\
\bottomrule
\end{tabular}
\tabnote{G2 and G3 share no papers with G1, and neither was used to revise the prompt or choose among versions. G2$'$ is a random development sample intended to
resemble the corpus, whereas G3$'$ includes larger numbers of papers using rare
designs.}
\end{table}

Stage~1 asks whether a paper is quantitative and empirical. This is Task~1 from the initial submission, run by GPT-4o, which we retain without rerunning. Of the 91,632 papers for which we have full text, Stage~1 classifies 42,317 as quantitative and empirical. We then drop records whose retrieved text did not match the article requested, along with review articles, leaving 37,163 papers for Stage~2. We evaluate Stage~1 against G1, a random sample of 200 papers. Two research assistants (RAs) read each paper, and a third RA settled the four cases on which they disagreed. The machine agrees with the settled label in more than 97\% of cases.

Stage~2 replaces everything the initial submission did after this first screen. Under the \emph{new classifier}, each of the 37,163 papers is sent to Claude Sonnet 5 with the instructions reproduced in Section~\ref{sisec:prompt_v8e}. The model first decides whether the paper is explanatory and, if so, identifies its research design. We froze the prompt before scoring either of the samples reported below, and made no changes afterward. The classifier evaluated here is therefore the same one used for the corpus-wide run in this revision.

We evaluate Stage~2 against two samples, G2 and G3. G2 is a simple random sample of 200 papers drawn from a pool of 35,297 eligible papers. We formed this pool by removing from the 37,163 papers the 1,525 used in earlier development or validation rounds, as well as papers with fewer than two thousand characters of retrieved text. G2 therefore measures overall accuracy among quantitative-empirical papers. G3 contains 244 papers, with at most eight papers for each design-period combination, and allows us to measure accuracy separately by design and over time.

Two additional samples, denoted with primes, were used to develop the new classifier rather than to evaluate it. G2$'$ is a random sample of 199 papers from the initial submission. G3$'$, collected later, contains 290 papers read by two RAs, 280 of which have labels that can be scored against the classifier. Neither sample contributes to the accuracy estimates reported from G2 or G3.

We also compare the new classifier directly with the one used in the original submission. Both classifiers are scored on the same 131 papers against the same reference labels. The new classifier identifies the exact research design correctly for 0.809 of the papers, compared with 0.718 for the original classifier. It places 0.893 on the correct side of the design-based distinction, compared with 0.748 for the original classifier. Because the new instructions were developed partly using these papers, these figures should not be read as out-of-sample accuracy estimates. Their main value is comparative: on the same papers and against the same labels, the new classifier performs substantially better than the original one.

\subsubsection{Accuracy of the two screens}

\paragraph*{Results from Stage 1.} Table~\ref{tab:g1} reports how often GPT-4o's classification matches the settled label. We recomputed both figures directly from the RAs' coding sheets. G2 provides an additional check because it is drawn from papers that Stage~1 classified as quantitative and empirical: any ineligible paper in this sample is therefore a false admission from Stage~1. We found one such case, which is a simulation paper that contained no data analysis.

\begin{table}[!htbp]\centering
\caption{Stage-1 Accuracy (G1)}
\label{tab:g1}
\begin{tabular}{@{}lcc@{}}
\toprule
 & $n$ & Agreement \\
\midrule
The two RAs agree with each other & 200 & 0.980 \\
GPT-4o agrees with the settled label & 200 & 0.975 \\
\bottomrule
\end{tabular}
\tabnote{The RAs gave the same answer on 196 of 200 papers, and GPT-4o
matched the settled label on 195 of 200. Both figures are recomputed from the
RAs' original coding sheets. }
\end{table}

\paragraph*{Results from Stage 2.} Before assigning a research design, Stage~2 applies two further screens. The first repeats Stage~1's classification using the full text. Papers that turn out not to be quantitative and empirical leave the pipeline at this point, whether they are methodological work, formal theory, review essays, or qualitative studies. The second screen classifies the paper's purpose. Papers that ask why something happened are classified as explanatory; those that measure or map without asking why are descriptive; and those that forecast are predictive. Only explanatory papers proceed to the research-design classification. Unlike the old prompt, the new list contains no \texttt{Other} category, so the model must assign a design. If it selects \texttt{Multiple Designs}, it must also identify the designs the paper combines.

In the same request, the model records ten additional attributes: subfield, whether the paper states an identification assumption, whether it makes an explicit causal claim, whether it runs a placebo test, whether it mentions a power analysis, whether it has a pre-analysis plan, whether it reports statistically significant results, its sample size, its estimand, and the source of its data. For four of these---the identification assumption, sample size, estimand, and data source---the model must also quote the sentence on which its answer is based.

We have less validation evidence for these ten attributes than for the two main Stage~2 classifications: whether a paper is explanatory and, if so, which research design it uses. For three attributes, we can compare the model with a 176-paper human reference set from the development round, using labels produced under the previous prompt. Accuracy is 0.807 for subfield, 0.594 for whether the paper states an identification assumption, and 0.834 for whether it makes an explicit causal claim. We also measure run-twice repeatability for all ten attributes. These rates range from 0.82 to 1.00; the two lowest are for whether an identification assumption is stated and for the estimand, both at 0.82.

\subsubsection{Accuracy by research design}

\paragraph*{The random sample (G2).} One of the authors constructed G2's gold standard with AI assistance. Two language models, \texttt{Claude Fable 5} and \texttt{Claude Opus 4.8}, independently read all 200 papers. Neither saw the new classifier's answer. They agreed on both scope and research design for 185 papers and disagreed on 15. One author then reviewed all 15 disagreements and a random audit sample of 20 agreements. We shuffled these 35 papers before review, and the author saw neither the AI readers' labels nor whether they had agreed until all decisions were complete. One of the 20 audited agreements was incorrect. Weighting this error by each paper's probability of selection implies an error rate of 0.016 for the full sample.

We scored the classifier against these labels once, without rerunning it or selecting among runs. Table~\ref{tab:g2} reports the result. Research-design accuracy is 0.974 among the 154 papers that both the classifier and the reference label classify as explanatory. Of these, 126 are model-based, the design that the classifier handles best. This estimate therefore reflects accuracy at the corpus's observed mix of designs and says less about performance on rarer designs. The composite accuracy rate of 0.905 uses all 200 papers and requires the classifier to identify both the paper's purpose and, when applicable, its research design correctly. Because every design other than Model-based appears nine times or fewer in G2, we use G3 to estimate accuracy separately by design.

\begin{table}[!htbp]\centering
\caption{Stage-2 Accuracy on the Random Sample (G2)}
\label{tab:g2}
\begin{tabular}{@{}lcc@{}}
\toprule
 & $n$ & Share correct \\
\midrule
Is the paper explanatory?                     & 200 & 0.925 \\
Exact design, where both call it explanatory  & 154 & 0.974 \\
Scope and design both right                   & 200 & 0.905 \\
\bottomrule
\end{tabular}
\tabnote{The second row includes the 154 papers that both the reference label and the
classifier call explanatory. The final row includes all 200 papers.}
\end{table}

\paragraph*{The design-balanced sample (G3).} G3 is drawn from 412 papers labeled independently by two AI readers, with one author resolving their disagreements. We retain at most eight papers of each design in each period. The resulting sample contains 244 papers with roughly balanced representation across designs, which lets us estimate accuracy for rare designs and for papers that combine several designs.

The classifier identifies 240 of the 244 papers as explanatory; the per-design results condition on this set. Among these papers, it assigns the exact research design correctly 87\% of the time. Table~\ref{tab:g3_design} reports accuracy for each design. Each of the ten individual designs reaches at least 71\% accuracy, while Multiple Designs reaches only 50\%.

\begin{table}[!h]\centering
\caption{Accuracy by Research Design (G3)}
\label{tab:g3_design}\small
\begin{tabular}{@{}lccccc@{}}
\toprule
 & \multicolumn{2}{c}{Exact design} & \multicolumn{2}{c}{Presence} & Design-based \\
\cmidrule(lr){2-3}\cmidrule(lr){4-5}
 & $n$ & Recall & $n$ & Recall & or not \\
\midrule
Survey Experiment                    & 24 & 0.96 & 28 & 0.89 & 0.96 \\
Lab Experiment                       & 24 & 0.88 & 29 & 0.86 & 0.92 \\
Field Experiment                     & 23 & 1.00 & 25 & 0.96 & 1.00 \\
Matching/Balancing/Weighting         & 24 & 0.96 & 30 & 0.93 & 0.96 \\
Instrumental Variables               & 24 & 0.79 & 30 & 0.83 & 0.92 \\
Difference-in-Differences            & 24 & 0.71 & 26 & 0.85 & 0.96 \\
Natural Experiment/Quasi Experiment  & 23 & 1.00 & 26 & 0.92 & 1.00 \\
Regression Discontinuity Design      & 18 & 0.94 & 21 & 0.90 & 1.00 \\
Synthetic Control$^{*}$              & 10 & 1.00 & 11 & 0.91 & 1.00 \\
Model-based                          & 22 & 0.95 & 44 & 0.84 & 0.95 \\
Multiple Designs                     & 24 & 0.50 & --- & --- & --- \\
\midrule
\multicolumn{6}{@{}l}{\emph{All 244 papers}}\\
Classifies the paper as explanatory  & 244 & 0.984 &     &       & \\
Names the exact design (of 240 kept) & 240 & 0.871 &     &       & \\
Design-based or not (of 240 kept)    & 240 &       &     &       & 0.933 \\
All design mentions                  &     &       & 270 & 0.885 & \\
\bottomrule
\end{tabular}
\tabnote{\emph{Exact design} recall is the share of a design's papers the classifier gives
that same design, among the 240 it keeps as explanatory; counting the four it excluded as
errors as well puts the rate at 0.857 on all 244. The \emph{Presence} columns count each
design separately. A multi-design paper therefore contributes one observation for every
design it uses. Across the 244 papers, this produces 270 paper-design observations, so
the denominator differs across rows.
$^{*}$Synthetic Control is exempt from the eight-paper cap and never reaches it, the corpus having none before 2010 and two in 2010--2016. Early regression-discontinuity papers are as scarce. }
\end{table}

Multiple Designs is the classifier's weakest category, with errors in both directions. It classifies eight multi-design papers as Model-based and seven single-design papers as Multiple Designs, often treating a baseline regression or a second estimate that uses the same variation as a separate design. Together, these fifteen cases account for nearly half of the sample's thirty-one design errors. Because the errors run in both directions, their net effect on the corpus count of multi-design papers is unclear. Overall, such papers are a very small proportion of the entire corpus ($<3\%$ of all the empirical quantitativepapers).

Two alternative outcomes are more directly relevant to the corpus estimates, and both appear in Table~\ref{tab:g3_design}. The first collapses the eleven labels into two categories, design-based and model-based. On this distinction, the classifier is correct for 0.933 of the 240 papers. It classifies fifteen design-based papers as model-based and only one model-based paper as design-based. Precision among papers classified as design-based is therefore 0.995.

The second outcome counts every design named for a paper rather than requiring the classifier to reproduce the exact combination. Under this rule, recall for Difference-in-Differences rises from 0.71 to 0.85, and recall for every individual design is at least 0.83. Most of the remaining errors are the fourteen papers classified as Model-based when they are not. The classifier therefore tends to understate design-based work and overstate model-based work. Counting each component of a multi-design paper improves the per-design estimates.

\paragraph*{What the classifier confuses.} Table~\ref{tab:g3_confusion} reports the full cross-tabulation behind these rates, with the reference label in rows and the classifier's answer in columns. The errors are concentrated in a small number of cells. Only fifteen of the 110 off-diagonal cells contain any papers, and nearly half of the thirty-one errors fall in a single column: the classifier labels fifteen design-based papers as Model-based, spanning six different reference categories. Multiple Designs accounts for most of the remaining errors in both directions. Eight multi-design papers are classified as Model-based, while seven single-design papers are classified as Multiple Designs---four Difference-in-Differences papers and three Instrumental Variables papers. The remaining errors occur along boundaries that the codebook itself treats as close: two Difference-in-Differences papers and one Lab Experiment are classified as Natural Experiments, and one Regression Discontinuity paper is classified as Instrumental Variables. Outside the Model-based column, no off-diagonal cell contains more than four papers.

\begin{table}[!htbp]\centering
\caption{What the Classifier Confuses (G3)}
\label{tab:g3_confusion}
{\footnotesize
\setlength{\tabcolsep}{3.5pt}
\resizebox{\textwidth}{!}{%
\begin{tabular}{@{}l ccccccccccc cc@{}}
\toprule
 & \multicolumn{11}{c}{Classifier} & & \\
\cmidrule(lr){2-12}
Reference label & SE & LE & FE & MW & IV & DD & NQ & RD & SC & MB & MD & $n$ & Recall \\
\midrule
Survey Experiment & \textbf{23} & . & . & . & . & . & . & . & . & 1 & . & 24 & 0.96 \\
Lab Experiment & . & \textbf{21} & . & . & . & . & 1 & . & . & 2 & . & 24 & 0.88 \\
Field Experiment & . & . & \textbf{23} & . & . & . & . & . & . & . & . & 23 & 1.00 \\
Matching/Balancing/Weighting & . & . & . & \textbf{23} & . & . & . & . & . & 1 & . & 24 & 0.96 \\
Instrumental Variables & . & . & . & . & \textbf{19} & . & . & . & . & 2 & 3 & 24 & 0.79 \\
Difference-in-Differences & . & . & . & . & . & \textbf{17} & 2 & . & . & 1 & 4 & 24 & 0.71 \\
Natural Experiment/Quasi Experiment & . & . & . & . & . & . & \textbf{23} & . & . & . & . & 23 & 1.00 \\
Regression Discontinuity Design & . & . & . & . & 1 & . & . & \textbf{17} & . & . & . & 18 & 0.94 \\
Synthetic Control & . & . & . & . & . & . & . & . & \textbf{10} & . & . & 10 & 1.00 \\
Model-based & . & . & . & . & . & . & . & . & . & \textbf{21} & 1 & 22 & 0.95 \\
Multiple Designs & 1 & 1 & . & 2 & . & . & . & . & . & 8 & \textbf{12} & 24 & 0.50 \\
\bottomrule
\end{tabular}}
}
\tabnote{Rows give the settled reference label and columns the classifier's answer for the 240 of 244 G3 papers that the classifier retains as explanatory. Bold marks the diagonal; a dot marks an empty cell. Column keys: SE Survey Experiment, LE Lab Experiment, FE Field Experiment, MW Matching/Balancing/Weighting, IV Instrumental Variables, DD Difference-in-Differences, NQ Natural Experiment/Quasi Experiment, RD Regression Discontinuity Design, SC Synthetic Control, MB Model-based, and MD Multiple Designs. Because G3 includes at most eight papers of each design in each period, the row totals reflect the sample design rather than the distribution of designs in the corpus. Table~\ref{tab:dsl} reports the corrected corpus shares.}
\end{table}

Observed accuracy is similar across the three periods in Table~\ref{tab:g3_period}. The largest difference in point estimates is 2.4 percentage points for exact research design and 1.8 percentage points for the design-based versus model-based distinction. These comparisons are descriptive, however, and do not support strong conclusions about the classifier's stability over time.

\begin{table}[!htbp]\centering
\caption{Stability over Time (G3)}
\label{tab:g3_period}
\begin{tabular}{@{}lcccc@{}}
\toprule
Period & $n$ & Exact design & Design-based or not & Explanatory \\
\midrule
2003--2009 & 74 & 0.861 & 0.944 & 0.973 \\
2010--2016 & 82 & 0.864 & 0.926 & 0.988 \\
2017--2023 & 88 & 0.885 & 0.931 & 0.989 \\
\bottomrule
\end{tabular}
\tabnote{Based on one run of the frozen classifier, Claude Sonnet 5 with the new prompt. The first two accuracy columns condition on the classifier keeping the paper as explanatory ($n$ = 72, 81, and 87); the last column is computed on every paper in the period. Because G3 holds the mix of designs roughly constant across periods, these comparisons are not mechanically driven by changes in the mix of labeled designs.}
\end{table}

\subsubsection{Four human coders on the same papers}

Four human coders independently reviewed 60 G3 papers using the same codebook and without access to the classifier's answer. We sampled these papers with equal probability from the 220 G3 papers that carry a single design. Each paper was assigned to two coders, and each of the six possible coder pairs reviewed ten papers, producing 120 coder-paper judgments in total.

Before coding began, one of the authors settled the reference label for each paper using the two AI readers' labels together with the paper itself. The coding interface concealed the classifier's answer. These labels serve as the gold standard.

\begin{table}[!htbp]\centering
\caption{Human Coders and the Classifier on the Same Papers (G3)}
\label{tab:coders60}
\resizebox{\textwidth}{!}{%
\begin{tabular}{@{}lccc@{}}
\toprule
 & $n$ & Exact design & Design-based or not \\
\midrule
Coder 1 & 30 & 0.900 & 0.967 \\
Coder 2 & 30 & 0.900 & 0.933 \\
Coder 3 & 30 & 0.967 & 1.000 \\
Coder 4 & 30 & 0.967 & 0.967 \\
\addlinespace
All four coders & 120 & 0.933 & 0.967 \\
Classifier, duplicated to match coder assignments & 120 & 0.933 & 0.950 \\
\bottomrule
\end{tabular}}
\tabnote{Each paper enters the pooled human coder row twice, once for each coder. The
classifier row scores the same 120 paper-coder pairs. Scored once per paper instead,
the classifier identifies the exact design on 56 of 60 and the correct side of the
design-based-versus-model-based distinction on 57 of 60, giving the same rates to
three decimals. A paper that a coder places outside the scope of the classification
counts as an error when its reference label gives a design.}
\end{table}

The pooled human and classifier results are identical on exact research design. Human coders make 112 correct decisions out of 120 coder-paper judgments. To put the classifier on the same denominator, we count its single prediction for each paper once for each of the two coders assigned to that paper; it is also correct on 112 of 120. At the paper level, the classifier is correct on 56 of 60 papers. For the coarser distinction between design-based and model-based work, the human coders are correct on 116 of 120 judgments and the classifier on 57 of 60 papers.

The remaining errors in this coarser classification all involve the Model-based category, the only label on that side of the distinction. Human coders assign Model-based on 16 readings, 13 of them correctly, and fail to assign it on one of the 14 readings whose reference label is Model-based. The classifier assigns Model-based to 9 of the 60 papers, 7 correctly, and misses none of the 7 papers whose reference label is Model-based. Both humans and the classifier therefore make more errors by assigning Model-based to design-based work than by failing to identify model-based work. Only the classifier's errors, however, enter the corpus estimates.

The two human coders assigned to a paper give the same design on 53 of the 60 papers, and all 53 shared answers match the reference label. All eight coder errors occur among the seven papers on which the two coders disagree. On six of those seven papers, one coder gives the reference label. This high level of agreement should be interpreted cautiously, because the two AI readers used to construct the reference labels had already agreed on 57 of the 60 papers. The sample may therefore contain relatively few of the hardest cases.

\subsubsection{Intrinsic ambiguity}

Some of the classification error documented above reflects ambiguity in the papers themselves rather than mistakes unique to the classifier. Whenever two readers independently assign research designs to the same papers, they disagree at similar rates. The two RAs who read the 290 papers in G3$'$ chose the same design for 254, an agreement rate of 0.876. The two AI readers agreed on scope and design for 185 of the 200 papers in G2, a rate of 0.925. The paired human coders in Table~\ref{tab:coders60} chose the same design for 53 of 60 papers, a rate of 0.883. Across three kinds of reader pairs, working at different times and under different instructions, agreement therefore falls between 0.87 and 0.93. Readers agree more readily on simpler classifications: the two RAs who read G1 agreed on whether a paper was quantitative and empirical for 196 of 200 papers. Research design, unlike that screen, includes cases on which reasonable readers do not converge.

The disagreements are concentrated in categories that require judgment rather than the detection of a visible feature. In G3$'$, the RA pair agreed on all sixteen Field Experiment papers and all seventeen Synthetic Control papers. In each case, a relatively clear feature settles the classification: researchers randomize a treatment, or the analysis constructs a donor pool. Agreement falls to 0.75 for Instrumental Variables, where the reader must decide whether the instrument constitutes the paper's design or serves only as a robustness check, and to 0.50 for Matching, where the reader must decide whether matching carries the identification claim or merely constructs the sample. Multiple Designs is the clearest example. The RA pair agreed on none of the three G3$'$ papers whose settled label combines designs, and the classifier likewise performs worst on Multiple Designs in G3 (Table~\ref{tab:g3_design}). The same few problems recur: a procedure fits more than one category, two designs appear side by side, or a paper names one design while implementing another.

The codebook is meant to resolve such cases by rule, and it does so in part. The classifier's instructions encode those rules, and the classifier reaches 0.96 accuracy on Matching in G3, the category on which the unaided RA pair agreed least in G3$'$. Because the two rates come from different samples, they should not be read as evidence that the classifier improves on the RAs. They instead illustrate the role of the codebook: a rule cannot remove ambiguity from a paper, but it can standardize how that ambiguity is resolved.

Two implications follow. First, benchmark accuracy on this task is partly limited by disagreement among readers. When a paper supports more than one reasonable interpretation, scoring still requires a single reference label, and that label necessarily reflects a coding convention. The comparison in Table~\ref{tab:coders60} is therefore especially informative: on the same papers and under the same codebook, the classifier assigns designs at roughly the same rate as trained human coders. Second, ambiguity that remains after applying the codebook enters the corpus estimates as classification error. Collapsing the labels into design-based and model-based categories resolves some disagreements, but not all. Some difficult cases, such as Matching versus Model-based, fall precisely on that boundary. Section~\ref{sisec:dslcorr} adjusts the corpus estimates for the remaining classification error, whether it reflects ambiguity in the papers or mistakes by the classifier, under the assumption that a single ground-truth label exists.

\subsubsection{Summary and limitations}

In summary, both screens perform well while accuracy varies by design. Stage~1 agrees with settled human labels on more than 97\% of G1, and Stage~2 identifies explanatory papers correctly in 0.925 of G2 and 0.984 of G3. Among papers that both the classifier and reference label treat as explanatory, exact design accuracy is 0.974 in G2. In G3, accuracy is at least 0.71 for every individual design, 0.933 for the design-based-versus-model-based distinction, and 0.50 for Multiple Designs. These rates are close to the task's practical ceiling: four human coders using the same codebook match the classifier's pooled accuracy on the same papers, and independent reader pairs agree on research design about 90\% of the time.

Our classification and validation scheme has three main limitations. First, we developed the reference labels with AI assistance. Two language
models labeled the papers in G2 and G3, after which one of us resolved their
disagreements without seeing the classifier's answer. We also audited labels on which
the two AI readers agreed. One of 20 audited labels in G2 was wrong, corresponding to a
weighted error rate of 0.016. One of 30 audited labels in G3 was wrong, for a weighted error rate of 0.015. Because both audits sample agreements, they measure error among consensus labels rather than among labels that required adjudication. The two AI readers and the classifier are all Claude models and may therefore share errors. The four-coder exercise adds human judgments and shows that the pooled human and classifier accuracy rates are similar (Table~\ref{tab:coders60}), but that similarity cannot rule out errors shared by the Claude models. 

Second, the classifier is not perfectly stable. When run twice on the same papers, it
gives the same design between 0.88 and 0.91 of the time. These results
show that some paper-level labels depend on the run. Each paper in the corpus was
labeled once, so this instability is a source of noise we do not model. Its effect on the
quantities we report appears small: comparing the corpus run with the separate runs used
for validation, the two agree on whether a paper is explanatory for 0.955 of the G2 papers
and 0.996 of the G3 papers, and on the research design for 0.974 and 0.942 of the papers
both runs treat as explanatory.

Third, authors' descriptions of their own designs form part of our classification
rule. Because the language used to describe research designs may have changed over
the two decades we study, this rule could create artificial trends in the assigned
labels. The point estimates in Table~\ref{tab:g3_period} differ little across the
three periods, but the sample does not support precise period comparisons and
therefore does not rule out changes over time.

\clearpage

\subsection{Correcting the corpus estimates for classification error}\label{sisec:dslcorr}

The shares and trends reported in the main text treat the classifier's labels as correct, so their confidence intervals reflect only sampling variation. But the classifier is not perfectly accurate. Some errors arise because papers themselves are ambiguous: as the preceding section shows, reasonable readers sometimes disagree about research design. Others arise from the classifier. We use design-based supervised learning \citep[DSL,][]{egami2024dsl} to incorporate both sources of classification error into the estimates. The main results are unchanged after correction.

The correction is intuitive. Classification error biases an estimated share when mistakes are asymmetric. The classifier sometimes labels design-based papers as model-based and sometimes makes the reverse error. If one mistake is more common, the observed share is biased downward or upward. The validation samples let us estimate this bias because, for each validated paper, we observe both the classifier's label and the reference label. DSL uses these observed errors to adjust the corpus estimate, weighting each validated paper by the number of corpus papers it represents. The resulting confidence interval reflects both classification error and uncertainty from the finite validation sample.

\begin{table}[!tbp]\centering
\caption{DSL-Corrected Corpus Trends}
\label{tab:dsl}
\begin{tabular}{@{}lccc@{}}
\toprule
Period & Design-based share, uncorrected & Corrected & 95\% CI \\
\midrule
2003--2009 & 0.104 & 0.115 & [0.096, 0.133] \\
2010--2016 & 0.161 & 0.179 & [0.153, 0.205] \\
2017--2023 & 0.249 & 0.271 & [0.240, 0.301] \\
\bottomrule
\end{tabular}
\tabnote{Share of papers that are explanatory and use a design-based method, among the 35,713 papers eligible for a validation draw. We compute the correction separately for each period, weighting each paper by the inverse of its recorded probability of selection. A small number of reference labels remain unresolved and are excluded; resolving them in every possible way changes no estimate materially. The replication materials record each paper's selection probability and these bounds.}
\end{table}

We pool 842 reference-labeled papers with known probabilities of selection: the 200 papers in G2 and 642 papers read while constructing and extending G3. Because rare designs were deliberately oversampled, we account for their unequal selection probabilities with weights. We correct three quantities: the share of explanatory papers, the share of papers that are both explanatory and design-based, and the share using each design in each period. We work by period rather than by year because the validation samples were drawn within these three periods, and finer splits would leave too few validated papers to estimate error reliably.

\begin{figure}[!htbp]\centering
\caption{Corrected Trend in Design-based Research}\label{fig:dsltrend}
\includegraphics[width=0.7\textwidth, trim={0cm 0.5cm 0cm 0.5cm}, clip]{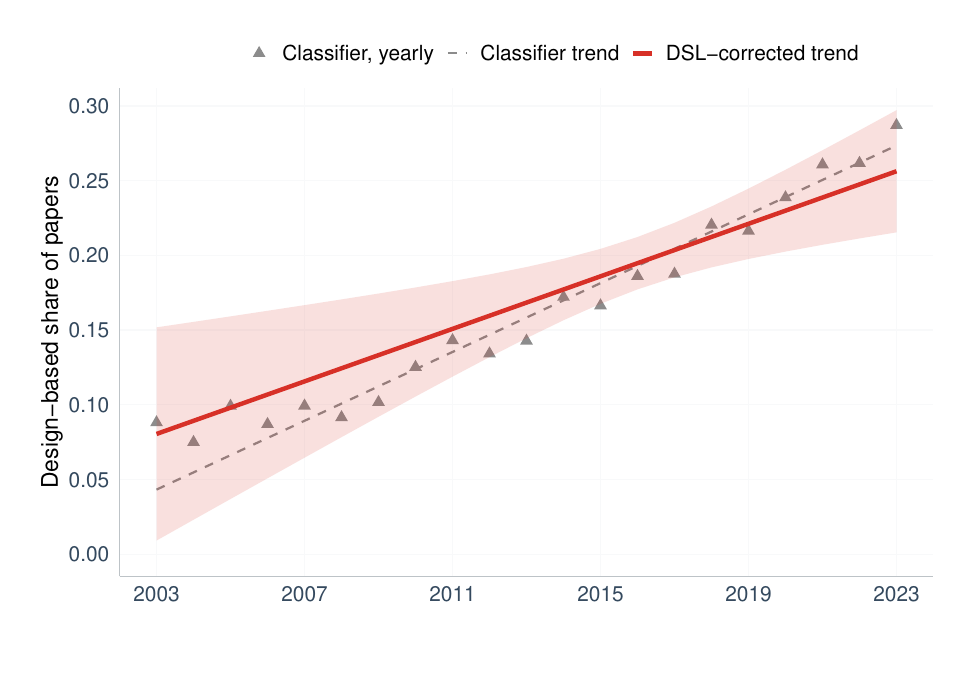}
\fignote{Yearly share of papers that are explanatory and use a design-based method. Gray
triangles show the classifier estimate; the dashed line shows its least-squares trend. The
solid line shows the DSL-corrected trend and its 95\% confidence band, estimated from the
200 papers in G2.}
\end{figure}

\begin{figure}[!tbp]\centering
\caption{Corrected Share of Each Design, by Period}\label{fig:dslera}
\includegraphics[width=0.9\textwidth]{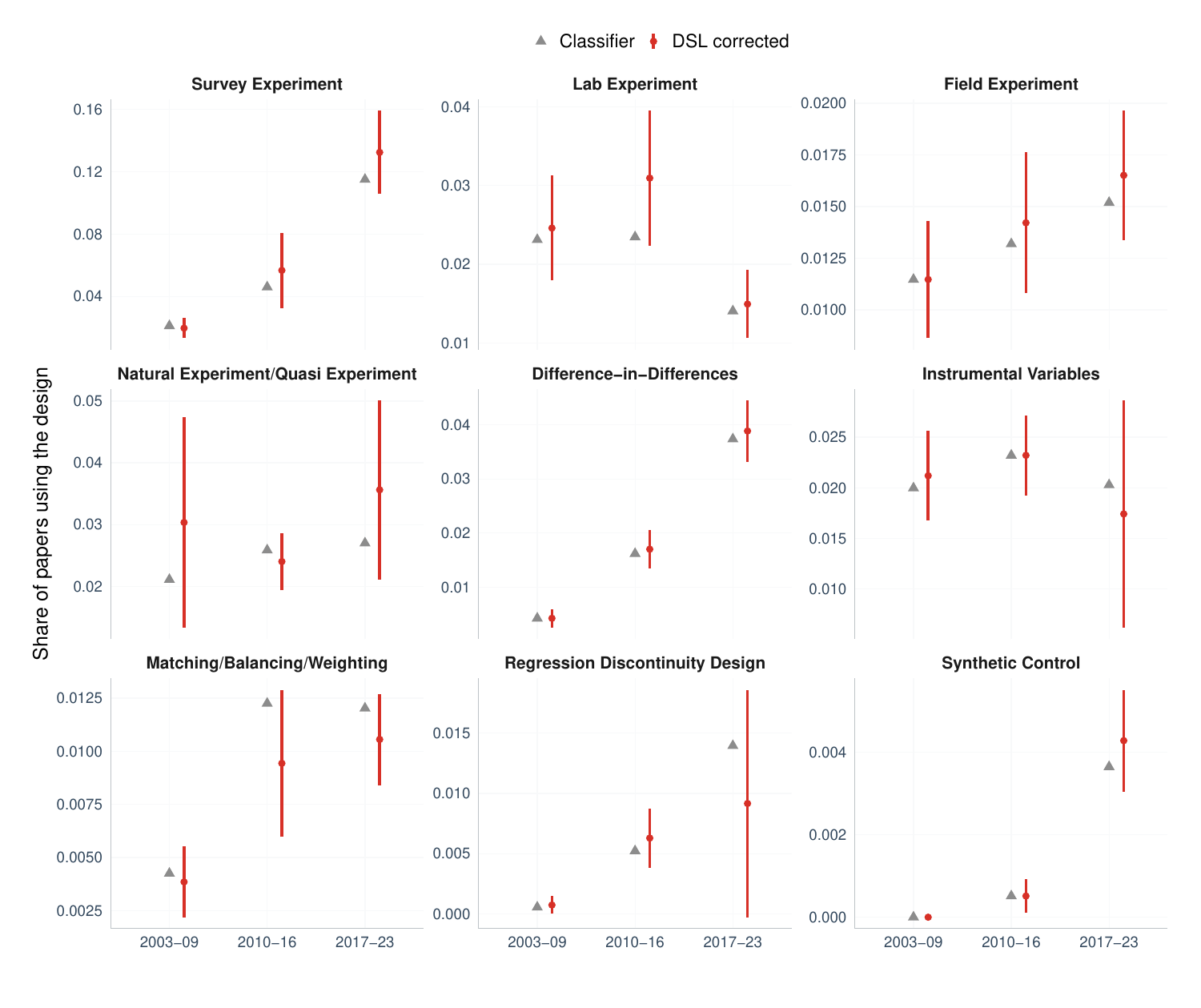}
\fignote{Share of papers using each design-based method, by period. Gray triangles show the
classifier estimate. Red points show the DSL-corrected estimate and its 95\% confidence
interval, calculated from 842 reference-labeled papers with recorded selection
probabilities.}
\end{figure}

The corrections are small and uniformly positive. The corrected explanatory share is 0.857, with a 95\% CI of [0.825, 0.889], compared with the classifier's 0.804, because in G2 the classifier misses more explanatory papers than it incorrectly adds (13 versus 2; Table~\ref{tab:g2}). The corrected share of papers that are both explanatory and design-based is 0.217 [0.199, 0.235], compared with the classifier's 0.198. Table~\ref{tab:dsl} reports the correction by period: the design-based share rises in every period under both the corrected and uncorrected estimates. Figure~\ref{fig:dsltrend} shows a corrected trend of 0.88 percentage points per year (standard error 0.27), compared with 1.15 points uncorrected. Figure~\ref{fig:dslera} applies the same correction to each of the nine design-based methods. The corrected estimates remain close to the classifier's throughout, and the growth of survey experiments, difference-in-differences, and regression discontinuity designs remains. Confidence intervals are widest where few validated papers use a design, especially for rare designs in the earliest period. Classification error therefore does not account for the main results.

\clearpage

\subsection{Institutional ranking}\label{rankingmeth}

In this section, we describe how we collected and processed institutional ranking data and linked these data to the affiliations of authors in our corpus. We rank academic institutions using the Academic Ranking of World Universities (ARWU, or Shanghai Ranking). Our goal is to match each author of each paper in our corpus to the ARWU ranking of their listed institution in the year the paper was published. We proceed in three steps. First, we process the ARWU rankings. Second, we extract all author-institution pairs from Scopus. Third, we match the institutions reported by Scopus to those appearing in ARWU.

Each year, ARWU scores about 2,500 research universities on six fixed indicators of research output and distinction: Nobel and Fields laureates among alumni and staff, Highly Cited Researchers, papers published in \textit{Nature} and \textit{Science}, and total indexed publications. Each indicator is expressed as a percentage of the value attained by the highest-scoring university on that indicator. From these scores, ARWU publishes an ordered ranking of the top 1,000 universities. Universities in the top 100 receive an exact rank, while the remainder are reported in bands (101--150, 151--200, and so on). We assign universities in the top 100 their published rank and universities reported in bands the midpoint of their band.

Scopus assigns each affiliation a stable identifier and links that affiliation to individual authors. The 91,632 papers in our corpus contain 158,227 paper-author pairs. Scopus records an institutional identifier for 149,714 of these pairs; the remaining 8,513 have no recorded affiliation. These identifiers correspond to 9,668 distinct institutions, each associated with the institution name reported by Scopus. Some Scopus identifiers refer to faculties or schools rather than entire universities, so some records have names such as ``Faculty of Arts'' without identifying the parent university.

We match these institutions to ARWU in two stages, using conservative matching rules in both. In the first stage, we match institution names exactly after normalization. We lowercase each Scopus and ARWU name, strip accents, and standardize translations of \textit{university} to a common token. For example, ``Universiteit van Amsterdam'' and ``University of Amsterdam'' reduce to the same normalized string. We then match a Scopus identifier to an ARWU institution when their normalized names are identical. This procedure matches 801 institutions, covering 82,870 paper-author pairs, or 52\% of the total.

In the second stage, we submit the remaining 8,867 institutions to Claude Sonnet 5 for matching. Each call includes the complete list of 1,424 university names that appear in the ARWU data, along with the name, city, and country of each unmatched Scopus institution. The model is instructed to return an ARWU name verbatim only when it is confident that the two records refer to the same institution. Otherwise, it returns no match and classifies the Scopus institution as an unranked university, a research institute, a non-academic organization, or an institution whose name is too fragmentary to identify. Providing the model with the complete set of candidate ARWU institutions and instructing it to prefer no match when uncertain is intended to limit false-positive matches, at the cost of potentially leaving some valid matches unresolved. This second stage matches another 868 institutions, covering 28,787 paper-author pairs. We reproduce the instructions given to the model below. To ensure accuracy, we show a higher capacity second model, \texttt{Claude Opus 4.8}, each institution name and the ARWU entry it was matched to and ask it to agree or disagree with the match. The second reader confirms the matches of institutions in 99.2\% of the matched paper-authors.

{\scriptsize
\begin{verbatim}
You match institutions to a world university ranking, holding the ENTIRE
ranking so there is no candidate shortlist and no hint.

For each institution you get its Scopus name, city, and country.
Do TWO things.

(1) MATCH to the ranking. Decide whether this institution is the SAME
institution as one entry in the ARWU list. Output the entry's EXACT text
if you are confident, otherwise NONE. Rules:
- Bias to NONE. Match only when sure it is the same institution. A missed
  match is cheap; a false match is not.
- Translations and word-order are fine: "Universiteit Leiden" = "Leiden
  University", "Technische Universitaet Muenchen" = "Technical University
  of Munich".
- A school/faculty/campus resolves to its PARENT only if the parent is
  named in the string and the parent is in the list: "Harvard Kennedy
  School" -> "Harvard University"; "Stanford Graduate School of Business"
  -> "Stanford University".
- Use city to separate same-named places: "City University of Hong Kong"
  is not "The University of Hong Kong"; a campus in one city is not the
  campus in another.
- Same name, wrong country/city -> NONE.
- "choice" must be verbatim one ARWU entry, or NONE.

(2) If NONE, CLASSIFY why, in two fields:
- status:
   "clear_outside" = you can identify the specific institution and it is
                     genuinely not in the ranking
   "chunk"         = a bare sub-unit fragment (a faculty, school,
                     department, centre) with NO institution named, so you
                     cannot say which institution it is ("Faculty of Arts")
   "vague"         = a name too generic or ambiguous to pin to one
                     specific institution ("Institute of Social Studies")
- kind (only when status is clear_outside; else leave empty):
   "university"          = a degree-granting university or college below
                           the ranking
   "research_institute"  = an academic research organization, not
                           degree-granting (Max Planck, CNRS, an academy
                           institute, a policy research centre inside no
                           university)
   "non_academic"        = a firm, bank, government body, IGO, think tank,
                           NGO, hospital, or publisher
If you MATCH, set status="matched", kind="", choice=the ARWU entry.
\end{verbatim}
}

Table~\ref{tab:affil_match_steps} summarizes the results of the two matching stages. Together, they match 1,669 institutions, covering 111,657 paper-author pairs, or 70.6\% of all paper-author pairs in the corpus. Table~\ref{tab:affil_disposition} reports the disposition of all 158,227 paper-author pairs, including those associated with institutions outside ARWU and those for which Scopus records no usable affiliation.

\begin{table}[!ht]
\centering
\caption{Institution Matching by Method}\label{tab:affil_match_steps}
\small
\begin{tabular}{lrr}
\toprule
Method & Institutions & Paper-author pairs \\
\midrule
Exact name, normalized & 801 & 82,870 \\
LLM match, full ARWU list & 868 & 28,787 \\
Total matched & 1,669 & 111,657 \\
\bottomrule
\end{tabular}
\end{table}

\begin{table}[!ht]
\centering
\caption{Disposition of Paper-Author Pairs}
\label{tab:affil_disposition}
\small
\begin{tabular}{lrrrr}
\toprule
& \multicolumn{2}{c}{All papers} & \multicolumn{2}{c}{Explanatory quant} \\
Affiliation status & Count & Share & Count & Share \\
\midrule
Total & 158,227 & 100.0\% & 62,629 & 100.0\% \\
matched to an ARWU institution & 111,657 & 70.6\% & 47,670 & 76.1\% \\
unranked university & 22,553 & 14.3\% & 8,634 & 13.8\% \\
research institute & 6,644 & 4.2\% & 2,238 & 3.6\% \\
non-academic organization & 4,442 & 2.8\% & 1,282 & 2.0\% \\
unidentified affiliation & 4,418 & 2.8\% & 1,378 & 2.2\% \\
no affiliation recorded & 8,513 & 5.4\% & 1,427 & 2.3\% \\
\bottomrule
\end{tabular}
\end{table}

Among unmatched paper-author pairs, the largest category consists of universities classified as outside the ARWU ranking. The universities with the most paper-author pairs in this category are the European University Institute (509), American University (467), the University of Mannheim (464), Central European University (338), and the Hertie School (239). Several of these institutions specialize heavily in the social sciences, while ARWU ranks universities using indicators that include Nobel and Fields laureates, Highly Cited Researchers, \textit{Nature} and \textit{Science} papers, and total indexed publications. Other unmatched affiliations belong to research institutes that ARWU does not rank, including the WZB Berlin Social Science Center, RAND, and the Peace Research Institute Oslo, or to organizations outside academia, including the World Bank, International Monetary Fund, and European Central Bank. A final group, accounting for less than 3\% of paper-author pairs, consists of affiliation strings too incomplete or ambiguous to identify.

A university can enter or leave the ARWU top 1,000 from year to year. We therefore define a paper as \textit{covered} when at least one author is affiliated with a university that appears in the ARWU ranking in the year the paper was published. By this definition, 67.3\% of papers in the full corpus are covered, as are 77.0\% of empirical quantitative papers and 77.6\% of the explanatory quantitative papers used in the institutional analysis (Table~\ref{tab:affil_coverage}). 
\begin{table}[!ht]
\centering
\caption{Coverage by Analytic Sample}
\label{tab:affil_coverage}
\small
\begin{tabular}{lrrr}
\toprule
Frame & Papers & Covered & Coverage \\
\midrule
Full-text sample & 91,632 & 61,698 & 67.3\% \\
Empirical quantitative & 35,349 & 27,226 & 77.0\% \\
Explanatory quantitative (institutional analysis) & 29,987 & 23,283 & 77.6\% \\
\bottomrule
\end{tabular}
\end{table}

Finally, we ask whether incomplete institutional coverage could explain the higher use of design-based methods at top-ranked institutions. Our baseline assigns each paper the mean ARWU rank of authors matched to a university ranked in the publication year. Missing matches could bias the gradient either by excluding papers with no ranked authors or by misplacing papers with only some authors matched. In the baseline, 38.9\% of explanatory quantitative papers at ranks 1--20 are design-based, compared with 24.5\% at ranks 41--100. We assess both concerns with three checks.

First, we restrict the analysis to the 6,913 papers for which every author is matched to a ranked university, so additional matches cannot change a paper's position. The gradient is nearly unchanged: 39.8\% design-based at ranks 1--20 versus 24.1\% at ranks 41--100.

Second, we assign universities absent from ARWU in the publication year their rank from the nearest available year. This adds 7,616 papers to the covered corpus, including 2,219 explanatory quantitative papers, but again leaves the gradient nearly unchanged: 39.5\% versus 24.7\% (Table~\ref{tab:affil_gradient_bounds}). Most newly covered papers fall outside the top 100, so the added information has little effect on the comparison shown in the main text.

\begin{table}[!ht]
\centering
\caption{Design-based Share by Institutional Rank}
\label{tab:affil_gradient_bounds}
\small
\begin{tabular}{lrrrr}
\toprule
\multicolumn{1}{c}{} & \multicolumn{2}{c}{In-year rule} & \multicolumn{2}{c}{Nearest-year rule} \\
Rank bin & Papers & Design-based & Papers & Design-based \\
\midrule
1--20 & 2,713 & 38.9\% & 2,539 & 39.5\% \\
21--40 & 1,943 & 31.7\% & 1,837 & 31.5\% \\
41--100 & 4,801 & 24.5\% & 4,617 & 24.7\% \\
\bottomrule
\end{tabular}
\end{table}

Third, we assign the remaining uncovered papers in the way most unfavorable to our result: every model-based paper to ranks 1--20 and every design-based paper to ranks 41--100. The gap falls from 14.7 to 8.9 percentage points when we include 402 papers with fragmentary affiliations, and to 2.4 points when we also include 578 papers with no recorded affiliation. Even under this extreme assignment, design-based research remains more common at top-ranked institutions.

\clearpage

\subsection{Additional empirical results}

\subsubsection{Alternative explanations}\label{sisec:altexps}

In this section, we assess whether the patterns documented in the main text regarding the increased adoption of design-based methods could be driven by systematic changes in the denominator rather than by changes in the relative adoption of design-based strategies. One concern is that, instead of an increase in the use of design-based methods, heightened scrutiny of causal claims may have encouraged authors relying on model-based approaches to frame similar analyses in non-causal terms, reclassifying their work as descriptive rather than explanatory. If so, the observed growth of design-based methods within explanatory research could reflect changes in labeling rather than substantive methodological change.

\begin{figure}[!ht]
    \centering
    \caption{Trends in the Share of Papers Coded as Descriptive}
    \label{fig:descptrends}
        \includegraphics[width=.9\linewidth]{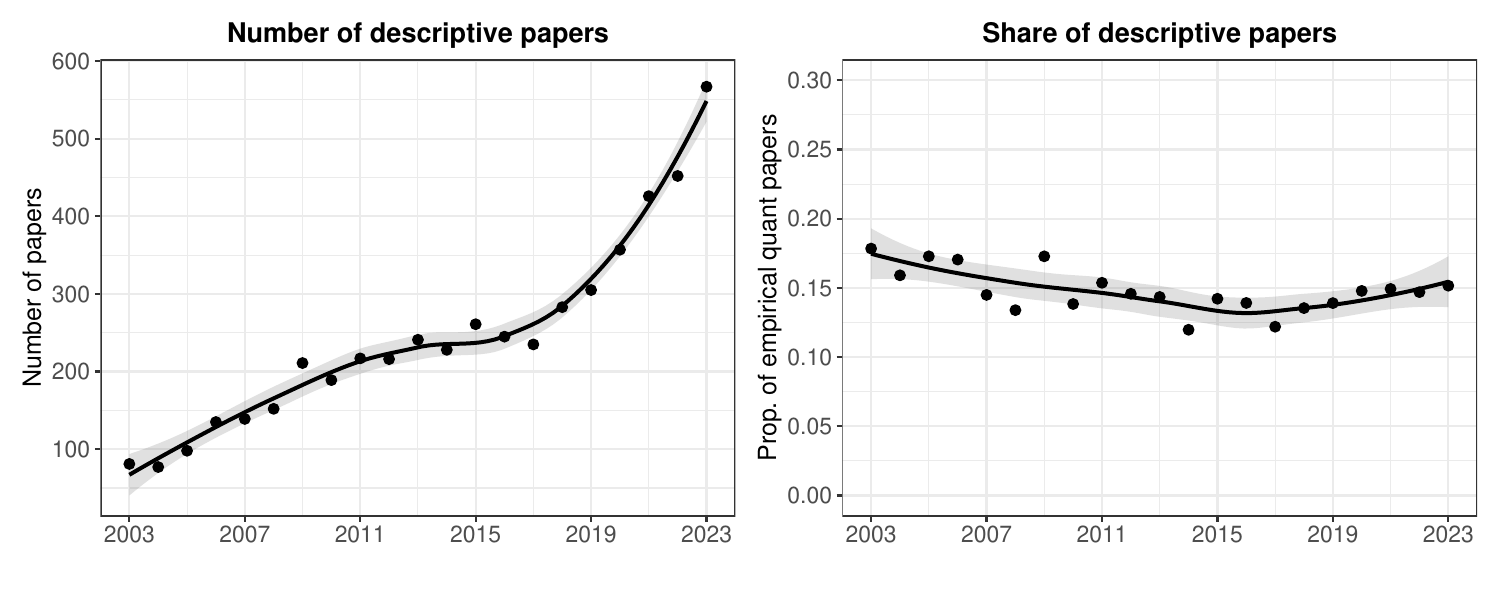}
    \fignote{The denominator is the number of empirical quantitative papers.}
\end{figure}

\begin{figure}[!ht]
    \centering
    \caption{Explicit Discussion of Identification Assumptions over Time}
    \label{fig:idassumps}
        \includegraphics[width=.7\linewidth]{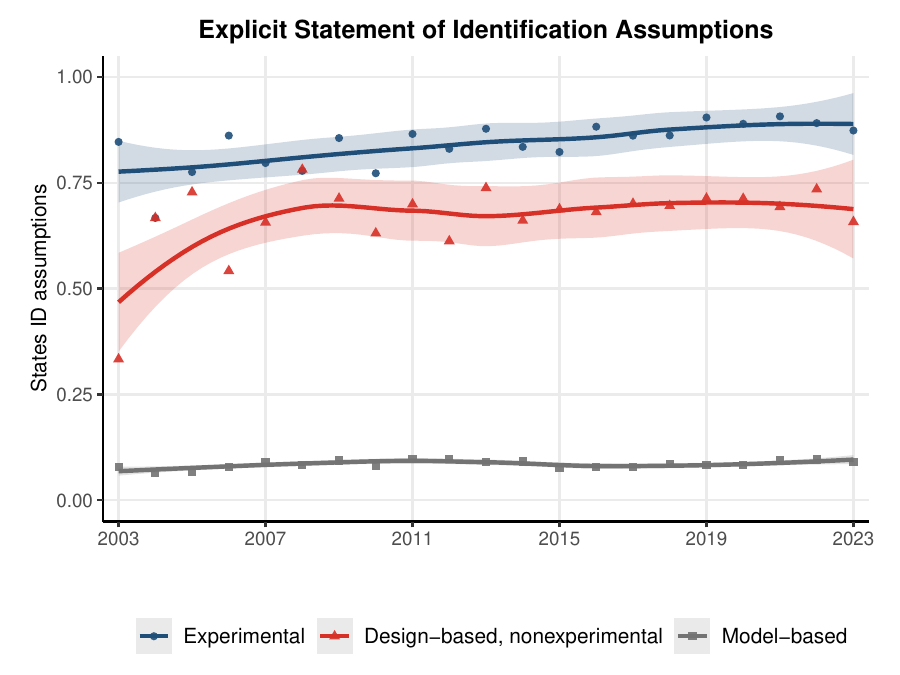}
    \fignote{The figure plots the share of explanatory papers that explicitly mention or discuss their identification assumptions, by research design category.}
\end{figure}

We find little evidence consistent with this account. First, the share of descriptive papers in our sample remains stable over the period we study, as shown in Figure~\ref{fig:descptrends}. It is 17.8\% in 2003 and 15.2\% in 2023, and it stays between 12\% and 18\% throughout. This stability suggests that explanatory work has not been systematically reclassified into descriptive categories that could mechanically account for our results. Second, explicit discussion of identification assumptions within explanatory papers shows a distinct and informative pattern. As shown in Figure~\ref{fig:idassumps}, such statements are standard in experimental work throughout the period, at 85\% to 88\% of papers. In design-based observational work, they rise from a third of papers in 2003 to about two thirds in 2023. In model-based papers, they are rare, at 8\% to 9\%, and flat across the whole period. The flat model-based series suggests that model-based research has not become more guarded about causal claims over time. Overall, the evidence suggests that the trends we document are unlikely to be artifacts of shifting denominators or changes in how authors describe their research aims.

\subsubsection{Results without survey experiments}

Survey experiments randomize the assignment of treatment and are therefore design-based. Yet they can also be conceptualized as strategies of measurement rather than of causal identification, a tradition with different epistemological roots. We therefore report the main trends on design-based and model-based adoption after excluding them.

\begin{figure}[!ht]
    \centering
    \caption{Trends in Design-based Methods, Excluding Survey Experiments}
    \label{fig:basictrends_nosurvey}
    \begin{subfigure}[b]{0.47\textwidth}
        \centering
        \includegraphics[width=\linewidth]{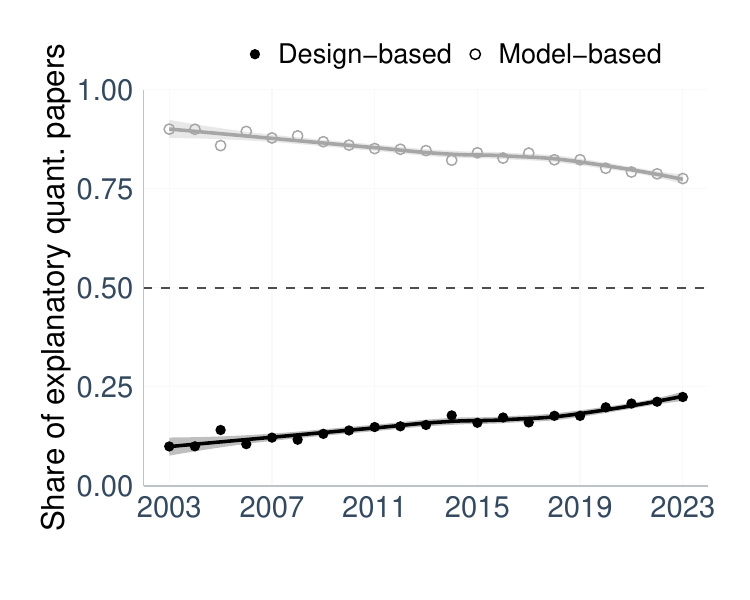}
        \caption{Design-based vs. model-based studies}
        \label{fig:design_overallnosurvey}
    \end{subfigure}    \hfill
    \begin{subfigure}[b]{0.47\textwidth}
        \centering
        \includegraphics[width=\linewidth]{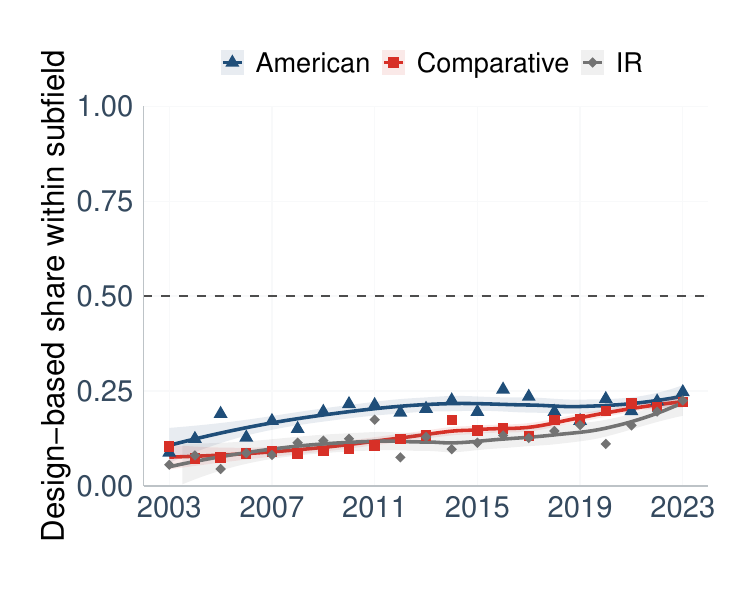}
        \caption{Design-based studies by subfield}
        \label{fig:design_subfieldsnosurvey}
    \end{subfigure}
    \fignote{The sample comprises all explanatory empirical quantitative papers, excluding survey experiments.}
\end{figure}

Figure~\ref{fig:basictrends_nosurvey} replicates the analysis of Figure~\ref{fig:basictrends} after excluding survey experiments from the sample. Panel (a) shows that design-based methods account for a smaller share of explanatory quantitative research under this restriction, rising from 10\% in 2003 to 22\% in 2023, while model-based methods account for more than three quarters of papers even in 2023. Panel (b) shows that the subfield differences narrow under this restriction and nearly close by the end of the period. American Politics still adopts design-based methods earlier, at 15\% of papers in 2003 to 2008 against 8\% in Comparative Politics and International Relations. By 2023, however, the design-based share excluding survey experiments are close for the three subfields  25\% for American Politics and 22\% for both International Relations and Comparative Politics. 

\subsubsection{Regression analyses}

This section presents tabular versions of selected figures from the main text. We also report a regression analysis of the citation gap associated with design-based methods, accounting for journal, subfield, and year fixed effects.

\paragraph*{Research design trends (Figure 5).} Table \ref{tab:design_trends} reports the data underlying Figure \ref{fig:bymethods2}, showing the distribution of research designs overall and for selected years. Design-based methods grew from 12.6\% of explanatory quantitative papers in 2003 to 36.5\% in 2023, while model-based methods declined from 87.4\% to 63.5\%.

\begin{table}[htbp]
\centering
\caption{Research Design Trends}
\label{tab:design_trends}
\footnotesize
\begin{tabular}{@{}lcccccccc@{}}
\toprule
Research design & \multicolumn{2}{c}{Overall} & \multicolumn{2}{c}{2003} & \multicolumn{2}{c}{2013} & \multicolumn{2}{c}{2023} \\
 & \multicolumn{2}{c}{(N=29,987)} & \multicolumn{2}{c}{(N=373)} & \multicolumn{2}{c}{(N=1,418)} & \multicolumn{2}{c}{(N=3,155)} \\
\cmidrule(lr){2-3} \cmidrule(lr){4-5} \cmidrule(lr){6-7} \cmidrule(lr){8-9}
 & N & \% & N & \% & N & \% & N & \% \\
\midrule
Model-based & 22,495 & 75.0 & 326 & 87.4 & 1,144 & 80.7 & 2,002 & 63.5 \\
All design-based & 7,492 & 25.0 & 47 & 12.6 & 274 & 19.3 & 1,153 & 36.5 \\
\midrule
Survey Experiment & 2,888 & 38.5 & 11 & 23.4 & 69 & 25.2 & 569 & 49.3 \\
Natural/Quasi Experiment & 985 & 13.1 & 12 & 25.5 & 45 & 16.4 & 118 & 10.2 \\
Difference-in-Differences & 958 & 12.8 & 1 & 2.1 & 25 & 9.1 & 193 & 16.7 \\
Instrumental Variables & 809 & 10.8 & 6 & 12.8 & 41 & 15.0 & 70 & 6.1 \\
Lab Experiment & 729 & 9.7 & 10 & 21.3 & 45 & 16.4 & 57 & 4.9 \\
Field Experiment & 526 & 7.0 & 5 & 10.6 & 17 & 6.2 & 50 & 4.3 \\
Matching and Weighting & 406 & 5.4 & 2 & 4.3 & 31 & 11.3 & 58 & 5.0 \\
RDD & 363 & 4.8 & 0 & 0.0 & 8 & 2.9 & 57 & 4.9 \\
Synthetic Control & 92 & 1.2 & 0 & 0.0 & 3 & 1.1 & 17 & 1.5 \\
\bottomrule
\end{tabular}
\tabnote{The first two rows are shares of the explanatory quantitative papers in each column. The per-method rows are shares of the design-based papers in each column, matching the denominator of Figure \ref{fig:bymethods2}. Papers combining two or more design-based methods count under each method they use, so the per-method shares can sum to slightly more than 100.}
\end{table}

\paragraph*{Design-based share by journal (Figure 7).} Table \ref{tab:byjournal} reports the share of design-based papers for each of the top 20 journals that underpin Figure \ref{fig:journals} in the main text, sorted by the proportion of design-based studies. Consistent with the figure, the rightmost columns exclude survey experiments from the sample.

\begin{table}[htbp]
\centering
\caption{Design-based Share by Journal (Top 20)}
\label{tab:byjournal}
\footnotesize
\resizebox{\textwidth}{!}{
\begin{tabular}{@{}lrrccccc@{}}
\toprule
Journal & SJR rank & N & \multicolumn{2}{c}{All designs} & \multicolumn{2}{c}{Excl.\ survey exp.} \\
\cmidrule(lr){4-5} \cmidrule(lr){6-7}
 & & & \% DB & SE & \% DB & SE \\
\midrule
Quarterly Journal of Political Science & 10 & 128 & 64.8 & (0.04) & 61.2 & (0.04) \\
Political Science Research and Methods & 16 & 218 & 62.4 & (0.03) & 49.4 & (0.04) \\
Political Analysis & 4 & 51 & 60.8 & (0.07) & 54.5 & (0.08) \\
American Political Science Review & 1 & 635 & 54.6 & (0.02) & 47.8 & (0.02) \\
Public Opinion Quarterly & 19 & 328 & 54.0 & (0.03) & 29.1 & (0.03) \\
Political Behavior & 13 & 774 & 47.7 & (0.02) & 28.2 & (0.02) \\
American Journal of Political Science & 2 & 929 & 42.6 & (0.02) & 33.9 & (0.02) \\
Journal of Politics & 7 & 1,322 & 42.6 & (0.01) & 33.0 & (0.01) \\
Political Communication & 18 & 378 & 41.8 & (0.03) & 23.4 & (0.02) \\
British Journal of Political Science & 9 & 650 & 31.8 & (0.02) & 20.8 & (0.02) \\
International Organization & 3 & 281 & 27.8 & (0.03) & 20.3 & (0.02) \\
World Politics & 5 & 181 & 27.6 & (0.03) & 22.4 & (0.03) \\
Journal of Conflict Resolution & 11 & 850 & 26.6 & (0.02) & 19.7 & (0.01) \\
Comparative Political Studies & 8 & 864 & 26.2 & (0.01) & 19.2 & (0.01) \\
Journal of Public Administration Research and Theory & 6 & 467 & 24.8 & (0.02) & 16.4 & (0.02) \\
Journal of Peace Research & 12 & 683 & 18.7 & (0.01) & 13.6 & (0.01) \\
International Studies Quarterly & 15 & 731 & 17.2 & (0.01) & 10.6 & (0.01) \\
European Journal of Political Research & 14 & 604 & 16.7 & (0.02) & 9.0 & (0.01) \\
European Union Politics & 17 & 417 & 16.1 & (0.02) & 9.7 & (0.01) \\
Journal of European Public Policy & 20 & 400 & 14.8 & (0.02) & 8.2 & (0.01) \\
\bottomrule
\end{tabular}}
\end{table}

\paragraph*{Citation Difference}\label{sisec:citeprem}

In Figure~\ref{fig:citemeansoverall}, we report evidence of a citation gap between design-based and model-based studies. Here we complement that evidence by constraining the comparison. We account for subfield and publication-year fixed effects, and estimate the following regression model:
\begin{equation}
\text{Citations}_i = \beta , D_i + \alpha_s + \alpha_t + \varepsilon_i
\end{equation}
where $D_i = 1$ if paper $i$ employs a design-based method and zero otherwise. We include fixed effects $\alpha_s$ and $\alpha_t$ for subfields and publication year. We do not condition on journal placement, because publication venue may itself be part of the professional rewards associated with methodological choice. Standard errors are clustered at the journal level. The coefficient $\beta$ captures the average difference in accumulated citations between design-based and model-based papers published in the same year and subfield. We estimate the model on the pooled sample and separately for the three publication bins of Figure~\ref{fig:citemeansoverall}(b). 

\begin{table}[!htbp]
\centering
\caption{\label{tab:citation_regression} Citation Difference between Design-based and Model-based Studies}
\footnotesize
\begin{tabular}{lcccc}
\midrule
             & 2003--2018 & 2003--2008 & 2009--2013 & 2014--2018 \\
Model:       & (1)        & (2)        & (3)        & (4)\\
\midrule
Design-based & 11.9       & 15.1       & 15.7       & 8.3\\
             & (4.63)     & (10.9)     & (5.77)     & (3.06)\\
\midrule \emph{Fixed-effects} &   &   &   &  \\
Subfield     & Yes        & Yes        & Yes        & Yes\\
Year         & Yes        & Yes        & Yes        & Yes\\
\midrule \emph{Fit statistics} &   &   &   &  \\
Observations & 17,883     & 3,673      & 6,008      & 8,202\\
R$^2$        & 0.067      & 0.012      & 0.027      & 0.024\\
\midrule
\end{tabular}
\tabnote{The dependent variable is the total number of citations a paper has accumulated by the end of the observation window. Design-based is an indicator equal to one for papers using a design-based method. Standard errors clustered by journal are in parentheses. The sample comprises explanatory quantitative papers published between 2003 and 2018.}
\end{table}

We report the results in Table~\ref{tab:citation_regression}. The estimates are positive in every column. In the pooled sample, design-based papers receive 11.9 more citations than model-based papers published in the same year and subfield. The estimate exceeds the raw difference of 6.2 citations because design-based papers are concentrated in recent years, which have had less time to be cited. The period estimates are 15.1 additional citations for papers published between 2003 and 2008, 15.7 for papers published between 2009 and 2013, and 8.3 for papers published between 2014 and 2018. The period estimates are not statistically distinguishable from each other, and the estimate for the earliest period is imprecise. 

\subsubsection{Other results}

\begin{figure}[!ht]
    \centering
    \caption{Share of Each Design-based Method by Journal Ranking}
    \label{fig:linestopjournals}
    \begin{subfigure}[b]{0.95\textwidth}
        \centering
        \includegraphics[width=\linewidth]{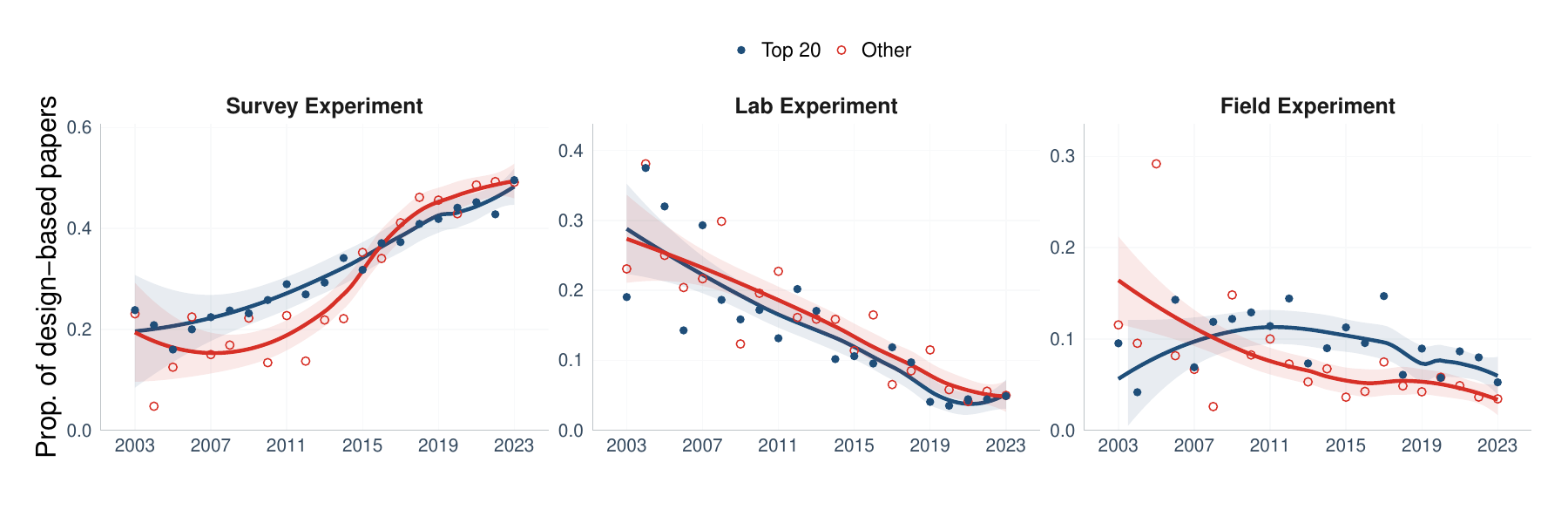}
        \caption{Experimental}
        \label{fig:experimentslinestop20}
    \end{subfigure}
    \hfill
    \begin{subfigure}[b]{0.95\textwidth}
        \centering
        \includegraphics[width=\linewidth]{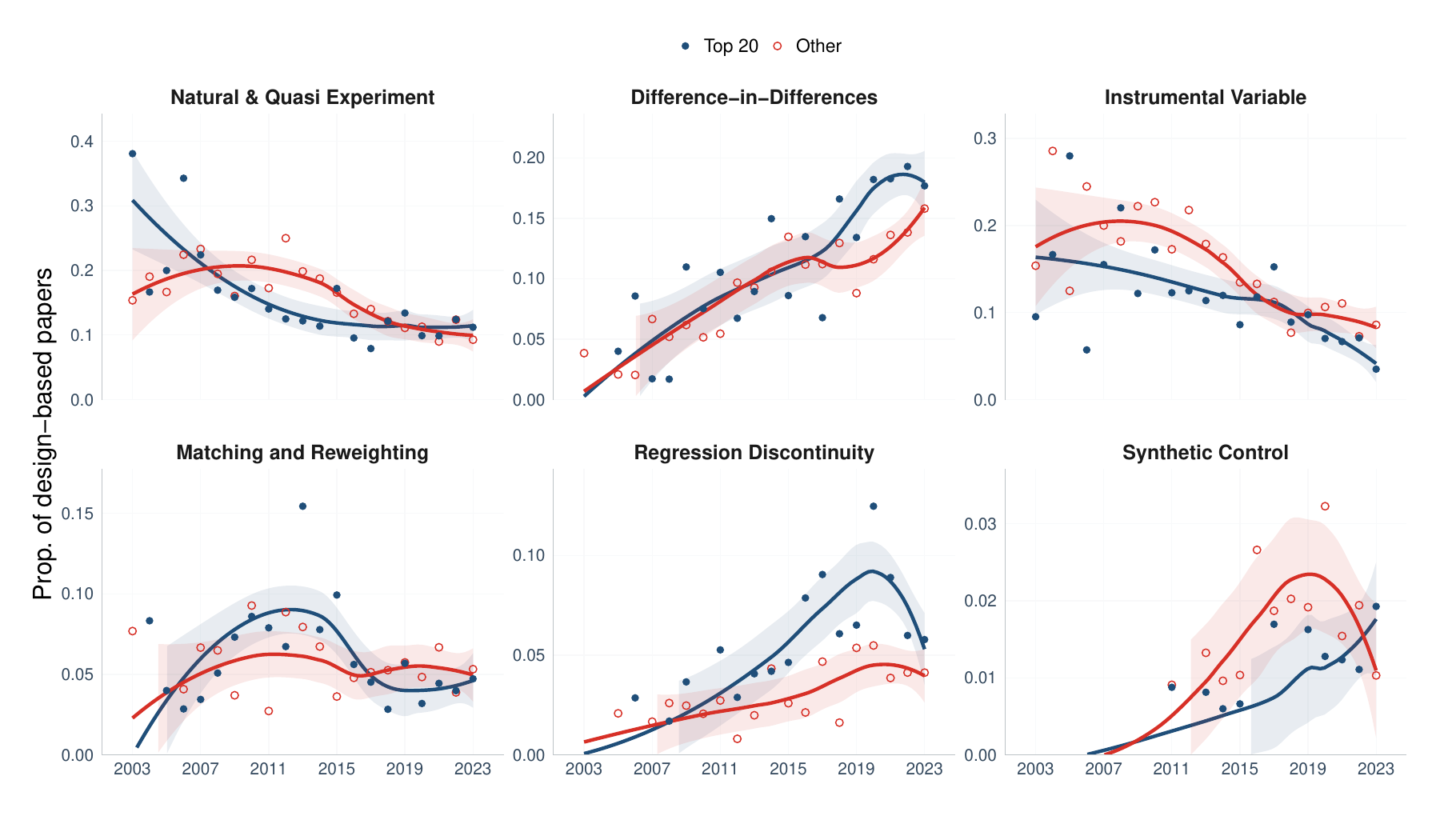}
        \caption{Non-experimental}
        \label{fig:notexperimentslinestop20}
    \end{subfigure}
    \fignote{The denominator is the total number of design-based papers published each year. Top 20 journals are shown in blue; all other journals in red.}
\end{figure}

In the main text, we show that design-based research has spread unevenly across journals and institutions. Figure~\ref{fig:linestopjournals} examines which methods drive this shift and how their adoption differs between the top 20 journals and all other outlets. For most design-based methods, top journals adopt them earlier and at higher rates, suggesting that influential outlets have played a disproportionate role in shaping causal inference practices in the discipline. The broader shift is driven especially by survey experiments and difference-in-differences, with high-impact journals consistently at the leading edge.

\subsubsection*{Journal tiers without weighting}

\begin{figure}[!ht]
    \centering
        \caption{Design-based and model-based studies by journal tier, unweighted}
    \begin{subfigure}[b]{0.47\textwidth}
        \centering
        \includegraphics[width=\linewidth]{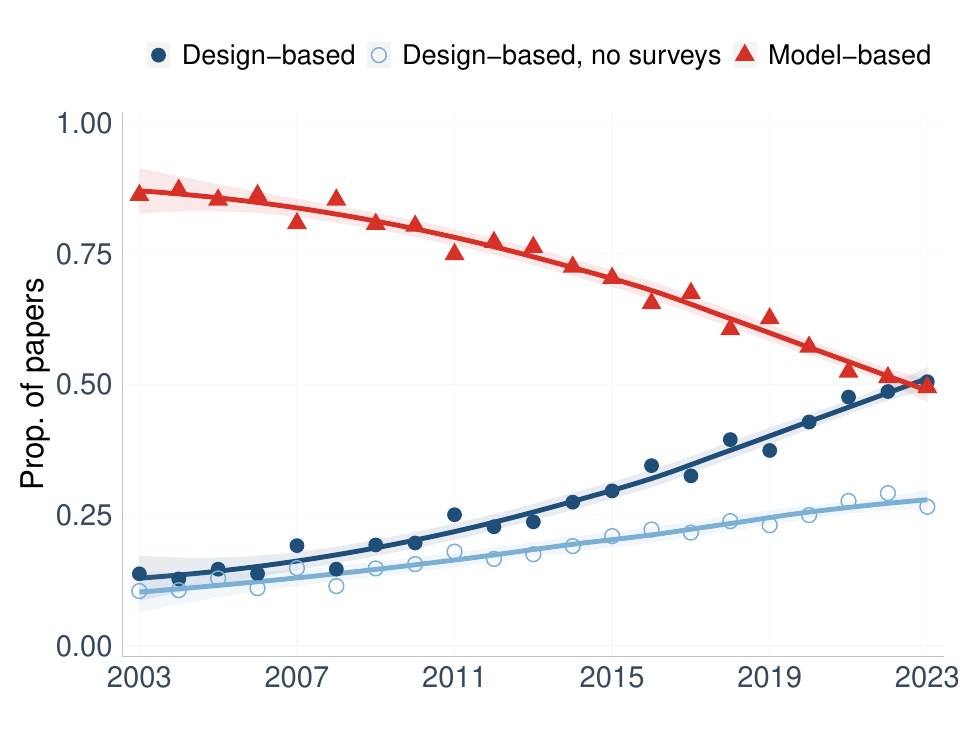}
        \caption{Studies published in top 20 journals}
        \label{fig:unw_top20}
    \end{subfigure}
    \hfill
    \begin{subfigure}[b]{0.47\textwidth}
        \centering
        \includegraphics[width=\linewidth]{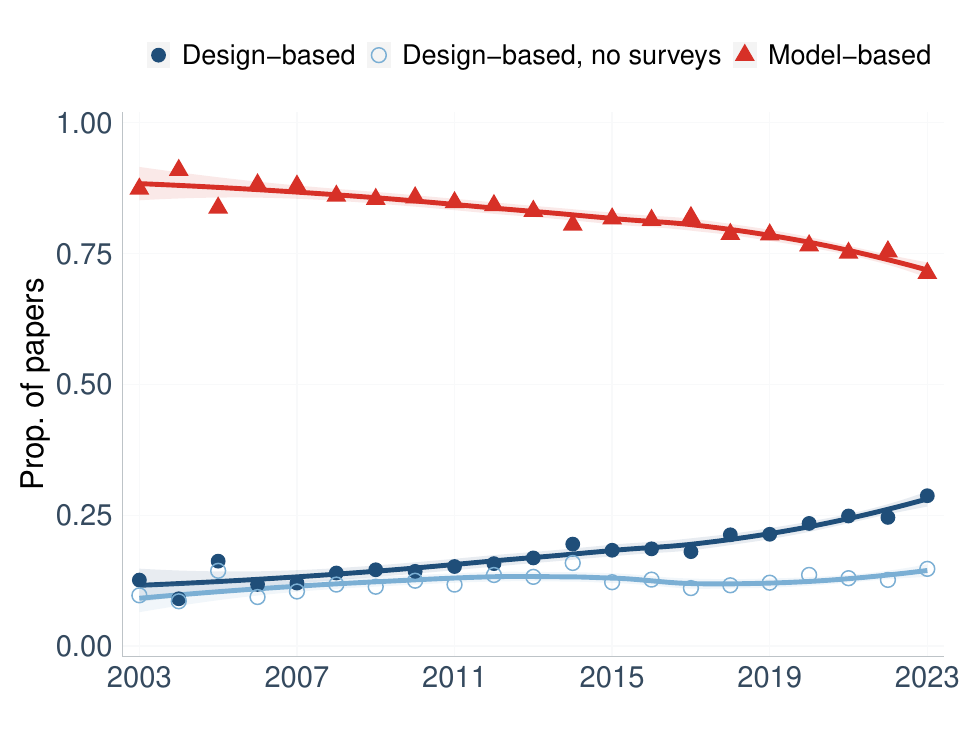}
        \caption{Studies published in other journals}
        \label{fig:unw_rest}
    \end{subfigure}
    \label{fig:unweightedtiers}
    \fignote{Shares are shown both including and excluding survey experiments. The denominator is the total number of explanatory empirical quantitative papers published each year. The left panel shows papers in the top 20 journals; the right panel shows papers in other journals. Curves with 95\% confidence intervals are loess fits.}
\end{figure}

Figure~\ref{fig:unweightedtiers} shows that the journal-tier pattern does not depend on SJR weighting. When every paper receives equal weight, design-based research still grows more quickly in the top 20 journals than elsewhere, rising from 14\% in 2003 to 51\% in 2023. It reaches parity with model-based work only at the end of the period, rather than in 2021 under SJR weighting. In the remaining journals, the design-based share rises from 13\% to 29\%; excluding survey experiments, it rises only from 10\% to 15\%. Weighting therefore accentuates an existing divide rather than creating it.

\end{document}